\documentclass[twocolumn,nofootinbib]{revtex4-2}
\usepackage{amsfonts}
\usepackage{amsmath}
\usepackage{mathrsfs}  
\usepackage{graphicx}
\usepackage{amssymb}
\usepackage{xcolor}
\usepackage{float}
\usepackage{siunitx}

\usepackage{tensor}

\newcommand{\ok}{\mbox{$k_{\perp}$}}
\newcommand{\pk}{\mbox{$k_{\parallel}$}}

\newcommand{\kp}{\mbox{$k_{\parallel}$}}

\newcommand{\ik}{\frac{ik_{\parallel}}{a_1}}

\newcommand{\okk}{\mbox{$k^2_{\perp}$}}
\newcommand{\pkk}{\mbox{$k^2_{\parallel}$}}
\newcommand{\eps}{\mbox{$\mathcal{E}$}}

\definecolor{blue}{rgb}{0,0,1}
\definecolor{green}{rgb}{0,0.8,0.3}
\definecolor{turkos}{rgb}{0,0.4,0.4}
\definecolor{red}{rgb}{1,0,0}
\definecolor{darkblue}{rgb}{0,0,0.5}
\definecolor{darkgreen}{cmyk}{1,0,1,0}
\definecolor{green0}{rgb}{0,1,0}

\newcommand{\rka}{\frac{\mathcal{R}a_2^2-k_{\perp}^2}{2a_2}}

\newcommand{\okt}{\mbox{$k_{\perp}^2$}}
\newcommand{\pkt}{\mbox{$k_{\parallel}^2$}}

\begin{document}

\title{General Perturbations of Homogeneous  and Orthogonal \\
Locally Rotationally Symmetric Class II Cosmologies\\
with Applications to Dissipative Fluids
}
\author{Philip Semr\'en$^{1}$ and Michael Bradley$^{1}$} 
\address{$^{1}$\textit{Department of Physics, Ume{\aa } University, Ume\aa, Sweden,} \\
philip.semren@umu.se, michael.bradley@umu.se}
\date{\today }

\begin{abstract}
First order perturbations of homogeneous and hypersurface orthogonal LRS (Locally Rotationally Symmetric) class II cosmologies with a cosmological constant are considered in the framework of the 1+1+2 covariant decomposition of spacetime. The perturbations, which are for a general energy-momentum tensor, include scalar, vector and tensor modes  and extend some previous works where matter was assumed to be a perfect fluid. Through a harmonic decomposition, the system of equations is then transformed to evolution equations in time and algebraic constraints. 
This result is then applied to dissipative one-component fluids, and on using the simplified acausal Eckart theory
the system is reduced to two closed subsystems governed by four and eight harmonic coefficients for the odd and even sectors respectively. The system is also seen to close in a simplified causal theory. It is then demonstrated, within the Eckart theory, how vorticity can be generated from viscosity. 
\end{abstract}

\maketitle

\section{Introduction}\label{Intro}

Even if most observations indicate that the universe on the large scale is homogeneous and isotropic,
and to high accuracy described by the  $\Lambda$CDM model \cite{WMAP9yr,Planck2,Pantheon}, the upper limits on anisotropy from redshift measurements are rather high  \cite{H1,H2,Dec}.
Hence it is of interest to see how different types of perturbations, like density fluctuations, gravitational waves, rotation of matter etc., evolve and interact on an anisotropic background. For some earlier works on perturbations in anisotropic universes see e.g. \cite{Doroschkevich,Perko,Tomita,Gumruk,Periera,Pitrou}.
As a first generalization away from isotropy, without introducing the full complexity of general anisotropic models, we will in the following consider backgrounds with one direction of anisotropy.
In some earlier papers  \cite{GWKS,BFK,TornkvistBradley} we studied perfect fluid perturbations of homogeneous and Locally Rotationally Symmetric (LRS) cosmologies of class II
using a perturbation method based on the 1+1+2 covariant split of spacetime \cite{1+1+2,Schperturb,LRSIIscalar}, which is an extension of the 1+3
covariant split  \cite{cov1,Cargese,cov3,cov5,EMM}. 

In this work we extend the analysis to a general form of the energy-momentum tensor
\begin{displaymath}\label{EMdiss}
T_{ab}=\mu u_au_b+ p h_{ab}+q_a u_b+q_b u_a + \pi_{ab} \, ,
\end{displaymath}
where $u^a$ is the 4-velocity of some timelike observer, $\mu$ is the energy density, $p$ is the isotropic stress, $\pi_{ab}$ is the anisotropic stress, and $q^a$ is the energy 
flux relative $u^a$. For their definitions see section \ref{1+3+1+1+2}.
Depending on the physical situation, $u^a$ can be specified in different ways. For example, for a one-component perfect fluid a natural choice would be the 4-velocity of the fluid, whereas for a one-component imperfect fluid one can choose either the restframe of the fluid or the system where the energy-flow $q_a$ vanishes. The 1+1+2 split will then be made with respect to $u^a$ and a spacelike unit vector $n^a$,
which coincides with the direction of anisotropy on the background. As variables, we use the kinematic quantities of the $u^a$ and $n^a$ vectors, the electric and magnetic parts of the Weyl tensor, and the energy-momentum tensor, where the latter, through Einstein's equations, is equivalent to the Ricci tensor and the cosmological constant $\Lambda$. 
These objects are then subject to the Ricci identities for $u^a$ and $n^a$ and the Bianchi identities. The covariant derivative is projected into three parts, one timelike derivative along $u^a$, a spacelike derivative along $n^a$, and one more along the perpendicular directions. These derivatives satisfy certain commutation relations, corresponding to the commutators of the basis vectors. 

We will now use this formalism to study general first order perturbations on homogeneous, but anisotropic, cosmological  models of LRS class II. For the perturbed quantities we use covariant variables which are zero on the background. In this way we avoid the gauge problem of identifying spacetime points on the background and the perturbed spacetime \cite{StewartWalker}.
On applying the formalism, a system of first order ordinary differential equations in time and a system of first order partial differential equations are obtained for the background and perturbation variables respectively. After a harmonic decomposition of the perturbed system, it is transformed to a system of ordinary differential equations in time together with algebraic constraints. Some of the variables can then be solved for algebraically in terms of the others. After eliminating these, the system is reduced to a system of ordinary differential equations for some of the remaining variables, but in general some variables are still freely specifiable. The system can be made to close by imposing some specific physical theory. For example, for a barotropic perfect fluid the system is closed by imposing  the equation of state $p=p(\mu)$.   

After finding the general equations we will apply the formalism to one-component imperfect fluids. First we consider the simplified Eckart theory \cite{Eckart,MaartensThermodynamics}, which is acausal but should be a fair approximation
when relaxations times are short compared to the typical macroscopic time scale, which in turn is set by the expansion rate of the universe. The Eckart theory also goes over into the standard theory for imperfect fluids  in the non-relativistic limit. The system will close by imposing, apart from the equation of state as above, also the coefficients of bulk and dynamic viscosity and heat conductivity. In general, these are functions of two variables, say energy density $\mu$ and particle density ${\cal{N}}$, which for the latter we also have to include an evolution equation, like the equation of continuity, to close the system. As an application, we then show how vorticity can be generated from the dissipative terms. For works on vorticity perturbations in isotropic universes see e.g. \cite{Hawking, LuAnandaClarksonMaartens,Christopherson1,ChristophersonMalik,EllisBruniHwang,Novelloetal} , and for the generation of vorticity in N-body simulations see \cite{CLAD}.
Finally, we show how the system can be closed also in a simplified causal theory \cite{MaartensThermodynamics}.

The paper is organized as follows: In section \ref{1+3+1+1+2} a short summary of the 1+3 and 1+1+2 covariant splits is given, and some basic elements of relativistic thermodynamics are collected in section  \ref{Thermodynamics}. The properties of the background metric are given in section \ref{sectionbackground}, and the procedure for the perturbation theory together with the harmonic decomposition is summarized  in section \ref{sectionperturbations}. In section \ref{EvolutionEquations}, the even and odd sectors of the reduced set of evolution equations for the perturbed quantities are presented, and in section \ref{Eckartsystem} the closed system obtained on imposing the Eckart theory is found. Similarly, the closed system for a simplified causal theory is given in section \ref{seccausal}. As an example of an application, the generation of vorticity in a dissipative fluid is discussed in section \ref{secvorticity}.
In Appendix A some relations in the 1+1+2 formalism are collected.
The commutation relations between the different differential operators are given in Appendix B,
and properties of the harmonics in Appendix C. The linearized equations in the 1+1+2 formalism are given in Appendix D, and their harmonic expansions in Appendix E. Then, in Appendix F, 
24 harmonic coefficients are solved for algebraically in terms of 11 coefficients determined from evolution equations and 9 freely specifiable coefficients.

\section{The 1 + 3 and 1 + 1 + 2 Covariant Splits of Spacetime}\label{1+3+1+1+2}

\label{sectioncovariant}

Our background will be characterized by two vector fields,  one timelike 4-velocity $u^a$ with $u_a u^a=-1$ and one preferred spatial direction $n^a$ with $n_a n^a =1$, and hence a 1+1+2 covariant split of spacetime, \cite{1+1+2, Schperturb}, will be useful. This is an extension of the 1+3 covariant split, which is with respect to a timelike vector field $u^a$ \cite{Cargese, cov1}. We here briefly summarize these formalisms and give the definitions of the quantities and derivativies used in the following text. For details the reader is referred to \cite{Cargese, cov1,1+1+2,Schperturb}, and in appendix A some useful relations are given.  

In the 1+3 formalism tensors are split into timelike and spatial parts with the projection  operators $U^a_b=-u^au_b$ and $h_{ab}=g_{ab}+u_a u_b$ respectively, where $g_{ab}$ is the 4-metric. Similarly, the covariant derivative, $\nabla_a$, is split into a covariant time derivative and a covariant projected spatial derivative as
\begin{equation}
\tensor{\dot{T}}{_{a\cdots b}^{c\cdots d}} \equiv u^e\nabla_e\tensor{T}{_{a\cdots b}^{c\cdots d}}\; ,
\end{equation}
and
\begin{equation}
D_e\tensor{T}{_{a\cdots b}^{c\cdots d}} \equiv \tensor{h}{_a^f}\cdots\tensor{h}{_b^g}\tensor{h}{_p^c}\cdots\tensor{h}{_q^d}\tensor{h}{^r_e}\nabla_r\tensor{T}{_{f\cdots g}^{p\cdots q}}\; ,
\end{equation}

\noindent respectively.

Instead of the metric, we now use decompositions of the curvature tensor and the kinematic quantities of $u^a$ as 
variables. The curvature tensor is given in terms of the electric and magnetic parts of the Weyl tensor,  $E_{ab} \equiv C_{acbd}u^cu^d$ and $H_{ab} \equiv \frac{1}{2}\varepsilon_{ade}\tensor{C}{^d^e_b_c}u^c$ respectively 
\footnote{Here, the volume element on the 3-surfaces is given by 
$\varepsilon_{abc}\equiv \eta _{dabc}u^{d}\equiv 4!\sqrt{-g}\delta _{\lbrack d}^{0}\delta _{a}^{1}\delta _{b}^{2}\delta _{c]}^{3}u^{d}$}, 
and by the Ricci tensor, which, through Einstein's equations, is given by the cosmological constant $\Lambda$ and the energy-momentum tensor. In general the energy-momentum tensor can be decomposed as
\begin{equation}\label{EMdiss}
T_{ab}=\mu u_au_b+ p h_{ab}+q_a u_b+q_b u_a + \pi_{ab} \, ,
\end{equation}
where
$\mu \equiv T_{ab}u^au^b$ is the energy density, $ p \equiv \frac{1}{3}T_{ab}h^{ab}$ is the isotropic stress, $\pi_{ab}$ is the anisotropic stress, satisfying $\pi_{ab}u^b=0$ and $\pi^a_a=0$, and $q^a$ is the energy flux relative $u^a$, satisfying $q^a u_a=0$. 
Due to the Bianchi identities, the energy-momentum tensor also satisfies $\nabla_b T^{ab}=0$.

The covariant derivative of $u^a$ is given in terms of the kinematic quantities of $u^a$ as
\begin{equation}\label{kinematic13}
	\nabla_au_b=-u_aA_b+\frac{1}{3}\Theta h_{ab}+\sigma_{ab}+\omega_{ab}\, ,
\end{equation}

\noindent 
where we have introduced the acceleration $A_a \equiv u^b\nabla_b u_a$, the expansion $\Theta \equiv D_a u^a$, the shear $\sigma_{ab} \equiv D_{\langle a}u_{b \rangle}$, and the vorticity $\omega_{ab} \equiv D_{[a}u_{b]}$,  which also can be given in terms of the corresponding vector $\omega^a \equiv \frac{1}{2}\varepsilon^{abc}\omega_{bc}$. Here $T_{\langle ab \rangle} \equiv \left(\tensor{h}{_{(a}^c}\tensor{h}{_{b)}^d}-\frac{1}{3}h_{ab}h^{cd} \right)T_{cd}$ has the properties  $T_{\langle ab \rangle}= T_{\langle ba \rangle}$ and $h^{ab}T_{\langle ab \rangle}=0$, and square brackets denote antisymmetrization, i.e. $A_{[ab]}=-A_{[ba]}$. The variables are now subject to integrability conditions given by the Ricci identity for $u^a$, the Bianchi identities, and the commutators between the derivatives $\dot\;\equiv u^a\nabla_a$ and $D_a$, see  e.g. \cite{Cargese}, whereas Einstein's equations already are imposed by expressing the Ricci tensor in terms of the energy momentum tensor (\ref{EMdiss}) and the cosmological constant $\Lambda$.

Similarly, when we have one more preferred vector field $n^a$, which is assumed to be spacelike and normalized, we can make a further split into a covariant 1+1+2 formalism. This is then done with the  projection operators $n^a n_b$ and $\tensor{N}{_a^b}=\tensor{h}{_a^b}-n_a n^b$. In this way, 3-vectors are split into a scalar along $n^a$ and perpendicular 2-vectors, and 3-tensors are split into a scalar, a 2-vector and a symmetric and trace-free 2-tensor. The quantities from the 1+3 split then decompose in the following way as
 $A^a =  \mathcal{A}n^a+\mathcal{A}^a$, $\omega^a = \Omega n^a + \Omega^a$,  $q^a = Qn^a+Q^a$ and $\sigma_{ab} = \Sigma\left(n_an_b - \frac{1}{2}N_{ab}\right)+ 2\Sigma_{(a}n_{b)}+\Sigma_{ab}$. The electric and magnetic parts of the Weyl tensor, $E_{ab}$ and $H_{ab}$, and the anisotropic part of the stress $\pi_{ab}$ are decomposed in the same way as the shear, in terms of the variables $\mathcal{E}$, $\mathcal{E}_a$, $\mathcal{E}_{ab}$, $\mathcal{H}$, $\mathcal{H}_a$, $\mathcal{H}_{ab}$ and $\Pi$, $\Pi_a$ and $\Pi_{ab}$ respectively. 

In a similar way which the covariant 4-derivative $\nabla_a$ was split, we can split $D_a$ into a derivative along $n^a$ and a projected derivative onto the 2-surfaces as
\begin{equation}
  \tensor{\widehat{T}}{_{a\cdots b}^{c\cdots d}} \equiv n^eD_e\tensor{T}{_{a\cdots b}^{c\cdots d}}\; ,
\end{equation}

\noindent and
\begin{equation}
  \delta_e\tensor{T}{_{a\cdots b}^{c\cdots d}}\equiv\tensor{N}{_e^j}\tensor{N}{_a^f}\cdots \tensor{N}{_b^g}\tensor{N}{_h^c}\cdots \tensor{N}{_i^d}D_j\tensor{T}{_{f\cdots g}^{h\cdots i}}\; ,
\end{equation}

\noindent respectively.

On decomposing $D_an_b$ and $\dot{n}_a$ analogously to (\ref{kinematic13}), new kinematic quantities are introduced as
\begin{equation}
	D_a n_b= n_a a_b + \frac{1}{2}\phi N_{ab} + \xi \varepsilon_{ab}+\zeta_{ab} \, ,
\end{equation}
and
\begin{equation}
	 \dot{n}^a = \mathcal{A}u^a+\alpha^a,
\end{equation}

\noindent where the acceleration of $n^a$, $a^a \equiv n^cD_cn^a$, the 2-sheet expansion $\phi \equiv \delta_a n^a$, the twisting of the 2-sheets $\xi \equiv \frac{1}{2}\varepsilon^{ab}\delta_an_b$, the shear of $n^a$, $\zeta_{ab} \equiv \delta_{\{ a} n_{b\}}$, and $\alpha_a \equiv \tensor{N}{_a^b}\dot n_b$.
Here $\varepsilon _{ab}\equiv \varepsilon _{abc}n^{c}$ and curly brackets denote the projected, trace-free and symmetric part of a 2-tensor in an analogous way to the  $\langle ab\rangle$ operation on 3-tensors. We also introduce the notation
$\psi_{\overline{a}}\equiv\tensor{N}{_a^b}\psi_{b}$ for projection onto the 2-sheets.

The previous integrability conditions for the 1+3 formalism now have to be completed with the Ricci identities  for $n^a$. For a full 1+1+2 split of the Ricci identities and the Bianchi identities into evolution equations and constraints, see e.g. \cite{1+1+2}, where also commutation relations between the different differential operators introduced can be found.

\section{Thermodynamics}\label{Thermodynamics}

To get an accurate description of relativistic non-equilibrium or dissipative thermodynamics the model should be based on a detailed kinetic theory. For some covariant and causal fluid descriptions valid close to equilibrium see, e.g. \cite{Israel, Carter, IsraelStewart,HiscockLindblom}. Even these are quite complicated and we will later turn to some simplifications, 
see e.g. \cite{MaartensThermodynamics,MaartensTriginer}.

Recall that the energy momentum tensor is given by equation (3) as
\begin{equation}\nonumber
T_{ab}=\mu u_au_b+ p h_{ab}+q_a u_b+q_b u_a + \pi_{ab} \, .
\end{equation}
In the following we will take $u^a$ to be the 4-velocity of matter (particle frame) when dealing with thermodynamics. Another possible choice is the so called energy frame, which makes $q^a=0$. 
For a dissipative fluid $p=\tilde p+p_{\zeta_B}$, where $\tilde p$ is the equilibrium pressure of the fluid and $p_{\zeta_B}$ represents the bulk viscosity.
Similarly $\pi_{ab}$ is the shear viscosity and the heat flow is given by $q^a$.

Now introduce also the particle flux ${\cal{N}}^a$ and entropy flux ${\cal{S}}^a$, given by
\begin{equation}
{\cal{N}}^a={\cal{N}}u^a\, ,
\end{equation}
and
\begin{equation}\label{entropyflux}
{\cal{S}}^a={\cal{S}}{\cal{N}}^a+\frac{R^a}{T}\, ,
\end{equation}
respectively, where ${\cal{N}}$ is the particle number density, ${\cal{S}}$ the specific entropy and $R^a$ a dissipative term satisfying $R^au_a=0$.
In line with \cite{IsraelStewart} we define the temperature $T$ from the Gibbs relation
\begin{equation}\label{Gibbs}
Td{\cal{S}}=d\left(\frac{\mu}{{\cal{N}}}\right)+\tilde p d\left(\frac{1}{{\cal{N}}}\right)\, ,
\end{equation}
which is subject to the integrability condition
\begin{equation}\label{ICT}
{\cal{N}}\left(\frac{\partial T}{\partial {\cal{N}}}\right)_\mu+\left(\mu+\tilde p\right)\left(\frac{\partial T}{\partial \mu}\right)_{\cal{N}}=T\left(\frac{\partial \tilde p}{\partial \mu}\right)_{\cal{N}}\, .
\end{equation}
The time evolution of the temperature $T$ is then given by  \cite{MaartensThermodynamics}
\begin{eqnarray}\nonumber
\dot T&=&-\left(\frac{\partial T}{\partial \mu}\right)_{\cal{N}}\left(\Theta p_{\zeta_B} +q^a_{;a}+A_a q^a+\sigma_{ab}\pi^{ab}\right) \\ \label{1lawTD}
&&-\Theta T\left(\frac{\partial \tilde{p}}{\partial \mu}\right)_{\cal{N}}\, .
\end{eqnarray}
From (\ref{Gibbs}) we also see that in general the equilibrium scalars $\mu$, $\tilde p$, ${\cal{N}}$, ${\cal{S}}$ and $T$ can be given in terms of just two of them, say $\mu$ and ${\cal{N}}$,
as $\tilde p(\mu, {\cal{N}})$, ${\cal{S}}(\mu, {\cal{N}})$ and $T(\mu, {\cal{N}})$.
Hence we will also need an evolution equation for ${\cal{N}}$, given by requiring particle conservation 
\begin{equation}\label{eqcontinuityTDb}
0=\nabla_a{\cal{N}}^a=\dot{\cal{N}}+\Theta{\cal{N}} \, .
\end{equation}
The evolution equation for the energy density is contained in the general set of equations obtained from the Ricci and Bianchi identities.
To get the theory consistent with the second law of thermodynamics one should also have
\begin{equation}\label{2lawTD}
\nabla_a{\cal{S}}^a\geq 0 \, ,
\end{equation}
which will put restrictions on the dissipative quantities $p_{\zeta_B}$, $\pi_{ab}$ and $q_a$.

\subsection{The Eckart Theory}\label{Eckartsection1}

A simple covariant, but acausal, theory was given by Eckart \cite{Eckart,MaartensThermodynamics}. It can give a good description when relaxation times are short compared to relevant time scales, such as the expansion rate of the universe.

With $R^a=q^a$, the heat flux, the expressions below are consistent with the second law of thermodynamics as stated in equation (\ref{2lawTD}).
The bulk viscosity is given by
\begin{equation}\label{bulk}
p_{\zeta_B}=-\zeta_B \Theta \, ,
\end{equation}
 in terms of the coefficient of bulk viscosity, $\zeta_B=\zeta_B(\mu,{\cal{N}})$.
Similarly $\pi_{ab}$, the shear viscosity, is given by 
\begin{equation}\label{visc}
\pi_{ab}=-2\eta\sigma_{ab} \, ,
\end{equation}
 in terms of the coefficient of dynamic viscosity, $\eta=\eta(\mu,{\cal{N}})$.  
 Finally, the heat flow is given by 
\begin{equation}\label{heat}
q^a=-\kappa \left(D^a T+T A^a\right) \, ,
\end{equation}
where $\kappa=\kappa(\mu,{\cal{N}})$ is the coefficient of thermal conductivity. 

These are natural relativistic generalizations of the corresponding non-relativistic laws. The main difference from them is that the acceleration appears in the heat equation.

\subsection{Simplified Causal Theory}\label{secsimplcausal}

A simplified, but causal, version of the theory developed in \cite{IsraelStewart,HiscockLindblom} can be obtained by neglecting couplings between viscosity and heat, as well as neglecting some terms which will be of higher order, see \cite{MaartensThermodynamics}. The equations (\ref{bulk})–(\ref{heat}) then become evolution equations, given by
\begin{eqnarray}\label{causaleq1}
\tau_{\zeta}\dot{ p}_{\zeta_B}+ p_{\zeta_B}&=&-\zeta_B \Theta \, , \\\label{causaleq2} 
\tau_{\pi}\dot{\pi}_{\langle ab\rangle }+\pi_{ab}&=&-2\eta\sigma_{ab} \, ,   \\\label{causaleq3} 
\tau_{q}\dot{q}^{\langle a\rangle}+q^a&=&-\kappa\left(D^a T+T A^a\right) \, ,
\end{eqnarray}
in terms of the relaxation times $\tau_{\zeta}$,  $\tau_{\pi}$ and $\tau_{q}$, which in general are assumed to be functions of $\mu$ and ${\cal{N}}$.

\section{Background spacetimes}
\label{sectionbackground}

To see the effects of anisotropy without going to the most general case, we consider for the backgrounds a class of cosmological models with local rotational symmetry (LRS),
 which has the flat Friedmann model as a special case. These are characterized by two vector fields, the 4-velocity $u^a$ and the symmetry axis $n^a$, and hence are suitable for 
a 1+1+2 split.
The models are homogeneous and hypersurface orthogonal and belong to the LRS class II, see e.g. \cite{LRS,MarklundBradley,BFK}, which is characterized by a vanishing 
magnetic part of the Weyl tensor $H_{ab}$, vanishing vorticity $\omega_{ab}$, and vanishing twist $\xi$ of the 2-sheets perpendicular to $n^a$. These sheets
are maximally symmetric with 2-curvature scalar ${\mathcal{R}}=2{\mathcal{K}}/a_2^2$, where $a_2(t)$ is the radius of curvature and ${\mathcal{K}}=\pm 1$ or 0 for spheres, pseudo-spheres or planes, respectively. 

If we also require $\phi=0$, which only excludes
the hyperbolic Friedmann Robertson-Walker models \cite{BFK}, all metrics in this class can, in terms of the scale factors $a_1(t)$ and  $a_2(t)$, be given as
\begin{equation}  \label{metric}
ds^{2}=-dt^{2}+a_{1}^{2}\left( t\right) dz^{2}+a_{2}^{2}\left( t\right)
\left( d\vartheta ^{2}+f_{\mathcal{K}}(\vartheta) d\varphi ^{2}\right) \ , 
\end{equation}%
where  $f_1(\vartheta)=\sin^2\vartheta$, $f_{-1}(\vartheta)=\sinh^2\vartheta$ and  $f_{0}(\vartheta)=1$  (or $f_0(\vartheta)=\vartheta^2$).

On using the covariant 1+1+2 split described in section \ref{1+3+1+1+2}, these spacetimes are specified by the following nonzero scalars
\begin{equation}
S^{(0)}= \{\mu,  p, {\mathcal{E}},\Theta,\Sigma,\Pi \} \, , 
\end{equation}
and the value of the  cosmological constant $\Lambda$.
The time evolutions of $\mu$, $\Theta$ and $\Sigma$ are given by 
\begin{equation}
\dot{\mu}=-\Theta \left( \mu + p\right)-\frac{3}{2}\Sigma\Pi \ ,  \label{continuity}
\end{equation}
\begin{equation}
\dot{\Theta}=-\frac{\Theta ^{2}}{3}-\frac{3}{2}\Sigma ^{2}-\frac{1}{2}\left(
\mu +3p\right) +\Lambda \ ,  \label{expansiondot}
\end{equation}
\begin{equation}
\dot{\Sigma}=\frac{2}{3}\left( \mu +\Lambda \right) +\frac{\Sigma ^{2}}{2}%
-\Sigma \Theta -\frac{2}{9}\Theta ^{2}+\Pi\ ,  \label{Sigmadot}
\end{equation}
whereas  ${\mathcal{E}}$ is given algebraically as
\begin{equation}
3\mathcal{E}=-2\left( \mu +\Lambda \right) -3\Sigma ^{2}+\frac{2}{3}\Theta
^{2}+\Sigma \Theta -\frac{3}{2}\Pi\ .  \label{EWeyl}
\end{equation}
The 2-curvature can be given
in terms of $S^{(0)}$ as
 \begin{equation}
\mathcal{R}=\frac{2{\mathcal{K}}}{a_{2}^{2}
}=2\left( \mu +\Lambda \right) +%
\frac{3}{2}\Sigma ^{2}-\frac{2\Theta ^{2}}{3}\, ,\   \label{2TimesGaussianCurv}
\end{equation}
and the scale factors are determined from the expansion and shear through
\begin{equation}
\Theta =\frac{\dot{a}_{1}}{a_{1}}+2\frac{\dot{a}_{2}}{a_{2}}\ ,
\label{Thetabackgroundexp}
\end{equation}
and
\begin{equation}
\Sigma =\frac{2}{3}\left( \frac{\dot{a}_{1}}{a_{1}}-\frac{\dot{a}_{2}}{a_{2}}%
\right) \; .  \label{Sigmabackgroundexp}
\end{equation}

A closed system is obtained if explicit expressions for $ p$ and $\Pi$ are given.  In the context of a dissipative one-component fluid, this is equivalent to specifying the equilibrium pressure $\tilde p$, the bulk viscosity $p_{\zeta_B}$, and shear viscosity $\Pi$, as functions of, e.g., the energy density $\mu$ and the number density ${\mathcal{N}}$. The evolution equation (\ref{eqcontinuityTDb})
\begin{displaymath}
\dot{\mathcal{N}}=-\Theta{\mathcal{N}}  \, ,
\end{displaymath}
for ${\mathcal{N}}$ then has to be added to the system. The time evolution of the corresponding  background temperature $T$ is  obtained from (\ref{1lawTD}) as
\begin{equation}\label{1lawTDb}
\dot T=-\Theta T\left(\frac{\partial \tilde{p}}{\partial \mu}\right)_{\cal{N}}-\left(\frac{\partial T}{\partial \mu}\right)_{\cal{N}}\left(\Theta p_{\zeta_B} +\frac{3}{2}\Sigma\Pi\right) \, .
\end{equation}

For the case of a dissipative fluid, we should also note that since we are allowing for non-zero $p_{\zeta_B}$ and $\Pi$, the background spacetime will not describe a thermodynamic equilibrium. Since the Eckart theory and its generalizations can be seen as mainly being valid close to thermodynamic equilibrium \cite{HiscockLindblom}, we should therefore keep in mind that we in principle ought to require that $p_{\zeta_B}$ and $\Pi$, although possibly non-zero, are small. On applying the Eckart theory on an expanding and anisotropic background, this is most naturally enforced by choosing suitably small coefficients of viscosity.

\section{Perturbations}
\label{sectionperturbations}

Perturbations on the backgrounds of section \ref{sectionbackground} were studied in \cite{GWKS,BFK,TornkvistBradley}. In these works the energy-momentum tensor was assumed to be that of a barotropic perfect fluid, both for the background and the perturbations.
 Here these studies will be extended 
to the most general first order perturbations, so  
at this stage we do not assume anything about the quantities $p$, $\pi_{ab}$ and $q_a$, but first later on specialize to dissipative fluids.

The covariant split of spacetime is advantageous for perturbative calculations. In this approach one uses covariant quantities which are zero on the background as the perturbed variables.
Hence, according to the  Stewart-Walker lemma \cite{StewartWalker}, one avoids the gauge problem, i.e. of how to identify points on the perturbed spacetime with points on the background, see e.g. \cite{Bardeen} for another method to construct gauge invariant objects and \cite{Lifshitz,Stewart,Hawking,Olson} for more perturbation methods in cosmology.

The set of covariant variables we will use are the scalars which specify the background, $G=\{\mu, p, {\mathcal{E}}, \Theta, \Sigma, \Pi\}$,
and the other 1+1+2 covariant quantities defined in section \ref{sectioncovariant}. The latter are zero on the background and hence are gauge-invariant
and suitable to describe the perturbations.
Later on, when specializing to dissipative fluids, we will also introduce the scalars ${\cal{N}}$ and $T$, which as well as the elements in $G$ are nonzero also to zeroth order.
When going to first order, the elements in $G$ can be expanded as $G=G^{(0)}+G^{(1)}$, where $G^{(0)}$ are the quantities to zeroth order. For example 
 $\mu=\mu^{(0)}+\mu^{(1)}$ etc. 
A convenient and covariant way to represent the $G^{(1)}$'s is by using the gradients of the $G$'s, i.e. by 
$\delta_a G$ and $\hat G$, which are zero on the homogeneous background. 
On performing a harmonic decomposition of the first order quantities, it is seen that 
the $\hat G$ derivatives may be expressed in terms of the $G_a\equiv\delta_a G$ \cite{GWKS,BFK,TornkvistBradley}. Hence, the perturbations of the 
background quantities will be represented by the $G_a$
\begin{gather}  \label{newvariables}\nonumber
\mu_a\equiv \delta_a\mu \, ,\;\; p_a\equiv \delta_a p \, ,\;\; W_a\equiv \delta_a\Theta, \\
 V_a\equiv \delta_a \Sigma \, ,\;\; X_a\equiv \delta_a{\mathcal{E}} \, ,\;\; Y_a\equiv \delta_a\Pi \, .
\end{gather}
The complete set of first order quantities which vanish on the background is then given by
\begin{equation}
\begin{aligned}
S^{(1)} & \equiv \{ X_{a},V_{a},W_{a},\mu _{a},p_{a}, Y_a, {\mathcal{A}},{
\mathcal{A}}_{a},\Sigma _{a},\Sigma _{ab},{\mathcal{E}}_{a},{\mathcal{E}}
_{ab}, \\
&  {\mathcal{H}},{\mathcal{H}}_{a},{\mathcal{H}}_{ab},a_{b},\alpha_a,\phi
,\xi ,\zeta _{ab},\Omega,\Omega_a, Q, Q_a, \Pi_a, \Pi_{ab}\} ~.
\end{aligned}
\end{equation}
When specializing to dissipative fluids, these quantities are supplemented with the first order quantities
\begin{equation}
Z_a\equiv \delta_a {\cal{N}}\, , \; T_a\equiv\delta_a T\, ,\; p_{\zeta a}\equiv\delta_a p_{\zeta_B} \, .
\end{equation}
So far there is a freedom in the choice of frame in the perturbed spacetime. For a one component fluid one could, for instance, let $u^a$ be the 4-velocity of matter.
 Nor is the direction $n^a$ of anisotropy well-defined in the perturbed spacetime, and we will later restrict the frame by requiring $a_a=0$. See  \cite{GWKS} for a discussion 
on how to fix the frame.

On imposing the Ricci identities for the vectors $u^a$ and $n^a$ and the Bianchi identities together with the commutator relations (see appendix \ref{commutation}),
evolution equations along $u^a$, propagation equations along $n^a$, and constraints are obtained for the covariant quantities. The exact 
equations are found in \cite{1+1+2}. The linearized equations for the first order quantities are given in Appendix \ref{sectionLE}.

The linearized partial differential equations are then transformed into ordinary differential equations by a harmonic decomposition
in terms of eigenfunctions $ P^{k_{\parallel }}=P^{k_{\parallel }}(z)$ and $Q^{k_{\perp }}=Q^{k_{\perp }}(\vartheta,\varphi)$,
where $k_{\parallel }$ are dimensionless comoving wave numbers in the $n^a$ direction and $k_\perp$ are the dimensionless comoving wavenumbers along
the 2-sheets. For their definitions and properties see Appendix \ref{subsectionharmonic}.
Scalars are decomposed as
\begin{equation}
\Psi =\displaystyle\sum\limits_{k_{\parallel },k_{\perp }}\Psi
_{k_{\parallel }k_{\perp }}^{S}\ P^{k_{\parallel }}\ Q^{k_{\perp }}\ ,
\end{equation}%
where $\Psi_{k_{\parallel }k_{\perp }}^{S}=\Psi_{k_{\parallel }k_{\perp }}^{S}(t)$.
From the scalar harmonics $Q^{k_{\perp }}$ even and odd vector and tensor harmonics are defined, see Appendix \ref{subsectionharmonic}.
In terms of them, vectors are expanded as
\begin{equation}
\Psi _{a}=\displaystyle\sum\limits_{k_{\parallel },k_{\perp
}}P^{k_{\parallel }}\ \left( \Psi _{k_{\parallel }k_{\perp
}}^{V}Q_{a}^{k_{\perp }}+\overline{\Psi }_{k_{\parallel }k_{\perp }}^{V} 
\overline{Q}_{a}^{k_{\perp }}\right) \, ,\   \label{harmexpV0}
\end{equation}
and tensors similarly as
\begin{equation}
\Psi _{ab}=\displaystyle\sum\limits_{k_{\parallel },k_{\perp
}}P^{k_{\parallel }}\ \left( \Psi _{k_{\parallel }k_{\perp
}}^{T}Q_{ab}^{k_{\perp }}+\overline{\Psi }_{k_{\parallel }k_{\perp }}^{T} 
\overline{Q}_{ab}^{k_{\perp }}\right) \, ,
\end{equation}
where all $\Psi^{V,T}_{k_{\parallel }k_{\perp }}$ and $\overline{\Psi}^{V,T}_{k_{\parallel }k_{\perp }}$ are functions of time.

We can now see how the hat derivatives, $\hat G$, of the elements $G$
may be expressed in terms of the $G_a$. Expand $\hat G$ and $G_a$ as
\begin{equation}\label{hatG}
\widehat{G} = \sum_{k_\parallel, k_\perp} \tilde{G}^S_{k_\parallel k_\perp} P^{k_\parallel}Q^{k_\perp} \, ,
\end{equation}
and
\begin{equation}
G_{a}=\displaystyle\sum\limits_{k_{\parallel },k_{\perp
}}P^{k_{\parallel }}\ \left( G_{k_{\parallel }k_{\perp
}}^{V}Q_{a}^{k_{\perp }}+\overline{G }_{k_{\parallel }k_{\perp }}^{V} 
\overline{Q}_{a}^{k_{\perp }}\right)\, , \   \label{harmexpV0}
\end{equation}
respectively.
From the even part of the commutation relation (\ref{commutatorA3}), the coefficients of $\hat G$ are then given in terms of the coefficients of $G_a$ as
\begin{equation}\label{hatGcoefficient}
	\tilde{G}^S_{k_\parallel k_\perp}=2a_2\dot{G}\overline{\Omega}^V_{k_\parallel k_\perp}+\frac{ia_2k_\parallel }{a_1}G^V_{k_\parallel k_\perp}  \ .
\end{equation}

Some other simplifying relations are also easy to find.
From the commutation relation (\ref{commutatorA4}) and the properties of the harmonics in appendix \ref{Qharmonics} it follows
that the odd coefficients of $G_a$ satisfy
\begin{equation}\label{oddgradcoeff}
\overline{G}^V_{k_\parallel k_\perp}=\frac{2a_2}{k_\perp ^2}\dot G\Omega^S_{k_\parallel k_\perp}\, , \ 
\end{equation}
and the degrees of freedom of the vorticity can also be decreased by substitution of the harmonic expansions of $\Omega$ and
$\Omega^a$ in propagation equation (\ref{OmegaVS}), which gives
\begin{equation}
\Omega^V_{k_\parallel k_\perp} = \frac{ia_2k_\parallel}{a_1k_\perp^2}\Omega^S_{k_\parallel k_\perp} \ .
\end{equation}

In the following sections and the appendices we will drop the indices $k_\parallel  k_\perp$ on harmonic coefficients, since their
meaning will be obvious due to the superscripts $S$, $V$ and $T$.

The harmonically decomposed system obtained from the first order system in Appendix \ref{sectionLE} can be found in Appendix \ref{LinearizedHarmonics}.
It is given in terms of evolution equations and constraints, and separates into an even and an odd sector. 
By solving the constraints for some of the harmonic coefficients in terms of the others and then plugging these back into the differential equations, new constraints or identities
are obtained. The procedure is continued until no more constraints are obtained. In this way
each sector can be reduced to evolution equations 
for a minimal set of harmonic coefficients, in terms of which the rest of the coefficients, together with some freely specifiable coefficients, are given algebraically,
see Appendix \ref{HarmonicCoefficients}.

\section{Evolution Equations}
\label{EvolutionEquations}

In this section we give the reduced systems, for the even and odd sectors respectively, before imposing the thermodynamic conditions from section \ref{Thermodynamics}. The systems obtained
on imposing the Eckart theory are given in the following section \ref{Eckartsystem} and the systems for a simplified causal thermodynamic theory are
found in section \ref{seccausal}.
The frame has been partially locked by requiring $a_a=0$. Later, when specializing to an imperfect fluid, we will choose $u^a$ to coincide with the 4-velocity of the fluid.

\subsection{Even parity sector}

Here the system is reduced to evolution equations for $\overline{\Omega}^V$, $\mu^V$, $\overline{\mathcal{H}}^{T}$, $\mathcal{E}^T$, $\Sigma^{T}$, $Q^V$ and $Q^S$. The coefficients $p^V$, $\mathcal{A}^V$,
$ \mathcal{A}^S$, $\Pi^V$, $\Pi^T$ and $Y^V$ are freely specifiable functions and the remaining quantities $\zeta^T$, $\mathcal{E}^V$, $\overline{\mathcal{H}}^{V}$,
$\Sigma^{V}$, $\alpha^V$, $W^V$, $V^V$, $X^V$ and $\phi^S$ are expressible in terms of the other coefficients, see section \ref{evencoeff}.

\begin{equation}\label{eqOmegaVgeneral}
    \dot{\overline{\Omega}}^V = -\left(\frac{\Sigma}{2}+\frac{2\Theta}{3} \right)\overline{\Omega}^V +\frac{1}{2a_2}\mathcal{A}^S-\frac{i\pk}{2a_1} \mathcal{A}^V,
\end{equation}

\begin{equation}
    \label{eq67}
    \begin{aligned}
&    \dot{\overline{\mathcal{H}}}^{T} =\frac{2}{a_2 B}\left(\tilde\mu+3\Pi 
\right)\left(\Sigma+\frac{\Theta}{3}\right)\overline{\Omega}^V-\frac{ik_{\parallel }}{a_{1}a_{2}B}\mu^{V} \\ 
    & -\frac{i a_{1}}{k_{\parallel}}\left(\frac{3\Pi}{2}+\frac{k_\parallel^2}{a_1^2}\left( 1-C\right)\right)\left( \mathcal{E}^{T}+\frac{\Pi^T}{2}\right)+\frac{ik_\parallel}{a_1}\Pi^T\\
    & -\frac{1}{3}\left( 4\Theta+3\Sigma- L\right) \overline{\mathcal{H}}^{T}-\frac{1}{a_2}\Pi^V+\frac{3ia_1}{2k_\parallel}\left(\Sigma+\frac{\Theta}{3}\right)\Pi\Sigma^T\\
    &-\frac{ik_\parallel}{a_1a_2B}\tilde F Q^V+\frac{1}{a_2^2B}\left(\Sigma+\frac{\Theta}{3}\right)Q^S ,
    \end{aligned}
\end{equation}

\begin{equation}\label{eqmu}
    \begin{aligned}
      &  \dot{\mu}^{V}  =-\left(\tilde\mu+3\Pi
\right)\left(\frac{2i\kp}{a_1}-\frac{a_1 \tilde\mu({\cal{R}}-\tilde k^2)
}{ik_{\parallel }B} \right)\overline{\Omega}^V \\
        &  + \left( \frac{\Sigma }{2}\left( 1-\frac{3}{B}\tilde\mu 
\right) -\frac{4\Theta }{3}\right) \mu^{V} + \\ 
        &  \frac{a_{2}}{2}\left(\tilde\mu  
\left( 1-C\right)+3\Pi \right)\left( B\Sigma^{T}-3\Sigma\left( \mathcal{E}^{T}+\frac{\Pi^T}{2}\right)\right)\vphantom{\frac{k_{\parallel}}{a_1}}  \\ 
          & +\frac{ia_1a_2}{2k_\parallel}\left(\tilde\mu
\left(\tilde k^2-\Sigma L\right)+6\Pi\frac{k_\parallel^2}{a_1^2}\right))\overline{\mathcal{H}}^{T}\\
&-\left(\Theta\left(\mu+p\right)+\frac{3\Sigma\Pi}{2}\right)\mathcal{A}^V-\Theta p^V-\frac{3\Sigma}{2}Y^V-\\
&\left(\tilde\mu+3\Pi
-\frac{k_\perp^2}{a_2^2}+\frac{3\Sigma}{2B}\tilde F
\tilde\mu
\right)Q^V-\\
&\frac{ia_1}{2a_2k_\parallel}\left(\tilde\mu\left(\frac{3\Sigma}{B}\left(\Sigma+\frac{\Theta}{3}\right)-1\right)+\frac{2k_\parallel^2}{a_1^2}\right)Q^S ,
    \end{aligned}
\end{equation}

\begin{equation}\label{eqE}
  \begin{aligned}
  &  \dot{\mathcal{E}}^{T}=-\frac{1}{2} \dot{\Pi}^T   + \frac{ia_1 }{a_2k_{\parallel}B}G\overline{\Omega}^V+\frac{3\Sigma }{2a_{2}B}\mu^{V}\\
    &-\frac{\tilde\mu
}{2}\Sigma^{T}+\left(\Sigma+\frac{\Theta}{3}\right)\Pi^T -\\
    &  \frac{3}{2}\left( F+\Sigma C\right)\left( \mathcal{E}^{T}+\frac{\Pi^T}{2}\right)-\frac{i a_{1}}{2k_{\parallel }}\left(\tilde k^2-\Sigma L\right)\overline{\mathcal{H}}^{T}+\\
    &\frac{3\Sigma\tilde F
-2B}{2a_2B}Q^V+\frac{ia_1\left(3\Sigma\left(\Sigma+\frac{\Theta}{3}\right)-B\right)}{2a_2 ^2k_\parallel B}Q^S,
    \end{aligned}
\end{equation}

\begin{equation}
	\dot{\Sigma}^T = \tilde F\Sigma^T+\frac{1}{a_2}\mathcal{A}^V-\mathcal{E}^T+\frac{1}{2}\Pi^T \, ,
\end{equation}

\begin{equation}\label{dotQV}
  \begin{aligned}
   & \dot Q^V=-p^V-\ik\Pi^V-\left(\tilde\mu+\frac{3}{2}\Pi\right)\mathcal{A}^V\\
   &+\frac{1}{2}Y^V-\left(\frac{4}{3}\Theta-\frac{1}{2}\Sigma\right)Q^V-\rka\Pi^T ,
  \end{aligned}    
\end{equation}

\begin{equation}\label{dotQS}
 \begin{aligned}
 &\dot Q^S =  -  \frac{i k_\parallel a_2}{a_1}\left(p^V +Y^V\right)- \left(\tilde\mu+3\Pi 
\right)\mathcal{A}^S+\\
&  a_2\left[3\Pi\left(\tilde F
+\frac{2}{B}\left(\Sigma+\frac{\Theta}{3}\right)\left(\tilde\mu+3\Pi
\right)\right)-2(\dot p+\dot\Pi)\right]\overline{\Omega}^V\\
&-\frac{3ik_\parallel a_2\Pi}{a_1 B}\left(\tilde F
Q^V+\mu^V\right)-\frac{ia_1a_2^2\Pi BL}{2k_\parallel}\Sigma^T\\
&+a_2^2\Pi L\overline{\mathcal{H}}^T+\left(\frac{3\Pi}{B}\left(\Sigma+\frac{\Theta}{3}\right)-\frac{4}{3}\Theta-\Sigma\right)Q^S\\
&+\frac{3ia_1a_2^2\Pi}{2k_\parallel}\left(\frac{k_\perp^2}{a_2^2}+2C\frac{k_\parallel^2}{a_1^2}\right)\left(\mathcal{E}^T+\frac{\Pi^T}{2}\right)
+\frac{k_\perp^2}{a_2}\Pi^V\, ,
 \end{aligned} 
\end{equation}

where the abbreviations

\begin{equation}
	\tilde{k}^2 \equiv \frac{\okt}{a_2^2} + 2\frac{\pkt}{a_1^2},
\end{equation}

\begin{equation}
    B\equiv \tilde{k}^2+\frac{9\Sigma ^{2}}{2}+3\mathcal{E}+\frac{3}{2}\Pi,
\end{equation}

\begin{equation}
  F\equiv \Sigma + \frac{2\Theta}{3} \; ,
\end{equation}

\begin{equation}
	\tilde{F} \equiv \Sigma - \frac{2\Theta}{3}\, ,
\end{equation}

\begin{equation}
	\tilde\mu \equiv \mu+p-2\Pi\, ,
\end{equation}

\begin{equation}
    CB\equiv -\frac{3}{2}\Sigma\tilde F-\frac{k_\perp^2}{a_2^2}\, ,
\end{equation}

\begin{equation}
	G\equiv\left(\tilde\mu+3\Pi 
\right)\left(\mathcal{R}-\tilde{k}^2 \right),
\end{equation}

\begin{equation}
	LB = 3\Sigma\left(\frac{\okk}{a_2^2} - \frac{\pkk}{a_1^2}\right) + \Theta\tilde{k}^2 \, ,
\end{equation}
have been used.

\subsection{Odd parity sector}

In the odd sector we obtain evolution equations for $\Omega^S$, $\mathcal{H}^{T}$, $\overline{\mathcal{E}}^T$, $\overline{Q}^V$. The coefficients $\overline{\mathcal{A}}^V$,
$\overline{\Pi}^V$ and $\overline{\Pi}^T$  are freely specifiable functions and the remaining quantities $\overline{\zeta}^T$, $\overline{\mathcal{E}}^V$, $\mathcal{H}^{S}$, $\mathcal{H}^{V}$,
$\overline{\Sigma}^{V}$, $\overline{\Sigma}^T$, $\overline{\alpha}^V$,  $\overline{V}^V$, $\overline{W}^V$, $\overline{X}^V$, $\overline{Y}^V$, $\xi^S$, $\Omega^V$,
$\overline{p}^V$ and $\overline{\mu}^V$ are expressible in terms of the others, see section \ref{oddcoeff}.

\begin{equation}\label{eqB55B}
  \begin{aligned}
   & \dot{\overline{Q}}^V=-\left(\frac{4}{3}\Theta-\frac{1}{2}\Sigma\right)\overline{Q}^V-\left(\tilde\mu+\frac{3}{2}\Pi
\right)\overline{\mathcal{A}}^V\\
   &+ \frac{a_2\left(\dot{\Pi}-2\dot p\right)}{k_\perp^2}\Omega^S+\rka\overline{\Pi}^T-\ik\overline{\Pi}^V ,
  \end{aligned}    
\end{equation}

\begin{equation}\label{dotOmegaS}
    \dot{\Omega}^S = \tilde F\Omega^S + \frac{\okt}{2a_2}\overline{\mathcal{A}}^V,
\end{equation}

\begin{equation}
\begin{aligned}
 & \dot{\mathcal{H}}^T=\frac{i a_1}{k_\parallel a_2 B}\left(\left(\Sigma+\frac{\Theta}{3}\right)\left(\tilde{k}^2+3\Pi\right)-3\Sigma
\frac{k_\parallel^2}{a_1^2}\right)Q^V+\\
&\frac{a_1}{2ik_\parallel B}\left(\left(\mathcal{R}-\tilde{k}^2\right)\left(\tilde{k}^2+3\Pi\right)-9\Sigma^2\frac{\pkk}{a_1^2}\right)
\left(\overline{\mathcal{E}}^T+\frac{1}{2}\overline{\Pi}^T\right)\\
&-\frac{i\pk}{a_1}\overline{\Pi}^T-\frac{1}{a_2}\overline{\Pi}^V
    -\frac{3}{2}\left(2E+F-2\frac{\Pi}{B}\left(\Sigma+\frac{\Theta}{3}\right)\right)\mathcal{H}^T\\
&+\left(S+U\right)\Omega^S,
\end{aligned}    
\end{equation}

\begin{equation}
  \begin{aligned}
   & \dot{\overline{\mathcal{E}}}^T=\frac{i\pk}{a_1}\left(1-D\right)\mathcal{H}^T-\frac{\dot{\overline\Pi}^T}{2} 
 +  P\Omega^S+\left(\Sigma+\frac{\Theta}{3}\right) \overline{\Pi}^T\\
& + \frac{1+D}{a_2}\overline{Q}^V
     -  \frac{3}{2}\left(F+\Sigma D\right)\left(\overline{\mathcal{E}}^T+\frac{\overline{\Pi}^T}{2}\right),
    \end{aligned}
\end{equation}

where
\begin{equation}
P\equiv  \frac{2}{k_\perp^2 B}\left(\tilde\mu+\frac{3}{2}\Pi \right)
\left(\tilde\mu-\frac{3}{2}\Sigma\tilde F+B
-\frac{k_\perp^2}{a_2^2}\right),
\end{equation}

\begin{equation}
S \equiv \frac{2ia_1}{3k_\parallel k_\perp^2 B}\left(\tilde\mu+\frac{3}{2}\Pi
\right)\left(3\Sigma \left( \frac{k_\perp^2}{a_2^2}-\frac{k_\parallel^2}{a_1^2}\right)+\Theta\tilde{k}^2 \right),
\end{equation}

\begin{equation}
U \equiv \frac{3ia_1\Pi}{k_\parallel k_\perp^2 B}\left(\Sigma+\frac{\Theta}{3}\right)\left(2\tilde\mu+3\Pi+B\right),
\end{equation}

\begin{equation}
    EB\equiv\frac{\Sigma}{2}\left(CB-\mathcal{E}-\frac{\Pi}{2}\right)+\frac{\Theta\left(\mathcal{E}+\frac{\Pi}{2}\right)}{3},
\end{equation}

\begin{equation}
    DB\equiv CB+ \tilde\mu 
\; .
\end{equation}

\section{Evolution equations using the Eckart theory}\label{Eckartsystem}

In this section we assume the simplified Eckart theory, as described in section \ref{Eckartsection1}. Hence from equations (\ref{bulk}), (\ref{visc}) and (\ref{heat})
we have
\begin{eqnarray}\nonumber
p_{\zeta_B}&=&-\zeta_B \Theta\, , \;\pi_{ab}=-2\eta\sigma_{ab}\, , \\\nonumber
 q^a&=&-\kappa \left(D^a T+T A^a\right)\, ,
\end{eqnarray}
 in terms of the temperature $T=T(\mu,{\cal{N}})$ and the coefficients of bulk viscosity $\zeta_B=\zeta_B(\mu,{\cal{N}})$,
 dynamic viscosity, $\eta=\eta(\mu,{\cal{N}})$, and conductivity
 $\kappa=\kappa(\mu,{\cal{N}})$. The isotropic stress is given by $p=\tilde p + p_{\zeta_B}$ where the equilibrium pressure $\tilde p=\tilde p(\mu,{\cal{N}})$. These equations also have to be supplemented with the equation of continuity (\ref{eqcontinuityTDb})
for the particle density ${\cal{N}}$
\begin{equation}\nonumber
0=\nabla_a{\cal{N}}^a=\dot{\cal{N}}+\Theta{\cal{N}} \, .
\end{equation}
We saw in section \ref{EvolutionEquations} that the harmonic coefficients 
$p^V$, $\mathcal{A}^S$,
$ \mathcal{A}^V$, $\Pi^V$, $\Pi^T$ and $Y^V$ in the even sector and 
$\overline{\mathcal{A}}^V$,
$\overline{\Pi}^V$ and $\overline{\Pi}^T$ in the odd sector, corresponding to
the quantities 
$p_a=\tilde p_a + p_{\zeta a}$, ${\cal{A}}$, ${\cal{A}}_a$, $\Pi_a$, $\Pi_{ab}$ and $Y_a\equiv\delta_a\Pi$,
were freely specifiable. By imposing equations (\ref{bulk}), (\ref{visc}) and (\ref{heat}), we can solve for these quantities in terms of the others as
\begin{equation}\label{eckartp}
p_a=\left(\varrho_0-\Theta \beta_0\right)\mu_a+\left(\varrho_1-\Theta \beta_1\right)Z_a-\zeta_B W_a
\end{equation}
\begin{equation}\label{eckartA}
{\cal{A}}=-\frac{1}{\kappa T}\left(Q+\kappa \widehat T\right)\, , \;
{\cal{A}}_a=-\frac{1}{\kappa T}\left(Q_a+\kappa T_a\right)
\end{equation}
\begin{equation}\label{eckartPi}
\Pi_a=-2\eta \Sigma_a \, , \;\; \Pi_{ab}=-2\eta \Sigma_{ab} 
\end{equation}
\begin{equation}\label{eckartY}
Y_a=-2\eta V_a-2\Sigma\left(\varsigma_0 \mu_a+\varsigma_1 Z_a\right)
\end{equation}
where we have introduced the first order variables
\begin{equation}\label{eckartT}
Z_a\equiv \delta_a {\cal{N}}\, , \;\; T_a\equiv \delta_a T=\mathscr{T}_0\mu_a+\mathscr{T}_1 Z_a
\end{equation}
and the coefficients
\begin{eqnarray}\nonumber\label{coefficientsEckart}
\varrho_0\equiv\frac{\partial \tilde p}{\partial \mu} , \; \beta_0\equiv\frac{\partial \zeta_B}{\partial \mu} , \;
\varsigma_0\equiv\frac{\partial \eta}{\partial \mu} , \; \mathscr{T}_0\equiv\frac{\partial T}{\partial \mu}\, ,\\
\varrho_1\equiv\frac{\partial \tilde p}{\partial {\cal{N}}} , \; \beta_1\equiv\frac{\partial \zeta_B}{\partial {\cal{N}}} , \;
\varsigma_1\equiv\frac{\partial \eta}{\partial {\cal{N}}} , \; \mathscr{T}_1\equiv\frac{\partial T}{\partial {\cal{N}}}\, .
\end{eqnarray}
We also have the zeroth order equality
\begin{equation}
\Pi=-2\eta\Sigma\, .
\end{equation}
From equation (\ref{eqcontinuityTDb}) and the commutator between the dot derivative and $\delta_a$, one obtains the projected time evolution equation for $Z_a$
\begin{equation}\label{eckartZ}
\dot Z_{\overline{a}}=\dot{\cal{N}}{\cal{A}}_a-{\cal{N}}W_a+\left(\frac{\Sigma}{2}-\frac{4\Theta}{3}\right)Z_a \, .
\end{equation}
The corresponding even parity harmonic coefficients of the quantities $p_a$, ${\cal{A}}$ and ${{\cal{A}}_a}$, ${\Pi_a}$ and ${\Pi_{ab}}$, $Y_a$ and $T_a$  are then given by (\ref{eckartp}),  (\ref{eckartA}), (\ref{eckartPi}), (\ref{eckartY}) and  (\ref{eckartT}) respectively as

\begin{equation}
p^V = \left(\varrho_0-\Theta\beta_0\right)\mu^V +\left(\varrho_1-\Theta\beta_1\right)Z^V - \zeta_B W^V\, ,
\end{equation}

\begin{equation}
\frac{1}{a_2} \mathcal{A}^S = -\frac{1}{\kappa T}\left( \frac{1}{a_2}Q^S + \frac{i\pk\kappa}{a_1}T^V + 2\kappa\dot{T}\overline{\Omega}^V \right)\, ,
\end{equation}

\begin{equation}
\mathcal{A}^V = -\frac{1}{\kappa T}\left(Q^V+ \kappa T^V\right)\, ,
\label{eq:AvEckart}
\end{equation}

\begin{equation}
Y^V = -2\eta V^V - 2\left(\varsigma_0\mu^V + \varsigma_1 Z^V \right)\Sigma\, ,
\end{equation}

\begin{equation}
\Pi^V = -2\eta\Sigma^V     \, , \;\, \Pi^T = -2\eta\Sigma^T\, ,
\end{equation}

\begin{equation}\label{eqTV}
T^V = \mathscr{T}_0\mu^V + \mathscr{T}_1 Z^V\, ,
\end{equation}
whereas the evolution equation for the even component of $Z_a$ is given by
\begin{equation}
\dot{Z}^V  = \dot{\mathcal{N}}\mathcal{A}^V -\mathcal{N}W^V + \left( \frac{\Sigma}{2} -\frac{4\Theta}{3}\right)Z^V\, .
\end{equation}
Similarly, the odd parity coefficients $\overline{\mathcal{A}}^V $, $\overline\Pi^V$ and $\overline\Pi^T$ are given by
\begin{equation}
\overline{\mathcal{A}}^V = -\frac{1}{\kappa T}\left(\overline{Q}^V+ \kappa  \frac{2a_2\dot T}{k_\perp^2}\Omega^S\right)\, ,
\end{equation}
\begin{equation}
\overline{\Pi}^V = -2\eta\overline{\Sigma}^V\, , \; \overline{\Pi}^T = -2\eta\overline{\Sigma}^T\, ,
\end{equation}
whereas $\overline{p}^V$, $\overline{T}^V$, $\overline{Y}^V$ and ${\overline{Z}}^V$ as previously are determined from (\ref{oddgradcoeff})
\begin{eqnarray}\nonumber
\overline{p}^V = \frac{2a_2\dot p}{k_\perp^2}\Omega^S\, , \;\overline{T}^V =  \frac{2a_2\dot T}{k_\perp^2}\Omega^S \, ,\\\nonumber
\overline{Y}^V =  \frac{2a_2\dot \Pi }{k_\perp^2}\Omega^S\, , \;{\overline{Z}}^V  =  \frac{2a_2\dot {\cal{N}}}{k_\perp^2}\Omega^S\, .
\end{eqnarray}
The closed systems of evolution equations for the even and odd sectors respectively are given below.

\subsubsection{Even sector}
For the even sector, the following closed system of evolution equations for the first order coefficients  $Q^V$, $Q^S$, $\Sigma^T$, ${\cal{E}}^T$, $\overline{\cal{H}}^T$, $\mu^V$, $\overline\Omega^V$ and $Z^V$ is obtained

\begin{equation}
\begin{aligned}
	\dot{\overline{\Omega}}^V =  \frac{i\pk}{2a_1\kappa T}Q^V-\frac{1}{2a_2\kappa T}Q^S  - \left( \frac{2\Theta}{3} + \frac{\Sigma}{2} + \frac{\dot{T}}{T}\right)\overline{\Omega}^V \, ,
\end{aligned}
\label{eq:barOmegav_dot_Eckart}
\end{equation}

\begin{equation}
\begin{aligned}
	&\dot{Q}^V =
 \Bigg[\frac{\mathscr{T}_0 }{T}\left(\tilde\mu+\frac{3\Pi}{2}
\right) -\varrho_0+\Theta\beta_0 - \varsigma_0\Sigma +\frac{\Sigma}{B}\tilde\alpha  \Bigg]\mu^V  \\
	~~
&+ \Bigg[ \frac{\mathscr{T}_1 }{T}\left(\tilde\mu+\frac{3\Pi}{2}\right)   -\varrho_1+ \Theta\beta_1 - \varsigma_1\Sigma \Bigg]Z^V +\\
	~~ 
	 &\frac{ia_1}{\pk}\Bigg[ \frac{2\pkk}{3a_1^2}\left(\eta + 3\zeta_B \right)  + \frac{2G}{3B}\tilde\alpha\Bigg]\overline{\Omega}^V 
	+ \frac{ia_2\pk}{3a_1}\left(J-2\right)\tilde\alpha
\overline{\mathcal{H}}^T\\
	~~
	& +\frac{a_2}{3}\tilde\alpha\left(1-C\right)\left(3\Sigma\eps^T- \left(B + 3\eta\Sigma\right)\Sigma^T\right)\\
~~
 & -\frac{ia_1}{3a_2\pk B}\left( \tilde k^2 - {\cal{R}} \right)\tilde\alpha
Q^S +    \Bigg[ \frac{\Sigma}{2} - \frac{4\Theta}{3} - 2\eta  \\
	~~
	& + \frac{1}{\kappa T}\left(\tilde\mu+\frac{3\Pi}{2}\right) + \frac{2\tilde\alpha}{3B}
 \left(B + \frac{3\Sigma}{2}\tilde F
 \right) \Bigg]Q^V\, ,
	\end{aligned} \label{eq:dotQv_Eckart}
\end{equation}

\begin{equation}
\begin{aligned}
	& \dot{\Sigma}^T = \left(\tilde F - \eta\right)\Sigma^T  - \eps^T \\
 & -\frac{1}{a_2\kappa T}\left(Q^V + \kappa\left(\mathscr{T}_0 \mu^V + \mathscr{T}_1  Z^V\right) \right) \, ,
\end{aligned}
\end{equation}

\begin{equation}
	\begin{aligned}
	&\dot{\overline{\mathcal{H}}}^T = \frac{1}{a_2^2B}\left(\Sigma + \frac{\Theta}{3}\right)Q^S- \frac{i\pk}{a_1a_2B}\mu^V  - \frac{
	i\pk}{a_1a_2B}\tilde F Q^V \\
	& -\frac{i\pk}{a_1} \left(1- C\right)\left(\eps^T-\eta\Sigma^T\right)  -\Bigg[ 2\eta + \frac{3}{2}\left(F + \frac{M}{B}\right)\Bigg] \overline{\mathcal{H}}^T  \\
		~~
 &+ \frac{2}{a_2}\Bigg[  \frac{1}{B}\left(\tilde\mu+3\Pi\right)\left(\Sigma + \frac{\Theta}{3} \right)+ \eta \Bigg]\overline{\Omega}^V\, , 
	\end{aligned}
\end{equation}
\begin{equation}
	\begin{aligned}
	&  \dot{\eps}^T =  \Bigg[\dot{\eta} + \frac{3\eta}{2}\left( \Sigma C +\frac{\tilde{F}}{3}\right)- \frac{1}{2}\tilde\mu  -\eta^2  \Bigg]\Sigma^T \\
	~~
&-\Bigg[ \eta + \frac{3}{2} \left( F + \Sigma C\right)\Bigg]\eps^T +\frac{i\pk}{2a_1}\left(J-2\right)\overline{\mathcal{H}}^T \\
	& + \Bigg[\frac{3\Sigma}{2a_2B} - \frac{\eta\mathscr{T}_0}{a_2 T}  \Bigg]\mu^V + \frac{ia_1G}{a_2\pk B}\overline{\Omega}^V  - \frac{\eta\mathscr{T}_1}{a_2 T}   Z^V\\
	&- \frac{
	ia_1}{2a_2^2\pk B}\left( \tilde k^2 - {\cal{R}}\right)Q^S \\
	&- \Bigg[\frac{1}{a_2B}\left(B - \frac{3\Sigma}{2}\tilde F\right) + \frac{\eta}{a_2\kappa T} \Bigg]Q^V \, ,
	\end{aligned}
\end{equation}
\begin{equation}
	\begin{aligned}
		&  \dot{Q}^S = \Bigg[ - \frac{4\Theta}{3} - \Sigma -2\eta +\frac{1}{\kappa T}(\tilde\mu+3\Pi) + \\
&\frac{1}{3B}\left( \tilde k^2 - {\cal{R}}\right)\tilde\alpha\Bigg]Q^S+ \frac{2ia_2\pk}{3a_1 B} \left(B +\frac{3\Sigma}{2}\tilde F\right)\tilde\alpha Q^V\\
		~~
		 &+\Bigg[ { -\frac{12a_2\Sigma}{B}\eta \left(\left(\tilde\mu+3\Pi\right)\left(\Sigma+\frac{\Theta}{3}\right)+\frac{\tilde F B}{2}\right) }\\
&- 2a_2\left(\dot{\Pi} + \dot{p} \right) + (\tilde\mu+3\Pi) \frac{2a_2\dot{T}}{T}- \frac{2 \okk\eta}{a_2 }\\
		 ~~
&-\frac{4a_2}{3}\left(\tilde\alpha\frac{\pkk}{a_1^2} - \frac{G}{B}\tilde\beta\right)  \Bigg]\overline{\Omega}^V  
- \frac{a_2^2\pkk}{3a_1^2}\left(J-2\right)\tilde\alpha\overline{\mathcal{H}}^T\\
	~~
	&  +\frac{i\pk a_2^2}{3a_1}\tilde\alpha
\left(1-C\right)\left(3\Sigma\eps^T- \left(B + 3\eta\Sigma\right)\Sigma^T\right) \\
		~~
& +\frac{ia_2\pk}{a_1}\Bigg[(\tilde\mu+3\Pi)\frac{\mathscr{T}_0}{ T} +  \frac{\Sigma}{B}\tilde\alpha
+ 2\Sigma\varsigma_0 - \varrho_0 + \Theta\beta_0 \Bigg]\mu^V \\
 ~~
		&  + \frac{ia_2\pk}{a_1}\Bigg[ (\tilde\mu+3\Pi
) \frac{\mathscr{T}_1}{ T} + 2\varsigma_1\Sigma   -\varrho_1+ \Theta\beta_1 \Bigg]Z^V \, , \\
	\end{aligned}  \label{eq:dotQs_Eckart}
\end{equation}

\begin{equation}
\begin{aligned}
	&  \dot{\mu}^V =  - \Bigg[  \frac{3\Sigma}{2B}\tilde\gamma\tilde F+\tilde\gamma+6\Pi 
 - \frac{\okk}{a_2^2} +\frac{\dot{\mu}}{\kappa T}  \Bigg] Q^V- \\
&\Bigg[ \frac{3\Sigma}{2B}\tilde\gamma
- \frac{\Sigma}{2} + \frac{4\Theta}{3}  +\frac{\dot{\mu}\mathscr{T}_0}{T}
 + \Theta\left(\varrho_0-\Theta\beta_0 \right) - 3\Sigma^2\varsigma_0 \Bigg] \mu^V- \\
	~~
	&\frac{ia_1}{\pk}\Bigg[\frac{G}{B}\tilde\gamma +  \frac{2\pkk}{a_1^2}\left(\tilde\gamma+6\Pi \right)\Bigg]\overline{\Omega}^V
 -\frac{ia_2\pk}{2a_1}\Bigg[\left(J -2\right)\tilde\gamma-12\Pi \Bigg] \overline{\mathcal{H}}^T \\
  ~~
 &+\frac{ia_1}{2a_2\pk B}\Bigg[\tilde\gamma\left( \tilde k^2 - {\cal{R}}\right) - \frac{2\pkk B}{a_1^2}  \Bigg]Q^S\\
~~
&+ \Bigg[-\frac{\dot{\mu}\mathscr{T}_1}{ T} - \Theta\left(\varrho_1-\Theta\beta_1\right) + 3\Sigma^2\varsigma_1\Bigg] Z^V\\
~~
& + \frac{a_2}{2}\Bigg[\left(C - 1\right)\tilde\gamma-6\Pi \Bigg]\left(3\Sigma\eps^T- (B+3\eta\Sigma)\Sigma^T\right)\, ,\\
	\end{aligned}
\end{equation}

\begin{equation}
\begin{aligned}
&  \dot{Z}^V = {\mathcal{N}} \left(\frac{\mathscr{T}_0\Theta}{T}- \frac{3\Sigma}{2B} \right)\mu^V + \left(\frac{\Sigma}{2} - \frac{4\Theta}{3} + \frac{\mathscr{T}_1\Theta{\mathcal{N}}}{T}\right)Z^V\\
~~
&- \mathcal{N}a_2\Bigg[\frac{B}{2}\left(C-1\right)\Sigma^T + \frac{i\pk}{2a_1}\left(J-2\right)\overline{\mathcal{H}}^T +\\
	&\frac{ia_1}{a_2\pk}\left(\frac{2\pkk}{a_1^2} + \frac{G}{B}\right)\overline{\Omega}^V + \frac{3\Sigma}{2}\left(1-C\right)\left( \eps^T -\eta\Sigma^T \right)- \\
	&\frac{ia_1}{2a_2^2\pk B}\left(  \tilde k^2 - {\cal{R}}\right)Q^S \Bigg]
-{\mathcal{N}}\left(1-\frac{\Theta}{\kappa T}+\frac{3\Sigma\tilde F}{2B}
\right)Q^V \, ,
\end{aligned}
\label{eq:ZvEckart}
\end{equation}

where

\begin{equation}
\tilde\alpha\equiv 2\eta+\frac{3\zeta_B}{2},
\end{equation}
\begin{equation}
\tilde\beta\equiv 2\eta-\frac{3\zeta_B}{4},
\end{equation}
\begin{equation}
\tilde\gamma\equiv \mu+p+p_{\zeta_B}-4\Pi,
\end{equation}

\begin{equation}
M = 2\left(\eps + \frac{\Pi}{2}\right)\left(\Sigma + \frac{\Theta}{3}\right)  +\frac{\Sigma}{a_2^2}\left(\mathcal{R}a_2^2 - \okk \right)\, ,
\end{equation}

\begin{equation}
	JB = \frac{a_1^2\okk }{a_2^2\pkk}\left(\mathcal{R} - \tilde{k}^2\right)  - 3\Sigma\left(\Sigma - \frac{2\Theta}{3}\right) \, .
\end{equation}

\subsubsection{Odd sector}
Similarly, for the odd sector we obtain a closed system for the coefficients $\Omega^S$, $\overline Q^V$, ${\overline{\cal{E}}}^T$ and ${\cal{H}}^T$
\begin{equation}\label{OmegaSEckart}
	\dot{\Omega}^S = -\frac{\okk}{2a_2\kappa T}\overline{Q}^V + \left(\tilde{F} - \frac{\dot{T}}{T}\right)\Omega^S\, ,
\end{equation}

\begin{equation}
	\begin{aligned}
	\dot{\overline{\eps}}^T    = & -\frac{3}{2}\left( F + \Sigma \left(D-\mathcal{O}\right) + \frac{2\eta}{3}  \right) \overline{\eps}^T  +\\
&  \frac{i\pk}{a_1}\left(1-D+\mathcal{O}\right)\mathcal{H}^T  + \left(P + \frac{2\eta\dot{T}}{\okk T} - \mathcal{P}\right)\Omega^S  \\
&+ \frac{1}{a_2}\left(1+D-\mathcal{O}+ \frac{\eta}{\kappa T}\right)\overline{Q}^V \, ,\\
	\end{aligned}
\end{equation}

\begin{equation}
	\begin{aligned}
	\dot{\mathcal{H}}^T = & \Bigg[ S + U + \frac{2ia_1\eta}{\pk\okk}\bigg\{\frac{ \mathscr{B}}{B+3\eta\Sigma}\left(\tilde\mu+\frac{3\Pi}{2}
+ \frac{B}{2} \right ) \\
	~~
      &+\left(\tilde\mu+\frac{3\Pi}{2}\right)\left(1+\frac{N}{\mu + p}\right)  
      + B+\mathcal{R}-k^2  \bigg\}   \Bigg]\Omega^S \\
& +\frac{ia_1}{2\pk}\Bigg( \frac{4\pkk}{a_1^2} +\frac{B\mathscr{B}}{B+3\eta\Sigma}
 \Bigg)\overline{\eps}^T \\
	 &  - \Bigg[3 E + \frac{3}{2}F + 2\eta\left(1-\frac{2\pkk}{a_1^2B}\right)  -\frac{\eta\mathscr{B} }{B+3\eta\Sigma}\Bigg] \mathcal{H}^T \\
	 &+ \frac{ia_1}{a_2\pk B}\Bigg[\left( \Sigma+ \frac{\Theta}{3}\right)\left(3\Pi + \frac{\okk}{a_2^2}\right)  -\frac{\pkk}{a_1^2}\tilde{F} \\
	 ~
	 &+ 2\eta\left( \mathcal{R}-\frac{\okk}{a_2^2} +B\right ) + \frac{\eta B \mathscr{B}}{B+3\eta\Sigma} \Bigg]\overline{Q}^V \, ,
	\end{aligned}
\end{equation}

\begin{equation}\label{QbarVEckart}
	\begin{aligned}
		\dot{\overline{Q}}^V = & - \frac{a_2}{\okk}\Bigg[ 2\dot{p} - \dot{\Pi} - \frac{2\dot{T}}{T}\left(\tilde\mu+\frac{3\Pi}{2}\right)  \\
		~~
&+ 2\eta\Bigg ( B-\frac{\pkk}{a_1^2} +2\tilde\mu+3\Pi\Bigg) \Bigg]\Omega^S  \\
		~~
&+ \Bigg[  \frac{\Sigma}{2} -\frac{4\Theta}{3} -2\eta + \frac{1}{\kappa T}\left(\tilde\mu+\frac{3\Pi}{2} \right)   \Bigg]\overline{Q}^V\, ,
	\end{aligned}
\end{equation} 
where

\begin{equation}
      N\equiv \left(\mu  + p \right)\left(1+\frac{2}{a_2^2B}\left(\mathcal{R}a_2^2-k_\perp^2 \right) \right).
\end{equation}

\begin{align} 
\mathscr{B} \equiv -\frac{2\pkk}{a_1^2}\left(1+\frac{6\eta\Sigma}{B}\right) - CB + 9\Sigma E  
\end{align}

\begin{equation}
\mathcal{P} \equiv \frac{2\mathcal{O}}{\okk} \left(\mu + p + \frac{B-\Pi}{2} \right )\, , \
\end{equation}

\begin{equation}
\mathcal{O} \equiv \frac{2\eta}{B + 3\eta\Sigma}\left(\frac{\dot{\eta}}{\eta} - \eta +\frac{\tilde{F}}{2} + \frac{3}{2}\Sigma D  \right) \, ,   
\end{equation}

\section{Causal theories}\label{seccausal}

In the previous section we saw how the two systems closed in the simplified acausal Eckart theory. We will here see how the system is changed from the point of view of closure when introducing a causal theory. Note that the main difference is that the Eckart equations (\ref{bulk})–(\ref{heat}) are transformed into evolution equations,  as (\ref{causaleq1})–(\ref{causaleq3}) 
\begin{eqnarray}\nonumber
\tau_{\zeta}\dot{ p}_{\zeta_B}+ p_{\zeta_B}&=&-\zeta_B \Theta \\\nonumber
\tau_{\pi}\dot{\pi}_{\langle ab \rangle}+\pi_{ab}&=&-2\eta\sigma_{ab}   \\\nonumber
\tau_{q}\dot{q}^{\langle a\rangle}+q^a&=&-\kappa \left(D^a T+T A^a\right)\, ,
\end{eqnarray}
for the simplified theory given in  \ref{secsimplcausal}.
Since this structure of the equations
is kept in the slightly more elaborate theories in \cite{MaartensThermodynamics,MaartensTriginer}, it is sufficient to consider the simplified theory given by (\ref{causaleq1})–(\ref{causaleq3}) for the purpose of deciding if the systems close, provided all introduced thermodynamical coefficients are given as functions of, e.g., $\mu$ and ${\cal{N}}$.

On making a 1+2 split of the equations (\ref{causaleq1})–(\ref{causaleq3}) we first note that they give two evolution equations
for the zeroth order scalars $p_{\zeta_B}$ and $\Pi$
\begin{equation}\label{eqpzeta}
\tau_{\zeta}\dot p_{\zeta_B}+p_{\zeta_B}+\zeta_B\Theta=0\, ,
\end{equation}
\begin{equation}\label{eqPib}
\tau_{\pi}\dot\Pi+ \Pi+2\eta\Sigma=0 \, ,
\end{equation}
respectively. Hence there are two more dynamical variables already to zeroth order. Similarly as with the other zeroth order variables we now introduce the 2-gradient of $p_{\zeta_B}$ 
($Y_a\equiv \delta_a \Pi$ was already introduced in section \ref{sectionperturbations}) 
as a new first order variable
\begin{equation}
p_{\zeta a}\equiv \delta_a p_{\zeta_B}\, .
\end{equation}
From $p=\tilde p + p_{\zeta_B}$ one then has that
\begin{equation}
p_a=\tilde p_a+p_{\zeta a}=\varrho_0\mu_a+\varrho_1 Z_a+p_{\zeta a} \, ,
\end{equation}
where $\varrho_0$ and $\varrho_1$ are defined in (\ref{coefficientsEckart}).
By acting with $\delta_a$ on equations (\ref{eqpzeta}) and (\ref{eqPib}) and using the commutation relation (\ref{commutatorA2}), the evolution equations for $p_{\zeta a}$ and $Y_a$ are then obtained as

\begin{eqnarray}\nonumber
&&\tau_{\zeta}\dot p_{\zeta \overline{a}}=-\dot p_{\zeta_B}\left(\tau_{\zeta\mu}\mu_a+\tau_{\zeta{\cal{N}}}Z_a\right)
+\frac{\tau_{\zeta}}{2}\tilde Fp_{\zeta a}+\\\label{dotbarpzeta}
&&\tau_{\zeta} {\cal{A}}_a\dot p_{\zeta_B}-p_{\zeta a}-\left(\beta_0\mu_a+\beta_1 Z_a\right)\Theta-\zeta_B W_a \, ,
\end{eqnarray}
\begin{eqnarray}\nonumber
&&\tau_{\pi} \dot Y_{\overline{a}}=-\dot\Pi\left(\tau_{\pi\mu}\mu_a+\tau_{\pi{\cal{N}}}Z_a\right)+\frac{\tau_{\pi}}{2}\tilde F
Y_a+\\\label{dotbarZ}
&&\tau_{\pi} {\cal{A}}_a\dot\Pi-Y_a-2\left(\varsigma_0\mu_a+\varsigma_1Z_a\right)\Sigma-2\eta V_a \, ,
\end{eqnarray}
where
\begin{equation}
\tau_{\zeta\mu}\equiv\frac{\partial \tau_{\zeta}}{\partial \mu}\, , \quad \tau_{\zeta{\cal{N}}}\equiv\frac{\partial \tau_{\zeta}}{\partial {\cal{N}}}\, , \quad
\tau_{\pi\mu}\equiv\frac{\partial \tau_{\pi}}{\partial \mu}\, , \quad \tau_{\pi{\cal{N}}}\equiv\frac{\partial \tau_{\pi}}{\partial {\cal{N}}}\, .
\end{equation}
From (\ref{causaleq2}) also the following evolution equations for $\Pi^a$ and $\Pi^{ab}$ 
\begin{equation}
\tau_{\pi}\dot\Pi^{\overline a}=-\Pi^a-2\eta\Sigma^a \, ,
\end{equation}
\begin{equation}
\tau_{\pi}\dot\Pi^{\{ab\}}=-\Pi^{ab}-2\eta\Sigma^{ab} \, ,
\end{equation}
are obtained. The heat equation (\ref{causaleq3}), splits as
\begin{equation}\label{dotQ}
\tau_{q} \dot Q=-Q-\kappa\widehat T-\kappa T {\cal{A}} \, ,
\end{equation}
\begin{equation}\label{dotQa}
\tau_{q} \dot Q^{\overline a}=-Q^a-\kappa T^a-\kappa T {\cal{A}}^a\, ,
\end{equation}
where $T^a$ is defined from (\ref{eckartT}) and (\ref{coefficientsEckart}).
Finally, note that the evolution equations (\ref{eqcontinuityTDb}) and  (\ref{eckartZ}) for ${\cal{N}}$ and $Z_a\equiv\delta_a{\cal{N}}$ still apply.

Since we already have evolution equations for $Q$ and $Q^a$, or equivalently for the corresponding harmonic coefficients  $Q^S$, $Q^V$ and $\overline{Q}^V$ in the general systems from section \ref{EvolutionEquations}, equations (\ref{dotQ}) and (\ref{dotQa}) will be used to solve for ${\cal{A}}$ and ${\cal{A}}^a$. The corresponding harmonic coefficients ${\cal{A}}^S$, ${\cal{A}}^V$ and $\overline{\cal{A}}^V$ are then obtained from 

\begin{equation}
 \begin{aligned}
 &\left(\tilde\mu+3\Pi
-\frac{\kappa}{\tau_{q}}T\right)\mathcal{A}^S = -  a_2\ik \left(p^V +Y^V\right)\\
&+ \frac{2\kappa a_2}{\tau_{q}}\dot T \overline\Omega^V+\frac{i\kappa a_2 k_\parallel}{\tau_{q} a_1}T^V-2a_2(\dot p+\dot\Pi)\overline \Omega^V\\
&+3\Pi a_2\left(\tilde F
+\frac{2}{B}\left(\Sigma+\frac{\Theta}{3}\right)\left(\tilde\mu+3\Pi
\right)\right)\overline{\Omega}^V\\
&-\frac{3ik_\parallel a_2\Pi}{a_1 B}\left(\tilde F
Q^V+\mu^V\right)-\frac{ia_1a_2^2\Pi BL}{2k_\parallel}\Sigma^T\\
&+a_2^2\Pi L\overline{\mathcal{H}}^T+\left(\frac{3\Pi}{B}\left(\Sigma+\frac{\Theta}{3}\right)-\frac{4}{3}\Theta-\Sigma+\frac{1}{\tau_{q}}\right)Q^S\\
&+\frac{3ia_1a_2^2\Pi}{2k_\parallel}\left(\frac{k_\perp^2}{a_2^2}+2C\frac{k_\parallel^2}{a_1^2}\right)\left(\mathcal{E}^T+\frac{\Pi^T}{2}\right)
+\frac{k_\perp^2}{a_2}\Pi^V\, ,
 \end{aligned} 
 \label{eq:117}    
\end{equation}

\begin{equation}
  \begin{aligned}
   &\left(\tilde\mu+\frac{3\Pi}{2}
-\frac{\kappa}{\tau_{q}}T\right)\mathcal{A}^V =\frac{1}{2}Y^V+\frac{\kappa}{\tau_{q}}T^V-p^V\\
   &-\left(\frac{4}{3}\Theta-\frac{1}{2}\Sigma-\frac{1}{\tau_{q}}\right)Q^V-\rka\Pi^T -\ik\Pi^V\, ,
  \end{aligned}
  \label{eq:118}    
\end{equation}
and
\begin{equation}
  \begin{aligned}
   &\left(\tilde\mu+\frac{3\Pi}{2}
-\frac{\kappa}{\tau_{q}}T\right)\overline{\mathcal{A}}^V
 =\frac{a_2}{k_\perp^2}\left(\frac{\kappa}{\tau_{q}}\dot T + \dot{\Pi}-2\dot p\right)\Omega^S\\
   & -\left(\frac{4}{3}\Theta-\frac{1}{2}\Sigma-\frac{1}{\tau_{q}}\right)\overline{Q}^V+\rka\overline{\Pi}^T-\ik\overline{\Pi}^V \, .
  \end{aligned}
  \label{eq:119}        
\end{equation}

\subsubsection{Even sector}
The even sector is now supplemented with the following evolution equations for $\Pi^V$, $\Pi^T$, $Z^V$, $Y^V$ and $p_\zeta^V$

\begin{equation}
\tau_{\pi} \dot \Pi^V=-\Pi^V-2\eta\Sigma^V\, ,
\end{equation}

\begin{equation}
\tau_{\pi} \dot \Pi^T=-\Pi^T-2\eta\Sigma^T\, ,
\end{equation}

\begin{equation}
\begin{aligned}
&\tau_{\pi} \dot Y^V=-\dot \Pi\left(\tau_{\pi\mu}\mu^V+\tau_{\pi{\cal{N}}}Z^V\right)+\frac{\tau_{\pi}}{2}\tilde F
Y^V+\\
&\tau_{\pi}\dot\Pi{\cal{A}}^V-Y^V-2\Sigma\left(\varsigma_0\mu^V+\varsigma_1 Z^V\right)-2\eta V^V\, ,
\end{aligned}
\end{equation}
\begin{equation}
\dot{Z}^V  = \dot{\mathcal{N}}\mathcal{A}^V -\mathcal{N}W^V + \left( \frac{\Sigma}{2} -\frac{4\Theta}{3}\right)Z^V\, 
\end{equation}
and
\begin{equation}
\begin{aligned}
&\tau_{\zeta} \dot p_\zeta^V=-\dot p_{\zeta_B}\left(\tau_{\zeta\mu}\mu^V+\tau_{\zeta{\cal{N}}}Z^V\right)+\frac{\tau_{\zeta}}{2}\tilde F
p_\zeta^V+\\
&\tau_{\zeta}\dot p_{\zeta_B}{\cal{A}}^V-p_\zeta^V-\Theta\left(\beta_0\mu^V+\beta_1 Z^V\right)-\zeta_B W^V \, ,
\end{aligned}
\end{equation}
from which $p^V$ is given algebraically by
\begin{equation}
p^V=\varrho_0\mu^V+\varrho_1 Z^V+p_\zeta^V\, .
\end{equation}

\subsubsection{Odd sector}
In the odd sector we obtain the following two evolution equations for $\overline{\Pi}^V$ and $\overline{\Pi}^T$

\begin{equation}
\tau_{\pi} \dot{\overline\Pi}^V=-\overline\Pi^V-2\eta\overline\Sigma^V\, ,
\end{equation}
and
\begin{equation}
\tau_{\pi} \dot{\overline\Pi}^T=-\overline\Pi^T-2\eta\overline\Sigma^T\, ,
\end{equation}
respectively.

$\overline{Y}^V$, ${\overline{Z}}^V$ and $\overline{p}_\zeta^V$ are as before given by (\ref{oddgradcoeff}) as

\begin{equation}
\overline Y^V=\frac{2a_2\dot\Pi}{k_\perp^2}\Omega^S\, ,
\end{equation}
\begin{equation}
{\overline{Z}}^V  =  \frac{2a_2\dot {\cal{N}}}{k_\perp^2}\Omega^S\, 
\end{equation}
and
\begin{equation}
\overline{p}_\zeta^V=\frac{2a_2\dot p_{\zeta_B}}{k_\perp^2}\Omega^S\, ,
\end{equation}
corresponding to
\begin{equation}
\overline p^V=\frac{2a_2\dot p}{k_\perp^2}\Omega^S\, ,
\end{equation}
where $\dot p =\dot{\tilde{p}} +\dot p_{\zeta_B}$. The corresponding differential equations for  $\overline{p}_\zeta^V$ and $\overline{Y}^V$, obtained from
 (\ref{dotbarpzeta}) and (\ref{dotbarZ}), will be identically satisfied.

By adding the evolution equations for $\Pi^V$, $\Pi^T$, $Y^V$, $Z^V$, $p_\zeta^V$ and $\overline{\Pi}^V$, $\overline{\Pi}^T$ to the even and odd general systems in section \ref{EvolutionEquations}  respectively, these are now seen to close.

\section{Vorticity}\label{secvorticity}
From the general evolution equations (\ref{eqOmegaVgeneral}) and (\ref{dotOmegaS}) for the vorticity we see that the acceleration coefficients act as source terms. For a perfect fluid with a barotropic equation of state, $p=p(\mu)$, these terms can be rewritten in terms of the vorticity.
Hence, vorticity cannot be generated, a result which holds to all orders in perturbation theory, see e.g. \cite{Raichoudhuri,Cargese,LuAnandaClarksonMaartens}. However, on applying the Eckart theory for a dissipative fluid, it can be seen from (\ref{eq:barOmegav_dot_Eckart}) and (\ref{OmegaSEckart}) that vorticity could be generated by heat flows and, as will be seen below, indirectly also by viscosity.

Thus, as an application of our equations, we now more thoroughly investigate the possibility of vorticity generation due to heat flows and viscous stresses.  For this purpose, we will focus on the Eckart system. Although it has been argued that the Eckart theory could be valid for normal materials within the limits of applicability of the hydrodynamic approximation, at least in certain cases, it is in general found to be plagued by severe instabilities when considering perturbations away from thermal equilibrium \cite{LindblomHiscock, HiscockLindblom, HiscockLindblom2}.

Since our main goal is to show that the inclusion of viscous stresses and heat flows could indeed provide a mechanism for generating fluid vorticity, we will not worry too much about the possible instabilities of the Eckart theory at this point. A more thorough investigation of the stability conditions of the Eckart theory on a homogeneous and anisotropic cosmological background could be interesting, but is outside the scope of this paper. Instead, when performing numerical calculations, we will specify initial conditions and parameter values as to not induce any noticeable instabilities. Although the numerical values may be unphysical, the solutions could still highlight the sought mechanisms. 

\subsection{The Importance of Dissipative Effects}\label{secwave}

Before turning to numerics, we investigate the importance of the dissipative effects analytically. As a first step, an additional time derivative is applied to equation (\ref{eq:barOmegav_dot_Eckart}), which results in the following relation,
\begin{equation}
\ddot{\overline{\Omega}}^V + C_1 \dot{\overline{\Omega}}^V + C_2 \overline{\Omega}^V = \mathscr{S},
\label{eq:barOmegav_wave}
\end{equation}
where
\begin{equation}
\begin{aligned}
C_1 = &~ 2\Theta + \frac{3\Sigma}{2}   +2\eta -\frac{\tilde{\mu}+3\Pi}{\kappa T} \\
~ 
&+\frac{d}{dt}\ln\Big(a_2\kappa T^2\Big),
\end{aligned}
\end{equation}
\begin{equation}
\begin{aligned}
C_2 = &\Bigg[\frac{d}{dt}\ln\left(a_2\kappa T\Bigg( \frac{2\Theta}{3} + \frac{\Sigma}{2} + \frac{\dot{T}}{T}\right)\Bigg) + \frac{4\Theta}{3} + \Sigma\\
	~ 
	& +2\eta  \Bigg] \left( \frac{2\Theta}{3} + \frac{\Sigma}{2} + \frac{\dot{T}}{T}\right)  -\frac{1}{\kappa T}\Bigg[ \eta\left(  k^2  - \frac{4}{3}\dot{\Theta} \right)   \\
	~ 
	& -2\dot{\eta}\Sigma + \dot{p}+(\tilde{\mu}+3\Pi)\left( \frac{2\Theta}{3} + \frac{\Sigma}{2} \right) \Bigg], 
\end{aligned}
\end{equation} 
and
\begin{equation}
\begin{aligned}
\mathscr{S} =&~ -\frac{3i\pk \eta\Sigma}{2a_1\kappa T}\Bigg[ \mathcal{A}^V + \frac{\eta^V}{\eta}  \Bigg]. 
\label{eq:barOmegav_source}
\end{aligned}
\end{equation}
Here we have defined the gradient of the shear viscosity 
\begin{equation}
\eta^V = \varsigma_0\mu^V + \varsigma_1 Z^V,
\end{equation}
and the square of the norm of the physical wave vector 
\begin{equation}
k^2 = \frac{\pkk}{a_1^2} + \frac{\okk}{a_2^2},
\end{equation} 
while $\mathcal{A}^V$ is determined by equations (\ref{eq:AvEckart}) and (\ref{eqTV}).
 
Looking at the source term $\mathscr{S}$ in (\ref{eq:barOmegav_source}), we can note the importance of the shear viscosity, as this source will vanish on taking the limit of vanishing shear viscosity, i.e.~$\eta,\varsigma_0, \varsigma_1 \rightarrow 0$. In this limit, $\overline{\Omega}^V = 0$ is a solution to equation (\ref{eq:barOmegav_wave}) with the initial conditions $\overline{\Omega}^V(t_0) =  \dot{\overline{\Omega}}^V(t_0) = 0$. Provided that these initial conditions are satisfied, for example by specifying  $ Q^S(t_0) = Q^V(t_0) = \overline{\Omega}^V(t_0) = 0$ as can be seen from (\ref{eq:barOmegav_dot_Eckart}), we would therefore expect that it is possible to have no vorticity generation. 
This absence of vorticity generation when setting $\eta = 0$ is due to that the curl-like combination 
\begin{equation}
\textnormal{curl}Q \equiv  \frac{i\pk }{a_1}Q^V - \frac{1}{a_2}Q^S \label{eq:curlQ},
\end{equation}
forms a closed subsystem together with $\overline{\Omega}^V$. This can be seen on studying the evolution equation for $\textnormal{curl}Q$ 
 \begin{equation}
 \begin{aligned}
&\frac{d}{dt}{\textnormal{curl}Q} = -\frac{3i\pk\Sigma T}{a_1}\left(\frac{\partial}{\partial \mu}\left(\frac{\eta}{T}\right)\mu^V + \frac{\partial}{\partial \mathcal{N}}\left(\frac{\eta}{T}\right)Z^V \right) \\
&+\Bigg[2\eta\left( k^2+3\Sigma\tilde{F}+2\tilde{\mu}+6\Pi\right) -2\left(\tilde{\mu}+3\Pi\right)\frac{\dot{T} }{T} \\
& +2\dot{\Pi}+2\dot{p}\Bigg]\overline{\Omega}^V - \frac{3\Pi}{2a_2\kappa T}Q^S\\
& -\left[\frac{\Sigma}{2}+\frac{5\Theta}{3} + 2\eta -\frac{1}{\kappa T}\left(\tilde{\mu}+\frac{3\Pi}{2}\right) \right]\textnormal{curl}Q 
 \end{aligned} \label{eq:curlQ_dot}
 \end{equation}
 which reduces to
  \begin{equation}
 \begin{aligned}
&\frac{d}{dt}{\textnormal{curl}Q} =  -\left(\frac{\Sigma}{2}+\frac{5\Theta}{3}  -\frac{1}{\kappa T}\left(\mu+p\right) \right)\textnormal{curl}Q \\
&-2\left(\mu+p\right)\left(\frac{\dot{T}}{T} - \frac{\dot{p}}{\mu+p}\right)\overline{\Omega}^V
 \end{aligned} \label{eq:curlQ_dot_NoViscosity}
 \end{equation}
 for $\eta=0$. Similarly, the evolution equation for $\overline{\Omega}^V$ can be written as 
 \begin{equation}
\begin{aligned}
	\dot{\overline{\Omega}}^V = \frac{1}{2\kappa T}\textnormal{curl}Q - \left( \frac{2\Theta}{3} + \frac{\Sigma}{2} + \frac{\dot{T}}{T}\right)\overline{\Omega}^V \, ,
\end{aligned} \label{eq:barOmegavEckartCurlQ}
\end{equation}
for all values of $\eta$. From the closed system constituted by equations  (\ref{eq:curlQ_dot_NoViscosity}) and (\ref{eq:barOmegavEckartCurlQ}), it can then be seen that $\overline{\Omega}^V = \textnormal{curl}Q = 0$ if $\overline{\Omega}^V(t_0) = \textnormal{curl}Q(t_0) = 0$ \footnote{ However, $Q^V$ and $Q^S$ may still evolve to become non-zero individually for $t>t_0$.}.

Hence, it is interesting to note that the generation of vorticity to first order with the Eckart theory, as illustrated through the source term in \eqref{eq:barOmegav_wave},  is crucially dependent on the fact that the background spacetime is both anisotropic and not in thermal equilibrium, so that the combination $\eta \Sigma = -\Pi/2$ is a non-vanishing quantity to zeroth order. However, it should also be noted that, even if $\eta = 0$, vorticity can be generated by a non-zero $\textnormal{curl}Q$ at $t_0$, as $\textnormal{curl}Q(t_0) \neq 0$ implies that $ \dot{\overline{\Omega}}^V(t_0) \neq 0$ if $\overline{\Omega}^V(t_0) = 0$ according to (\ref{eq:barOmegavEckartCurlQ}).

In a similar manner as for $\overline{\Omega}^V$, we can derive a homogeneous second order differential equation for the scalar part $\Omega^S$ in the odd sector
\begin{equation}
\ddot{\Omega}^S + D_1 \dot{\Omega}^S + D_2 \Omega^S = 0,
\label{eq:OmegaS_wave}
\end{equation}
where
\begin{equation}
D_1= 2\Theta - \frac{3\Sigma}{2}   +2\eta -\frac{(\tilde\mu+\frac{3\Pi}{2})}{\kappa T} +\frac{d}{dt}\ln\Big(a_2\kappa T^2\Big)
\end{equation}
and
\begin{equation}
\begin{aligned}
D_2=&-\Bigg[\frac{d}{dt}\ln\left(a_2\kappa T\left(\tilde F - \frac{\dot T}{T}\right)\right)-\frac{\Sigma}{2}+\frac{4\Theta}{3}\\
~
&+2\eta\Bigg]\left(\tilde F -\frac{\dot T}{T}\right)-\frac{1}{2\kappa T}\Bigg[2\dot p +2\dot{\eta}\Sigma \\
~
&-2\tilde{F}\left(\tilde\mu+\frac{3\Pi}{2}\right)+2\eta\left(k^2-\frac{4}{3}\dot{\Theta}\right)\Bigg]\, .
\end{aligned}
\end{equation}
Hence, with the initial conditions $\overline{Q}^V(t_0)=\Omega^S(t_0)=0$, which via equation (\ref{OmegaSEckart}) imply that $\dot{\Omega}^S(t_0)=0$, a nonzero component $\Omega^S$ cannot be created even if $\eta\neq 0$.

For a discussion of a similar connection between vorticity and heatflow in the context of axially symmetric systems with dissipation, see \cite{Herrera1,Herrera2}.

\subsection{Numerical results}

To get even more tangible results, we now perform some simple numerical calculations. In doing so, we fixate the remaining length and temperature scales by setting $\Lambda = 1$, and $k_B = 1$, where $k_B$ is Boltzmann's constant. Hence, all quantities are treated as being dimensionless in the following. 

The numerical calculations are performed using \texttt{ode45} in \texttt{MATLAB} to solve the even and odd sectors, given by (\ref{eq:barOmegav_dot_Eckart})–(\ref{eq:ZvEckart}) and (\ref{OmegaSEckart})–(\ref{QbarVEckart}) respectively, together with the background equations from section \ref{sectionbackground}. To avoid some numerical issues due to subtraction of very similar values, we determine the evolution of the vorticity $\overline{\Omega}^V$ numerically by writing equation (\ref{eq:barOmegav_dot_Eckart}) as in (\ref{eq:barOmegavEckartCurlQ}), adding (\ref{eq:curlQ_dot}) to the system as an additional equation. In principle, one could use (\ref{eq:curlQ}) to eliminate either $Q^V$ or $Q^S$, rendering either (\ref{eq:dotQv_Eckart}) or (\ref{eq:dotQs_Eckart}) unnecessary. However, for simplicity, we solve all three equations (\ref{eq:dotQv_Eckart}), (\ref{eq:dotQs_Eckart}), and  (\ref{eq:curlQ_dot}), using  (\ref{eq:curlQ}) to set consistent initial conditions. Any deviations between the left and right hand sides in (\ref{eq:curlQ}) should then be due to numerical errors. 

In addition to the aforementioned equations, it is also necessary to specify suitable equations of state to get a closed and numerically solvable system, For this purpose, we use that the temperature, $T=T(\mu,{\cal{N}})$, and equilibrium pressure, $\tilde p=\tilde p(\mu,{\cal{N}})$, should satisfy the integrability condition (\ref{ICT}).
A possible choice is then to let the pressure satisfy the ideal gas law, i.e.
\begin{equation}
\tilde p={\cal{N}}T
\end{equation}
and the energy density to be given by a rest-energy term ${\cal{N}}m$ and a thermal energy $f{\cal{N}}T/2$, so that
\begin{equation}
\mu={\cal{N}}\left(m+\frac{f}{2}T\right)\, ,
\end{equation}
where $m$ is the particle mass and $f$ is the number of degrees of freedom. From these equations we can solve for $T=T(\mu,{\cal{N}})$ and $\tilde p=\tilde p(\mu,{\cal{N}})$ 
as
\begin{equation}
\tilde p=\left(\gamma-1\right)\left(\mu-{\cal{N}}m\right)
\end{equation}
and
\begin{equation}
T=\left(\gamma-1\right)\left(\frac{\mu}{{\cal{N}}}-m \right) \, ,
\end{equation}
where $\gamma-1 = 2/f$.

With the basic thermodynamic quantities specified, we also need models for the dissipative coefficients of conductivity and viscosity. For this purpose, we assume that
\begin{equation}
	\begin{aligned}
		\kappa &\propto \sqrt{T}, \\
		\eta & \propto \sqrt{T}, \\
	\end{aligned}
\end{equation}
which holds quite generally for normal gases \cite{LifshitzPhysicalKinetics}. As for the bulk viscosity, we assume $\zeta_B= 0$ for simplicity, as the results from the previous section suggest that $\zeta_B$ is less important than $\eta$ when considering vorticity generation. 

On performing some preliminary numerical calculations using the system specified above, we note that the instabilities of the Eckart theory also seem to  emerge here. These instabilities are encountered when increasing the wave numbers for a fixed set of initial conditions and fluid properties, or when the fixed set of  parameters is not specified carefully to begin with.  A possible explanation for some of the latter type of instabilities can be seen when considering the expressions for the accelerations, which couple into many of the evolution equations. In the Eckart theory, these expressions involve terms behaving as $\sim Q/(\kappa T)$. Therefore, a problem seems to arise when $\kappa T$ becomes very small. This specific problem could possibly be solved when using the simplified causal theory, since $\kappa T$ does not appear alone in the denominator when solving equations (\ref{eq:117})-(\ref{eq:119}) for the accelerations. 

To avoid the instability problems we will, as mentioned previously, specify parameters and initial conditions as to not induce any noticeable instabilities, even though the values may not be physical. In doing so, we assume that the order of magnitude of the variables on the background  manifold do not vary drastically. The background spacetime will thus, for our purposes, be taken to be the one determined by the following parameters and initial conditions 
\begin{alignat}{2}
\Sigma(t_0) &= 0.1, \quad &  \eta(t_0) &= 2 \nonumber \\
\Theta(t_0) &= 0.6 , \quad & \kappa(t_0)&=100, \nonumber  \\
 \mu(t_0) &= 0.36, \quad & \mathcal{N}(t_0) &= 1,  \label{eq:Background_initial_conditions} \\
  a_1(t_0) &= 1, \quad & \gamma &= 1.1, \nonumber  \\
   a_2(t_0) &= 1, \quad& m &= 0.9\frac{\mu(t_0)}{\mathcal{N}(t_0)}, \nonumber
\end{alignat} 
which ends at an expanding de Sitter solution. 

\begin{figure}[h]
	\includegraphics[scale=1]{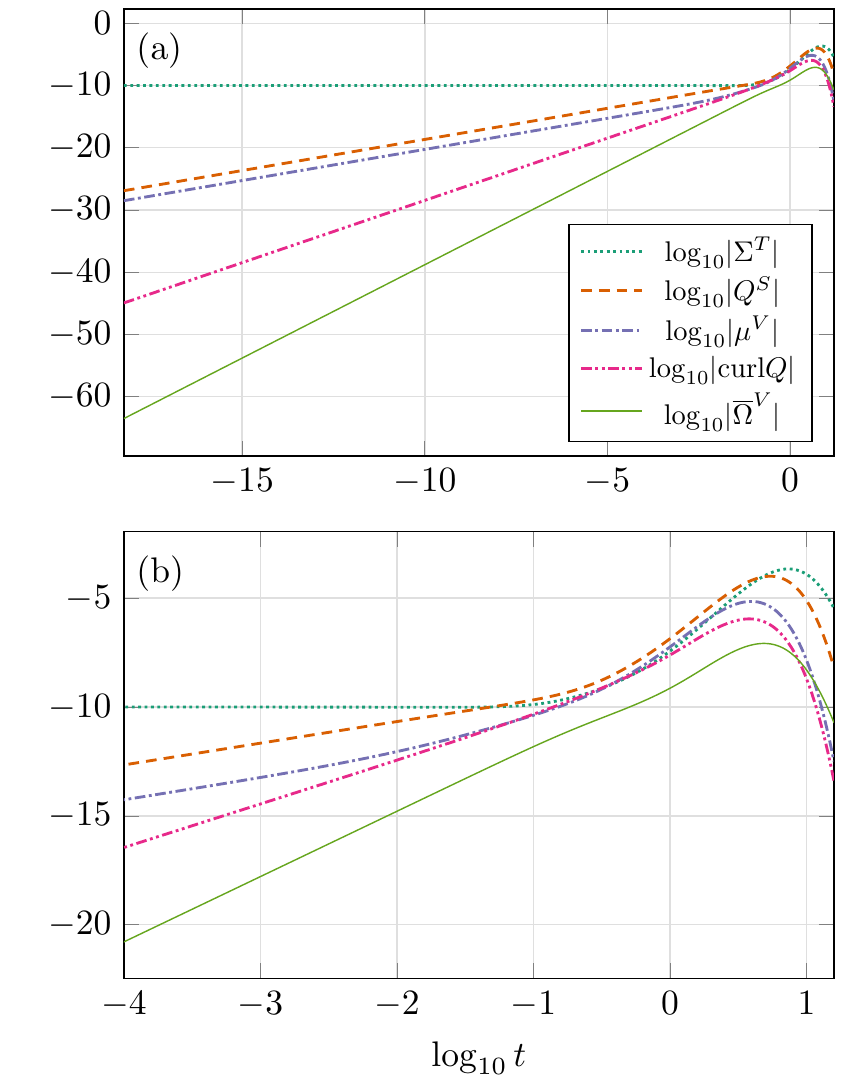}
	\caption{ First order perturbations on the background specified by the initial conditions in equation \eqref{eq:Background_initial_conditions}. The calculations were performed using an initial shear perturbation $\Sigma^T(t_0)= 10^{-10}$ with the wave numbers $\pk = 2$ and $\ok = 1$. All other perturbations were set to zero initially. a) displays the evolution starting from the first non-zero integration time, while b) highlights the evolution starting from $t=10^{-4}$. Colors are based on \cite{Colorbrewer}.}
	\label{fig:Viscosity}
\end{figure}

Equipped with a background spacetime, we move on to investigate the effect of the shear viscosity when considering the first order perturbations. For this purpose, we focus on the even sector, as the results from the previous section suggest that the viscosity is more important for the vorticity generation in this sector. On assuming that all perturbations except $\Sigma^T$ vanish initially, we get the results shown in Fig.~\ref{fig:Viscosity}.  

 In Fig.~\ref{fig:Viscosity}, it can be seen that the vorticity then reaches a maximum amplitude that is approximately three orders of magnitude larger than the initial shear perturbation. Thus, it would seem as though we have some new mechanism for creating vorticity. However, the precise quantitative results are questionable due to the unphysical nature of the chosen parameters and initial conditions.  
 
It can be argued that the initial vorticity perturbation in Fig.~\ref{fig:Viscosity} is actually generated due to numerical errors in the calculations, which were performed with a relative and absolute tolerance of $10^{-9}$ and $10^{-34}$ respectively. Although this might be the case, the dissipative effects can still be seen to play an important role. We can illustrate this importance by performing the same calculations as before, but setting  $\eta(t_0) = 0 $, which implies $\eta= 0$.  The numerical calculations then show that both $\textnormal{curl}Q$ and  $\overline{\Omega}^V$ remain identically zero throughout the whole simulation. This agrees with the differential equations from the previous section, as our choice of initial conditions imply that both ${\overline{\Omega}}^V$ and $\dot{\overline{\Omega}}^V$ are set to zero initially through 
equation {(\ref{eq:barOmegav_dot_Eckart})}. With $\eta$ put to zero, equation {(\ref{eq:barOmegav_wave})} becomes homogeneous, so with the given initial conditions ${\overline{\Omega}}^V$ should remain identically zero. 

 Finally, it should be noted that a non-zero vorticity  ${\overline{\Omega}}^V$ is obtained numerically when specifying  $\textnormal{curl}Q(t_0) \neq 0$, even if $\eta = 0$. This is in agreement with the discussion below (\ref{eq:barOmegavEckartCurlQ}).

\subsection{Perfect fluids with non barotropic equations of state}

In this section we close the general first order system in section \ref{EvolutionEquations} by requiring a perfect fluid form of the energy momentum tensor, but with 
a non barotropic equation of state
\begin{equation}
\tilde{p}=p=p(\mu,{\cal{N}}) \, ,
\end{equation}
so that
\begin{equation}
p^V=\varrho_0\mu^V+\varrho_1 Z^V \, .
\end{equation}
The perfect fluid form is obtained by setting the dissipative terms $Q^V$, $Q^S$, $\Pi$, $\Pi^V$, $\Pi^T$ and $Y^V$ in the even sector and $\overline{Q}^V$, $\overline{\Pi}^V$
and $\overline{\Pi}^T$ in the odd sector, respectively, all to zero. From equations (\ref{dotQV}) and (\ref{dotQS}) in the even sector and equation (\ref{eqB55B}) in the odd sector we
can now solve for the components of the acceleration as
\begin{equation}
{\cal{A}}^V=-\frac{p^V}{\mu+p}\, ,\quad {\cal{A}}^S=-\frac{a_2}{\mu+p}\left(\frac{i k_\parallel}{a_1}p^V-2\dot{p}\overline{\Omega}^V\right)
\end{equation}
and
\begin{equation}
\overline{{\cal{A}}}^V=-\frac{2a_2\dot{p}}{k_\perp^2(\mu+p)}\Omega^S \, ,
\end{equation}
respectively. Substitution into the evaluation equations (\ref{eqOmegaVgeneral}) and (\ref{dotOmegaS}) for $\overline{\Omega}^V$ and $\overline{\Omega}^S$
now gives
\begin{equation}
\dot{\overline{\Omega}}^V=-\left(\frac{\Sigma}{2}+\frac{2\Theta}{3}+\frac{\dot{p}}{\mu+p}\right)\overline{\Omega}^V
\end{equation}
and
\begin{equation}
\dot\Omega^S=\left(\tilde F - \frac{\dot p}{\mu+p}\right)\Omega^S \, .
\end{equation}
Hence there are no source terms in the equations, so that vorticity cannot be generated to first order in perturbation theory. 
This result was also obtained in \cite{Christopherson1,ChristophersonMalik} for perturbations on isotropic Friedmann backgrounds.

They also found that vorticity can be generated to second order if the density and entropy gradients are not parallel. Similarly,
starting from the exact evolution equation for the vorticity in the 1+3 covariant split,
\begin{equation}
h^{ab}\dot\omega_b=-\frac{2}{3}\Theta\omega^a+\sigma^a_b\omega^b+\frac{1}{2}\epsilon^{abc}D_bA_c \, ,
\end{equation}
\cite{Cargese},  an evolution equation for $\omega^a$ which holds to all ordered can be derived.
Together with the twice contracted Bianchi identities for a perfect fluid
\begin{equation}
D_a p+(\mu+p)A_a=0\, ,\quad \dot\mu+\Theta(\mu+p)=0\, ,
\end{equation}
and the commutator
\begin{equation}
D_{[a}D_{b]}\Psi=\epsilon_{abc}\omega^c\dot\Psi \, ,
\end{equation}
for a scalar $\Psi$, one then obtains
\begin{eqnarray}\nonumber
h^{ab}\dot\omega_b&=&-\left(\frac{2}{3}-\varrho_0-\varrho_1\frac{{\cal{N}}}{\mu+p}\right)\Theta\omega^a+\sigma^a_b\omega^b\\
&&-\frac{\varrho_1}{2}\epsilon^ {abc}D_c(\mu) D_b({\cal{N}})\, .
\end{eqnarray}
Hence vorticity can be generated if the last term is nonzero, which can happen if $\varrho_1\neq 0$ and the density and number density gradients
are non parallel. This term is of second order if both gradients are considered small. For an example of how second order perturbation theory can be done using the
covariant and gauge invariant approach, see e.g. \cite{Osano2017}.

\section{Conclusions}\label{conclusions}
In this paper we have presented results for general first order perturbations on homogeneous and hypersurface orthogonal LRS class II cosmologies. In doing so, we have extended previous works to general energy-momentum tensors, including effects from the energy flows $q_a$ and anisotropic pressures $\pi_{ab}$. Using the $1+1+2$ covariant formalism and harmonically decomposing the variables, we have shown that, after specifying the details of the energy-momentum tensor, it is possible to obtain closed systems of ordinary differential equations in time. Due to the generality of the results from section \ref{EvolutionEquations}, these can be combined with any detailed description of the energy momentum tensor. As an example, we have shown how the equations can be used to describe one-component dissipative fluids after imposing a suitable thermodynamic theory to close the system. In an upcoming article, we will also describe how the results from section \ref{EvolutionEquations} can be used to study a cosmological plasma and its associated electromagnetic fields \cite{Semren}. 

When specializing the theory to dissipative one-component fluids, we have observed possible mechanisms for generating fluid vorticity to first order in the perturbations, which was seen to be a result of the inclusion of viscosity, heat flows, and a background anisotropy. This is in contrast to the barotropic perfect fluid case, where it is well known that fluid vorticity cannot be generated. However, when studying the standard Eckart theory numerically, we encountered some problems with stability. Thus, the commonly observed problems of the Eckart theory also seem to plague the theory in this context, at least with our choice of equations of state and models for the coefficients of thermal conductivity and viscosity. A more thorough stability analysis could be interesting, but our numerical observations seem to favor the idea that the Eckart theory ought to be neglected in favor of causal and stable descriptions.

\appendix

\section{The 1 + 3 and 1 + 1 + 2 Covariant Splits of Spacetime}

Here we state some relations in the 1+3 and 1+1+2 covariant splits of spacetime which are useful when deriving the equations. When deriving the basic equations from the Ricci and Bianchi identities, one often encounters the following covariant derivatives for the preferred vector fields $n^a$ and $u^a$, the projection tensor $\tensor{N}{_a_b}$, the 2-volume element  $\tensor{\epsilon}{_a_b}$, general scalars $\Psi$, 2-vectors $\tensor{\Psi}{_a}$, and 2-tensors $\tensor{\Psi}{_a_b}$.

\begin{equation}
\begin{aligned}
 \tensor{\nabla}{_a}\tensor{u}{_b} = &- \mathcal{A}\tensor{u}{_a}\tensor{n}{_b} - \tensor{u}{_a}\tensor{\mathcal{A}}{_b}  +  2\tensor{n}{_{[a}}\tensor{\epsilon}{_{b]}^c}\tensor{\Omega}{_c} + \Omega\tensor{\epsilon}{_a_b}   \\
   &+ 2\tensor{n}{_{(a}}\tensor{\Sigma}{_{b)}} + \tensor{\Sigma}{_{a}_{b}}  - \frac{1}{2} \left( \Sigma - \frac{2\Theta}{3}\right)\tensor{N}{_a_b} \\
   ~
   &+ \left(\Sigma + \frac{\Theta}{3}\right)\tensor{n}{_a}\tensor{n}{_b},
 \end{aligned}
 \label{1+1+2_cov_u}
\end{equation}
\begin{equation}
\begin{aligned}
\tensor{\nabla}{_a}\tensor{n}{_b} = &-\tensor{u}{_a} \left(\mathcal{A}\tensor{u}{_b} +  \tensor{\alpha}{_b}\right) + \tensor{n}{_a}\tensor{a}{_b} + \frac{1}{2}\phi\tensor{N}{_a_b} \\
&+ \tensor{u}{_b}\left (- \tensor{\epsilon}{_a_c}\tensor{\Omega}{^c} +\tensor{\Sigma}{_a} + \left (\Sigma + \frac{\Theta}{3}\right )\tensor{n}{_a}\right ) \\
& +  \xi\tensor{\epsilon}{_a_b} + \tensor{\zeta}{_a_b}.
\end{aligned}
\label{eq:CovDn}
\end{equation} 
\begin{equation}
\begin{aligned}
 &\tensor{\nabla}{^a}\tensor{N}{_b_c} =~ 2\tensor{u}{^a}\tensor{n}{_{(b}}\tensor{\alpha}{_{c)}} - 2\tensor{n}{^a}\tensor{n}{_{(b}} \tensor{a}{_{c)}} - \phi\tensor{n}{_{(b}} \tensor{N}{^a_{c)}}  
  \\
  &- 2\xi\tensor{n}{_{(b}}\tensor{\epsilon}{^a_{c)}} - 2\tensor{n}{_{(b}}\tensor{\zeta}{^a_{c)}} - 2\tensor{u}{^a}\tensor{u}{_{(b}}\tensor{\mathcal{A}}{_{c)}}  \\
  ~
  &+  2\tensor{n}{^{a}}\tensor{u}{_{(b}}\tensor{\epsilon}{_{c)}^d}\tensor{\Omega}{_d} + 2\Omega\tensor{u}{_{(b}}\tensor{\epsilon}{^a_{c)}} + 2\tensor{n}{^{a}}\tensor{u}{_{(b}}\tensor{\Sigma}{_{c)}}  \\
  ~
  &+ 2\tensor{u}{_{(b}}\tensor{\Sigma}{^{a}_{c)}}  - \tensor{u}{_{(b}}\tensor{N}{^a_{c)}} \left( \Sigma - \frac{2\Theta}{3}\right) ,
 \end{aligned}
\end{equation}
\begin{equation}
    \begin{aligned}
    &\tensor{\nabla}{_a}\tensor{\epsilon}{_b_c} = ~ 2\tensor{n}{_{[b}}\tensor{\epsilon}{_{c]}_d} \bigg(-\tensor{u}{_a}\tensor{\alpha}{^d} + \tensor{n}{_a}\tensor{a}{^d} + \frac{1}{2}\phi\tensor{N}{_a^d} \\
    ~
    & + \xi\tensor{\epsilon}{_a^d} + \tensor{\zeta}{_a^d}\bigg)- 2\tensor{u}{_{[b}}\tensor{\epsilon}{_{c]}_e} \bigg(- \tensor{u}{_a}\tensor{\mathcal{A}}{^e}  +   \tensor{n}{_{a}}\tensor{\epsilon}{^{e}^f}\tensor{\Omega}{_f}  \\ 
   ~
   & + \Omega\tensor{\epsilon}{_a^e} 
   + \tensor{n}{_{a}}\tensor{\Sigma}{^{e}} +\tensor{\Sigma}{_{a}^{e}}  - \frac{1}{2} \left( \Sigma - \frac{2\Theta}{3}\right)\tensor{N}{_a^e} \bigg), 
     \end{aligned} 
\end{equation}
\begin{equation}
    \tensor{\nabla}{_a}\Psi = -\tensor{u}{_a}\dot{\Psi} + \tensor{n}{_a}\hat{\Psi} + \tensor{\delta}{_a}\Psi,
\end{equation}
\begin{equation}
\begin{aligned}
&\tensor{\nabla}{_a}\tensor{\Psi}{_b} = - \tensor{u}{_a}\tensor{u}{_b}\tensor{\Psi}{_d}\tensor{\mathcal{A}}{^d} + \tensor{u}{_a}\tensor{n}{_b}\tensor{\Psi}{_d}\tensor{\alpha}{^d} - \tensor{u}{_a}\tensor{\dot{\Psi}}{_{\bar{b}}}  \\
    &- \frac{1}{2} \left(\Sigma -\frac{2\Theta}{3}\right)\tensor{\Psi}{_a}\tensor{u}{_b}+ \Omega\tensor{u}{_b}\tensor{\Psi}{^d}\tensor{\epsilon}{_a_d} + \tensor{n}{_a}\tensor{u}{_b}\tensor{\Psi}{^d}\tensor{\Omega}{^c}\tensor{\epsilon}{_d_c}  \\
       &+ \tensor{n}{_a}\tensor{u}{_b}\tensor{\Psi}{^d}\tensor{\Sigma}{_d} + \tensor{u}{_b}\tensor{\Psi}{^d}\tensor{\Sigma}{_a_d}- \tensor{n}{_a}\tensor{n}{_b}\tensor{\Psi}{_d}\tensor{a}{^d} + \tensor{n}{_a}\tensor{\hat{\Psi}}{_{\bar{b}}}    \\
       &-\tensor{n}{_b}\left( \frac{1}{2}\phi\tensor{\Psi}{_a} +  \left(\xi\tensor{\epsilon}{_a_d}+ \tensor{\zeta}{_a_d} \right)\tensor{\Psi}{^d}\right) + \tensor{\delta}{_a}\tensor{\Psi}{_b},
\end{aligned}
\end{equation}
\begin{equation}
\begin{aligned}
    &\tensor{\nabla}{_a}\tensor{\Psi}{_b_c} = 2\tensor{u}{_{(b}}\bigg( \tensor{\Psi}{_{c)}^f}\big(- \tensor{u}{_a}\tensor{\mathcal{A}}{_f}  +  \tensor{n}{_{a}}\tensor{\epsilon}{_{f}^c}\tensor{\Omega}{_c} + \Omega\tensor{\epsilon}{_a_f}   \\
    &+ \tensor{n}{_{a}}\tensor{\Sigma}{_{f}} + \tensor{\Sigma}{_{a}_{f}} \big)- \frac{1}{2} \left( \Sigma - \frac{2\Theta}{3}\right)\tensor{\Psi}{_{c)}_a} \bigg) \\
    & -2\tensor{u}{_a}\tensor{n}{_{(b}}\tensor{\Psi}{_{c)}^f}\tensor{\alpha}{_f} -\tensor{u}{_a}\tensor{\dot{\Psi}}{_{\{b}_{c\}}} +\tensor{n}{_a}\tensor{\hat{\Psi}}{_{\{b}_{c\}}} \\
    &-2\tensor{n}{_{(b}}\left( \tensor{\Psi}{_{c)}^f}\left(\tensor{n}{_a}\tensor{a}{_f}  +  \xi\tensor{\epsilon}{_a_f} + \tensor{\zeta}{_a_f}\right) + \frac{1}{2}\phi\tensor{\Psi}{_{c)}_a}\right) \\
    &+\tensor{\delta}{_a}\tensor{\Psi}{_b_c}.
\end{aligned}
\end{equation}

\section{Commutation Relations}

\label{commutation} 

\noindent For a zeroth-order scalar field $\Psi$ on orthogonal and homogenous LRS class II backgrounds, with $\phi=0$, the following first order commutation relations hold:

\begin{equation}
    \widehat{\dot{\Psi}}-\dot{\widehat{\Psi }}=-\mathcal{A}\dot{\Psi}+\left(\Sigma +\frac{\Theta }{3}\right) \widehat{\Psi },
\end{equation}

\begin{equation}\label{commutatorA2}
    \delta _{a}\dot{\Psi}-N_{a}^{\,\,\,b}\left( \delta _{b}\Psi \right) ^{\cdot}=-\mathcal{A}_{a}\dot{\Psi}-\frac{1}{2}\left( \Sigma -\frac{2\Theta }{3}\right) \delta _{a}\Psi,
\end{equation}

\begin{equation}\label{commutatorA3}
    \delta _{a}\widehat{\Psi }-N_{a}^{\,\,\,b}\left( \widehat{\delta _{b}\Psi }\right) =-2\varepsilon _{ab}\Omega ^{b}\dot{\Psi} , 
\end{equation}

\begin{equation}\label{commutatorA4}
    \delta _{\lbrack a}\delta _{b]}\Psi =\varepsilon _{ab}\Omega \dot{\Psi}\ .
\end{equation}

For a first-order 2-vector $\Psi _{a}$ the following commutation relations, to first order, hold:

\begin{equation}
    \widehat{\dot{\Psi}}_{\bar{a}}-\dot{\widehat{\Psi }}_{\bar{a}}=\left( \Sigma+\frac{\Theta }{3}\right) \widehat{\Psi }_{\bar{a}}\ ,
\end{equation}

\begin{equation}
    \delta _{a}\dot{\Psi}_{b}-N_{a}^{\,\,\,c}N_{b}^{\,\,\,d}\left( \delta_{c}\Psi _{d}\right) ^{\cdot}=-\frac{1}{2}\left( \Sigma -\frac{2\Theta }{3}\right) \delta _{a}\Psi _{b}\ ,
\end{equation}

\begin{equation}
    \delta _{a}\widehat{\Psi }_{b}-N_{a}^{\,\,\,c}N_{b}^{\,\,\,d}\left( \widehat{\delta _{c}\Psi _{d}}\right)=0 ,
\end{equation}

\begin{equation}
    \delta _{\lbrack a}\delta _{b]}\Psi _{c}=\frac{1}{2}\mathcal{R}N_{c[a}\Psi_{b]}\ ,
\end{equation}

\noindent and for a first-order trace-free and symmetric 2-tensor $\Psi _{ab}$ it holds that:

\begin{equation}
    \widehat{\dot{\Psi}}_{\{ab\}}-\dot{\widehat{\Psi }}_{\{ab\}}=\left( \Sigma +\frac{\Theta }{3}\right) \widehat{\Psi }_{\bar{a}\bar{b}}\ ,
\end{equation}

\begin{equation}
    \delta _{a}\dot{\Psi}_{bc}-N_{a}^{\,\,\,d}N_{b}^{\,\,\,e}N_{c}^{\,\,\,f}\left( \delta _{d}\Psi _{ef}\right) ^{\cdot }=-\frac{1}{2}\left( \Sigma -\frac{2\Theta }{3}\right) \delta _{a}\Psi _{bc}\ ,
\end{equation}

\begin{equation}
    \delta _{a}\widehat{\Psi }_{bc}-N_{a}^{\,\,\,d}N_{b}^{\,\,\,e}N_{c}^{\,\,\,f}\left( \widehat{\delta _{d}\Psi _{ef}}\right) =0 ,
\end{equation}

\begin{equation}
    2\delta _{\lbrack a}\delta _{b]}\Psi _{cd}=\mathcal{R}\left( N_{c[a}\Psi_{b]d}+N_{d[a}\Psi _{b]c}\right).
\end{equation}

\section{Harmonics}\label{Qharmonics}

\label{subsectionharmonic}

The properties of the harmonics associated with the background metric (\ref{metric})
\begin{equation}  \nonumber
ds^{2}=-dt^{2}+a_{1}^{2} dz^{2}+a_{2}^{2}
\left( d\vartheta ^{2}+f_{\mathcal{K}}(\vartheta) d\varphi ^{2}\right) \ , 
\end{equation}
where the scale factors  $a_1\equiv a_1(t)$ and $a_2\equiv a_2(t)$, 
are here briefly summarized. For more details
the reader is referred to \cite{1+1+2,Schperturb,Schperturb2,BFK,LRSIItensor}.  

First order scalars are decomposed as 
\begin{equation}
\Psi =\displaystyle\sum\limits_{k_{\parallel },k_{\perp }}\Psi
_{k_{\parallel }k_{\perp }}^{S}\ P^{k_{\parallel }}\ Q^{k_{\perp }}\ ,
\end{equation}%
where the harmonic coefficients $\Psi _{k_{\parallel }k_{\perp }}^{S}$ are functions of time.
The eigenfunctions $P^{k_{\parallel }}$, which may be represented with $e^{ik_{\parallel }z}$, satisfy
\begin{equation}
\widehat{\Delta }P^{k_{\parallel }}\equiv n^{a}\nabla _{a}n^{b}\nabla _{b}P^{k_{\parallel }}=-\frac{k_{\parallel }^{2}}{a_{1}^{2}}%
P^{k_{\parallel }}\ ,\ \delta _{a}P^{k_{\parallel }}=\dot{P}^{k_{\parallel
}}=0\ ,
\end{equation}
in terms of the  dimensionless comoving wave numbers $k_{\parallel }$ along the $n^a$ direction, whence the physical wavenumbers are given by $k_{\parallel }/a_{1}$.
The eigenfunctions $Q^{k_\perp}$ satisfy: 
\begin{equation}  \label{Beltrami}
\delta^2 Q^{k_\perp}\equiv \delta _{a}\delta ^{a} Q^{k_\perp}=-\frac{k_\perp^2}{a_2^2} Q^{k_\perp}\, ,\ \widehat{Q}%
^{k_{\perp }}=\dot{Q}^{k_{\perp }}=0\ ,
\end{equation}
where $a_{2}$ is the scale factor of
the 2-sheets, and $k_\perp$ are the dimensionless comoving wavenumbers along
the 2-sheets.
When ${\mathcal{R}}$, the 2-curvature, is positive the 2-sheets are spheres and the eigenfunctions can be
represented by the spherical harmonics $Y_{l}^{m}$
with $k_{\perp }^{2}=l(l+1)$, $l=0,1,2,...$. Due to the background symmetry
the $m$-values will not appear explicitly in the equations.
When ${\mathcal{R}}\leq 0$, and the 2-sheets are open, the $k_\perp$ take continuous values.
For continuous values the sums go over into integrals.

Vectors $\Psi _{a}$ are expanded in terms of the even and odd vector harmonics  \cite{Schperturb,Schperturb2,LRSIItensor}
\begin{equation}\label{vectorharmonics}
Q_{a}^{k_{\perp }}=a_{2}\delta _{a}Q^{k_{\perp }}\ ,\ \ \overline{Q}%
_{a}^{k_{\perp }}=a_{2}\varepsilon _{ab}\delta ^{b}Q^{k_{\perp }}\ ,
\end{equation}%
as
\begin{equation}
\Psi _{a}=\displaystyle\sum\limits_{k_{\parallel },k_{\perp
}}P^{k_{\parallel }}\ \left( \Psi _{k_{\parallel }k_{\perp
}}^{V}Q_{a}^{k_{\perp }}+\overline{\Psi }_{k_{\parallel }k_{\perp }}^{V} 
\overline{Q}_{a}^{k_{\perp }}\right) \ .  \label{harmexpV0}
\end{equation}%

Similarly, a tensor $\Psi _{ab}$ can be expanded in terms of 
the even and odd tensor harmonics%
\begin{equation}\label{tensorharmonics}
Q_{ab}^{k_{\perp }}=a_{2}^{2}\delta _{\{a}\delta _{b\}}Q^{k_{\perp }}\ ,\ 
\overline{Q}_{ab}^{k_{\perp }}=a_{2}^{2}\varepsilon _{c\{a}\delta ^{c}\delta
_{b\}}Q^{k_{\perp }}\ ,
\end{equation}%
as
\begin{equation}
\Psi _{ab}=\displaystyle\sum\limits_{k_{\parallel },k_{\perp
}}P^{k_{\parallel }}\ \left( \Psi _{k_{\parallel }k_{\perp
}}^{T}Q_{ab}^{k_{\perp }}+\overline{\Psi }_{k_{\parallel }k_{\perp }}^{T} 
\overline{Q}_{ab}^{k_{\perp }}\right) \ .
\end{equation}

Some useful relations between the harmonics are listed below.
 The vector harmonics satisfy the following orthogonality relation

\begin{equation}
	\tensor{N}{^a^b}Q_a^{k_\perp}\overline{Q}_b^{k_\perp}=0,
\end{equation}

\noindent and the even and odd harmonics are related through 

\begin{equation}
	Q_a^{k_\perp}=-\tensor{\varepsilon}{_a^b}\overline{Q}_b^{k_\perp}, \quad \overline{Q}_a^{k_\perp}=\tensor{\varepsilon}{_a^b}Q_b^{k_\perp} \, .
\end{equation}

\noindent The harmonics also satisfy the following differential relations 

\begin{equation}
	\dot{Q}_a^{k_\perp}=\widehat{Q}_a^{k_\perp}=0, \quad \dot{\overline{Q}}_a^{k_\perp}=\widehat{\overline{Q}}_a^{k_\perp}=0,
\end{equation}

\begin{equation}
	\delta^2Q_a^{k_\perp}=\frac{\mathcal{R}a_2^2-2\okt}{2a_2^2} Q_a^{k_\perp}, \quad \delta^2\overline{Q}_a^{k_\perp}=\frac{\mathcal{R}a_2^2-2\okt}{2a_2^2} \overline{Q}_a^{k_\perp},
\end{equation}

\begin{equation}
	\delta^a Q_a^{k_\perp} = - \frac{\okt}{a_2}Q^{k_\perp}, \quad \delta^a\overline{Q}_a^{k_\perp}=0,
\end{equation}

\begin{equation}
	\varepsilon^{ab}\delta_a Q_b^{k_\perp}=0, \quad \varepsilon^{ab}\delta_a \overline{Q}_b^{k_\perp}=\frac{\okt}{a_2}Q^{k_\perp}.
\end{equation}

For the tensor harmonics the orthogonality relation is

\begin{equation}
	N^{ab}N^{cd}Q_{ac}^{k_\perp}\overline{Q}_{bd}^{k_\perp}=0,
\end{equation}

\noindent the relation between even and odd harmonics are

\begin{equation}
	Q_{ab}^{k_\perp}=\tensor{\varepsilon}{_{\{a}^c}\overline{Q}_{b\}c}^{k_\perp}, \quad \overline{Q}_{ab}^{k_\perp}=-\tensor{\varepsilon}{_{\{a}^c}Q_{b\}c}^{k_\perp}, 
\end{equation}

\noindent and the differential relations are

\begin{equation}
	\dot{Q}_{ab}^{k_\perp}=\widehat{Q}_{ab}^{k_\perp}=0, \quad \dot{\overline{Q}}_{ab}^{k_\perp}=\widehat{\overline{Q}}_{ab}^{k_\perp}=0,
\end{equation}

\begin{equation}
	\delta^2 Q_{ab}^{k_\perp}=\frac{2\mathcal{R}a_2^2-k_\perp^2}{a_2^2}Q_{ab}^{k_\perp}, \quad \delta^2\overline{Q}_{ab}^{k_\perp}=\frac{2\mathcal{R}a_2^2-k_\perp^2}{a_2^2}\overline{Q}_{ab}^{k_\perp},
\end{equation}

\begin{equation}
	\delta^b Q_{ab}^{k_\perp} = \frac{\mathcal{R}a_2^2-k_\perp^2}{2a_2}Q_a^{k_\perp}, \quad \delta^b \overline{Q}_{ab}^{k_\perp} = -\frac{\mathcal{R}a_2^2-k_\perp^2}{2a_2}\overline{Q}_a^{k_\perp},
\end{equation}

\begin{equation}
	\tensor{\varepsilon}{_a^c}\delta^bQ_{bc}^{k_\perp}=\frac{\mathcal{R}a_2^2-k_\perp^2}{2a_2}\overline{Q}_a^{k_\perp},  \tensor{\varepsilon}{_a^c}\delta^b\overline{Q}_{bc}^{k_\perp}=\frac{\mathcal{R}a_2^2-k_\perp^2}{2a_2}Q_a^{k_\perp},
\end{equation}

\begin{equation}
	\tensor{\varepsilon}{^b^c}\delta_bQ_{ac}^{k_\perp}=\frac{\mathcal{R}a_2^2-k_\perp^2}{2a_2}\overline{Q}_a^{k_\perp},  \tensor{\varepsilon}{^b^c}\delta_b\overline{Q}_{ac}^{k_\perp}=\frac{\mathcal{R}a_2^2-k_\perp^2}{2a_2}Q_a^{k_\perp}.
\end{equation}

The vector and tensor harmonics, $Q_a^{k_\perp}$, $\overline{Q}_a^{k_\perp}$ and $Q_{ab}^{k_\perp}$, $\overline{Q}_{ab}^{k_\perp}$ defined in (\ref{vectorharmonics})
and (\ref{tensorharmonics}) respectively are defined with factors $a_2$ and $a_2^2$.
However, since
the derivative $\delta_a$ picks out factors of order $k_\perp/a_2$, an alternative could be to define vector and tensor harmonics as
\begin{equation}\label{Starvector}
Q_{a}^{*k_{\perp }}=\frac{a_{2}}{k_\perp}\delta _{a}Q^{k_{\perp }}\ ,\ \ \overline{Q}%
_{a}^{*k_{\perp }}=\frac{a_{2}}{k_\perp}\varepsilon _{ab}\delta ^{b}Q^{k_{\perp }}\ ,
\end{equation}
and
\begin{equation}\label{Startensor}
Q_{ab}^{*k_{\perp}}=\frac{a_{2}^{2}}{k_\perp^2}\delta _{\{a}\delta _{b\}}Q^{k_{\perp }}\ ,\ 
\overline{Q}_{ab}^{*k_{\perp }}=\frac{a_{2}^{2}}{k_\perp^2}\varepsilon _{c\{a}\delta ^{c}\delta
_{b\}}Q^{k_{\perp }}\ ,
\end{equation}
to keep their norm to the order of unity. Correspondingly the vector and tensor harmonic coefficients change by factors $k_\perp$ and $k^2_\perp$ respectively as
\begin{equation}
\Psi^{V*}_{k_\parallel k_\perp}=k_\perp \Psi^V_{k_\parallel k_\perp}, \Psi^{T*}_{k_\parallel k_\perp}=k_\perp^2 \Psi^T_{k_\parallel k_\perp}
\end{equation}
and analogous for the odd coefficients. 
Hence, in equations with a mixture of  different types of objects (scalars, vectors, tensors), different factors of $k_\perp$ have to be 
introduced to get the right relative magnitudes in powers of $k$ between the terms. This can be of importance, for example if one want to study the equations
in the limit of large wavenumbers $k$, the so called optical limit.

The harmonic expansion of the linearized equations given in appendix \ref{LinearizedHarmonics} is easily 
transfered to the one  in the basis (\ref{Starvector}) and (\ref{Startensor}) by noting that each factor $a_2$ should be associated with a factor $1/k_\perp$.
For example, in equation (\ref{E1}) the term $\frac{k_\perp^2}{a_2}\alpha^V$ should be replaced by $\frac{k_\perp}{a_2}\alpha^V$ (on dropping the stars $*$).  

\section{Linearized Equations}\label{sectionLE}

The evolution equations are

\begin{equation}
    \dot{\phi} = \left(\Sigma-\frac{2\Theta}{3} \right)\left(\frac{1}{2}\phi-\mathcal{A} \right)+\delta_a \alpha^a + Q,
\end{equation}

\begin{equation}
    \dot{\xi} = \frac{1}{2}\left(\Sigma-\frac{2\Theta}{3} \right) \xi+\frac{1}{2}\varepsilon_{ab}\delta^a\alpha^b+\frac{1}{2}\mathcal{H},
\end{equation}

\begin{equation}
    \dot{\zeta}_{\{ab\}} = \frac{1}{2}\left(\Sigma-\frac{2\Theta}{3} \right)\zeta_{ab}+\delta_{\{a}\alpha_{b\}}-\varepsilon_{c\{a}\tensor{\mathcal{H}}{_{b\}}^c},
\end{equation}

\begin{equation}
    \dot{\Omega} = \left(\Sigma-\frac{2\Theta}{3} \right)\Omega+\frac{1}{2}\varepsilon_{ab}\delta^a\mathcal{A}^b,
\end{equation}

\begin{equation}
    \dot{\Sigma}_{\{ab\}} = \left(\Sigma-\frac{2\Theta}{3} \right)\Sigma_{ab}+\delta_{\{a}\mathcal{A}_{b\}}-\mathcal{E}_{ab}+\frac{1}{2}\Pi_{ab},
\end{equation}

\begin{equation}
    \dot{\mathcal{H}} =\left(\frac{3}{2}\Sigma-\Theta \right)\mathcal{H} -\varepsilon_{ab}\delta^a\left(\mathcal{E}^b-\frac{1}{2}\Pi^b\right) 
-3\left(\mathcal{E}-\frac{1}{2}\Pi\right)\xi,
\end{equation}

\begin{equation}
  \begin{aligned}
    \dot{\mu}_{\overline{a}} &  = \frac{1}{2}\left(\Sigma-\frac{2\Theta}{3} \right)\mu_a+\dot{\mu}\mathcal{A}_a-\frac{3}{2}\left(V_a\Pi+\Sigma Y_a \right) \\
    & \quad -(\mu+p)W_a-\Theta\left( \mu_a+p_a\right)-\delta_a\left(\hat Q + \delta_b Q^b\right),
    \end{aligned}
\end{equation}

\begin{equation}
	\begin{aligned}
        \dot{X}_{\overline{a}}+ \frac{1}{2}\dot{Y}_{\overline{a}}& =2 \left(\Sigma-\frac{2\Theta}{3} \right)X_a  - \frac{1}{3}\Theta Y_a +
 \left(\dot{\mathcal{E}}+\frac{1}{2}\dot\Pi\right)\mathcal{A}_a\\
& -\frac{1}{2}\left(\mu+p-3\mathcal{E}+\frac{1}{2}\Pi\right)V_a-\frac{1}{3}\delta_a\left(\hat Q -\frac{1}{2}\delta_b Q^b\right) \\ 
        & \quad -\frac{1}{2}(\mu_a+p_a)\Sigma -\left(\mathcal{E}+\frac{1}{6}\Pi\right)W_a+\varepsilon_{bc}\delta_a\delta^b\mathcal{H}^c, \\   
    \end{aligned}
\end{equation}

\begin{eqnarray}\nonumber
        \dot{V}_{\overline{a}}-\frac{2}{3}\dot{W}_{\overline{a}} & =&\left( \frac{3}{2}\Sigma-\Theta \right)\left(V_a-\frac{2}{3}W_a \right) +\frac{1}{3}(\mu_a+3p_a)-\\
        &&  \delta_a\delta_b\mathcal{A}^b +\left( \dot{\Sigma}-\frac{2\dot{\Theta}}{3}\right)\mathcal{A}_a-X_a +\frac{1}{2} Y_a \; .
\end{eqnarray}

\noindent The equations containing a mixture of evolution and propagation contributions are

\begin{equation}
    \widehat{\alpha}_{\overline{a}}-\dot{a}_{\overline{a}} = \left( \Sigma + \frac{\Theta}{3}\right)\left( \mathcal{A}_a + a_a\right)+\frac{1}{2} Q_a  -\varepsilon_{ab}\mathcal{H}^b,
\end{equation}

\begin{equation}
    \widehat{\mathcal{A}}-\dot{\Theta} = -\delta_a\mathcal{A}^a+\frac{\Theta^2}{3}+\frac{3\Sigma^2}{2}+\frac{1}{2}(\mu+3p)-\Lambda,
\end{equation}

\begin{equation}
    \dot{\Omega}_{\overline{a}}+\frac{1}{2}\varepsilon_{ab}\widehat{\mathcal{A}}^b = -\left( \frac{\Sigma}{2} + \frac{2\Theta}{3} \right)\Omega_a +\frac{1}{2}\varepsilon_{ab}\delta^b\mathcal{A},
\end{equation}

\begin{equation}
    \dot{\Sigma}_{\overline{a}}-\frac{1}{2}\widehat{\mathcal{A}}_{\overline{a}} = -\left( \frac{\Sigma}{2} + \frac{2\Theta}{3} \right)\Sigma_a + \frac{1}{2}\delta_a\mathcal{A} - \frac{3\Sigma}{2}\alpha_a-\mathcal{E}_a + \frac{1}{2}\Pi_a,
\end{equation}

\begin{equation}
	\begin{aligned}
        \dot{\mathcal{E}}_{\overline{a}}+&\frac{1}{2}\dot\Pi_{\overline{a}}+\frac{1}{2}\varepsilon_{ab}\widehat{\mathcal{H}}^b + \frac{1}{4}\widehat Q_{\overline{a}}  = \frac{3}{4}\varepsilon_{ab}\delta^b\mathcal{H} +\frac{1}{2}\varepsilon_{bc}\delta^b\tensor{\mathcal{H}}{^c_a} \\
&- \frac{1}{2}\left(\mu+p-\frac{3\mathcal{E}}{2}+\frac{1}{4}\Pi\right)\Sigma_a-\frac{1}{4}\delta_a Q \\
        & \quad +\left(\frac{3\Sigma}{4}-\Theta \right)\mathcal{E}_a+\frac{3}{4}\left(\mathcal{E}+\frac{1}{2}\Pi\right)\varepsilon_{ab}\Omega^b \\
        & \quad  -\frac{3}{2}\left(\mathcal{E}+\frac{1}{2}\Pi\right)\alpha_a
-\frac{1}{2}\left(\frac{1}{3}\Theta+\frac{1}{4}\Sigma\right)\Pi_a\ ,
    \end{aligned}
\end{equation}

\begin{equation}
  \begin{aligned}
    \dot{\mathcal{E}}_{\{ab\}}+&\frac{1}{2} \dot{\Pi}_{\{ab\}}-\varepsilon_{c\{a}\tensor{\widehat{\mathcal{H}}}{_{b\}}^c}   = -\varepsilon_{c\{a}\delta^c\mathcal{H}_{b\}}
-\frac{1}{2}\delta_{\{a}Q_{b\}}\\
& -\frac{1}{2}(\mu+p+3\mathcal{E}-\frac{1}{2}\Pi)\Sigma_{ab} \\
    & \quad -\left( \frac{3\Sigma}{2} + \Theta\right) \mathcal{E}_{ab}-\frac{1}{6}\left(\Theta-\frac{3}{2}\Sigma\right)\Pi_{ab},
    \end{aligned}
\end{equation}

\begin{equation}
  \begin{aligned}
    \dot{\mathcal{H}}_{\overline{a}}&-\frac{1}{2}\varepsilon_{ab}\left(\widehat{\mathcal{E}}^b-\frac{1}{2}\widehat{\Pi}^b\right)
   = -\frac{1}{2}\varepsilon_{bc}\delta^b\left(\tensor{\mathcal{E}}{^c_a}-\frac{1}{2}\tensor\Pi{^c_a}\right)+\\
&\left( \frac{3\Sigma}{4}-\Theta\right) \mathcal{H}_a 
     +\frac{3}{4}\varepsilon_{ab}a^b\left(\mathcal{E}-\frac{1}{2}\Pi\right)\\
&-\frac{3\mathcal{E}}{2}\varepsilon_{ab}\mathcal{A}^b  -\frac{3}{8}\varepsilon_{ab}\left(2X^b-Y^b+\Sigma Q^b\right),
    \end{aligned}
\end{equation}

\begin{equation}
  \begin{aligned}
    \dot{\mathcal{H}}_{\{ab\}}+&\varepsilon_{c\{a}\tensor{\widehat{\mathcal{E}}}{_{b\}}^c} -\frac{1}{2}\varepsilon_{c\{a}\tensor{\widehat{\Pi}}{_{b\}}^c} = \varepsilon_{c\{a}\delta^c\left(\mathcal{E}_{b\}}-\frac{1}{2}\Pi_{b\}}\right)+\\
&\frac{3}{2}\left(\mathcal{E}-\frac{1}{2}\Pi\right)\varepsilon_{c\{a}\tensor{\zeta}{_{b\}}^c} 
      -\left( \frac{3\Sigma}{2} + \Theta\right)\mathcal{H}_{ab} \; ,
    \end{aligned}
\end{equation}

\begin{equation}
 \dot Q+   \widehat{p} + \widehat{\Pi}= -(\mu+p)\mathcal{A}-\delta_a \Pi^a-\left(\frac{3}{2}\phi+\mathcal{A}\right)\Pi-\left(\frac{4}{3}\Theta+\Sigma\right)Q \; .
\end{equation}

\begin{eqnarray}\nonumber
&& \dot Q_{\overline{a}}+\widehat{\Pi}_{\overline{a}}+   p_a = 
\frac{1}{2}Y_a-\delta^b\Pi_{ab}-\frac{3}{2}\Pi a_a\\
&&-\left(\frac{4}{3}\Theta-\frac{1}{2}\Sigma\right)Q_a-(\mu+p-\frac{1}{2}\Pi)\mathcal{A}_a \; .
\end{eqnarray}

\noindent The equations containing only propagation contributions are

\begin{equation}
    \widehat{\phi} = \frac{2\Theta^2}{9}+\frac{\Theta\Sigma}{3} +\delta_a a^a -\frac{2}{3}(\mu+\Lambda)-\mathcal{E}-\frac{1}{2}\Pi-\Sigma^2,
\end{equation}

\begin{equation}
    \widehat{\xi} = \left( \Sigma +  \frac{\Theta}{3} \right)\Omega + \frac{1}{2}\varepsilon_{ab}\delta^a a^b,
\end{equation}

\begin{equation}
    \widehat{\zeta}_{\{ab\}} = \delta_{\{a}a_{b\}}+\left( \Sigma +  \frac{\Theta}{3} \right)\Sigma_{ab}-\mathcal{E}_{ab}-\frac{1}{2}\Pi_{ab},
\end{equation}

\begin{equation}
  \begin{aligned}
    \widehat{V}_{\overline{a}}-\frac{2}{3}\widehat{W}_{\overline{a}}  & = -\delta_a\delta_b\Sigma^b-\varepsilon_{bc}\delta_a\delta^b\Omega^c-\frac{3\Sigma}{2}\delta_a\phi \\
    & \quad +2\left( \dot{\Sigma}-\frac{2}{3}\dot{\Theta} \right)\varepsilon_{ab}\Omega^b-\delta_a Q,
    \end{aligned}
\end{equation}

\begin{equation}
    \widehat{\Sigma}_{\overline{a}}-\varepsilon_{ab}\widehat{\Omega}^b = \frac{1}{2}V_a +\frac{2}{3}W_a-\varepsilon_{ab}\delta^b\Omega - \delta^b\Sigma_{ab} - \frac{3\Sigma}{2} a_a -Q_a,
\end{equation}

\begin{equation}\label{OmegaVS}
    \widehat{\Omega} = -\delta_a\Omega^a,
\end{equation}

\begin{equation}
    \widehat{\Sigma}_{\{ab\}} = \delta_{\{a}\Sigma_{b\}}-\varepsilon_{c\{a}\delta^c\Omega_{b\}} - \varepsilon_{c\{a}\tensor{\mathcal{H}}{_{b\}}^c} + \frac{3\Sigma}{2}\zeta_{ab},
\end{equation}

\begin{eqnarray}\nonumber
&&    \widehat{X}_{\overline{a}}+\frac{1}{2} \widehat{Y}_{\overline{a}} -\frac{1}{3}\widehat{\mu}_{\overline{a}}= -\delta_a\delta_b\left(\mathcal{E}^b+\frac{1}{2}\Pi^b\right)\\\nonumber
&&-\frac{3}{2}\left(\mathcal{E}+\frac{1}{2}\Pi\right)\delta_a\phi+\left(2 \dot{\mathcal{E}}+\dot\Pi-\frac{2}{3}\dot{\mu} \right)\varepsilon_{ab}\Omega^b\\
&&+\left(\frac{1}{2}\Sigma-\frac{1}{3}\Theta\right)\delta_a Q \, ,
\end{eqnarray}

\begin{eqnarray}\nonumber
 &&   \widehat{\mathcal{E}}_{\overline{a}} +\frac{1}{2} \widehat{\Pi}_{\overline{a}}= \frac{1}{2}X_a+\frac{1}{4}Y_a+\frac{1}{3}\mu_a-\delta^b\left(\mathcal{E}_{ab}+\frac{1}{2}\Pi_{ab}\right)-\\
&&\frac{3}{2}\left(\mathcal{E}+\frac{1}{2}\Pi\right)a_a-\frac{3\Sigma}{2}\varepsilon_{ab}\mathcal{H}^b-\left(\frac{1}{3}\Theta+\frac{1}{4}\Sigma\right)Q_a,
\end{eqnarray}

\begin{equation}
    \widehat{\mathcal{H}} = -\delta_a\mathcal{H}^a -(\mu+p+3\mathcal{E}-\frac{1}{2}\Pi)\Omega-\frac{1}{2}\varepsilon_{ab}\delta^a Q^b,
\end{equation}

\begin{equation}
  \begin{aligned}
  &  \widehat{\mathcal{H}}_{\overline{a}} -\frac{1}{2}\varepsilon_{ab}\widehat Q^b  = \frac{1}{2}\delta_a\mathcal{H}-\delta^b\mathcal{H}_{ab}-\left( \mu+p -\frac{3\mathcal{E}}{2}+\frac{1}{4}\Pi\right) \Omega_a \\
    & \quad -\frac{3}{2}\left(\mathcal{E}+\frac{1}{2}\Pi\right)\varepsilon_{ab}\Sigma^b+\frac{3\Sigma}{2} \varepsilon_{ab}\left(\mathcal{E}^b+\frac{1}{2}\Pi^b\right)-\frac{1}{2}\varepsilon_{ab}\delta^b Q\, .
    \end{aligned}
\end{equation}

\noindent Lastly, the constraints are

\begin{equation}
    \delta_a\Omega^a+\varepsilon_{ab}\delta^a\Sigma^b = \mathcal{H}-3\Sigma\xi,
\end{equation}

\begin{equation}
    \frac{1}{2}\delta_a\phi - \varepsilon_{ab}\delta^b\xi-\delta^b\zeta_{ab} = \left( \frac{\Sigma}{2} - \frac{\Theta}{3} \right) \left( \varepsilon_{ab}\Omega^b-\Sigma_a \right) -\mathcal{E}_a-\frac{1}{2}\Pi_a,
\end{equation}

\begin{equation}
    V_a-\frac{2}{3}W_a+2\varepsilon_{ab}\delta^b\Omega+2\delta^b\Sigma_{ab} = -2\varepsilon_{ab}\mathcal{H}^b-Q_a\, .
\end{equation}

\section{Harmonic Expansion of Linearized Equations}
\label{LinearizedHarmonics}

Here the harmonic expansion of the first order equations given in section \ref{sectionLE} is presented.

\subsection{Even Parity}

\noindent The evolution equations for the scalars are

\begin{equation}\label{E1}
    \dot{\phi}^S = \left(\Sigma-\frac{2\Theta}{3} \right)\left(\frac{1}{2}\phi^S-\mathcal{A}^S \right)-\frac{\okt}{a_2}\alpha^V+Q^S,
\end{equation}

\begin{equation}
 \begin{aligned}
 &\dot Q^S  +  a_2\ik \left(p^V +Y^V\right)= -\left(\mu+p+\Pi \right)\mathcal{A}^S\\
&-2a_2\overline{\Omega}^V\left(\dot{p}+\dot\Pi\right)-\frac{3}{2}\Pi\phi^S-\left(\frac{4}{3}\Theta+\Sigma\right)Q^S+\frac{k_\perp^2}{a_2}\Pi^V,
 \end{aligned} 
\end{equation}

\bigskip

\noindent for the 2-vectors are

\begin{equation}
  \begin{aligned}
	\dot{\overline{\mathcal{H}}}^V &  = \frac{i\pk}{2a_1}\left(\mathcal{E}^V-\frac{1}{2}\Pi^V\right)+\frac{3}{4}\left(\Sigma-\frac{4\Theta}{3} \right)\overline{\mathcal{H}}^V\\
	&-\frac{3}{8}\left(2X^V-Y^V+\Sigma Q^V\right) - \frac{3\mathcal{E}}{2}\mathcal{A}^V\\
	&+\frac{3}{4}\left(\mathcal{E}-\frac{1}{2}\Pi\right)a^V-\frac{\mathcal{R}a_2^2-\okt}{4a_2}\left(\mathcal{E}^T-\frac{1}{2}\Pi^T\right),
    \end{aligned}
\end{equation}

\begin{equation}
    \dot{\overline{\Omega}}^V = -\left(\frac{\Sigma}{2}+\frac{2\Theta}{3} \right)\overline{\Omega}^V +\frac{1}{2a_2}\mathcal{A}^S-\frac{i\pk}{2a_1} \mathcal{A}^V,
\end{equation}

\begin{eqnarray}\nonumber
	&&\dot{\mu}^V=\frac{1}{2}\left(\Sigma-\frac{2\Theta}{3} \right)\mu^V-\Theta\left( \mu^V+p^V \right)-(\mu+p)W^V  \\
&&+\dot{\mu}\mathcal{A}^V-\frac{3}{2}\left(\Pi V^V+\Sigma Y^V\right)-\frac{ik_\parallel}{a_1 a_2}Q^S+\frac{k_\perp ^2}{a_2^2}Q^V\, ,
\end{eqnarray}

\begin{equation}
	\begin{aligned}
        &\dot{X}^V+\frac{1}{2}\dot{Y}^V  =2\left(\Sigma-\frac{2\Theta}{3} \right)X^V+ \left(\dot{\mathcal{E}}+\frac{1}{2}\dot\Pi\right)\mathcal{A}^V\\ 
        &  -\frac{1}{2}\left(\mu+p-3\mathcal{E}+\frac{1}{2}\Pi\right)V^V  -\frac{\Sigma}{2}\left( \mu^V+p^V \right)+\frac{\okt}{a_2^2}\overline{\mathcal{H}}^V\\
        &-\frac{1}{3}\Theta Y^V-\left(\mathcal{E}+\frac{1}{6}\Pi \right)W^V-\frac{1}{3}\frac{ik_\parallel}{a_1 a_2}Q^S-\frac{1}{6}\frac{k_\perp ^2}{a_2^2}Q^V,
    \end{aligned}
\end{equation}

\begin{equation}
	\begin{aligned}
        &\dot{V}^V-\frac{2}{3}\dot{W}^V  =\frac{1}{3}\left( \mu^V+3p^V \right) + \left(\dot{\Sigma}-\frac{2\dot{\Theta}}{3} + \frac{\okt}{a_2^2} \right)\mathcal{A}^V \\
        & \quad +\frac{3}{2}\left(\Sigma-\frac{2\Theta}{3} \right)\left(V^V-\frac{2}{3}W^V \right) - X^V+\frac{1}{2}Y^V,
    \end{aligned}
\end{equation}

\begin{equation}
    \dot{a}^V=\frac{i\pk}{a_1}\alpha^V-\left(\Sigma+\frac{\Theta}{3} \right)\left( \mathcal{A}^V+a^V \right)-\overline{\mathcal{H}}^V-\frac{1}{2}Q^V,
\end{equation}

\begin{equation}
  \begin{aligned}
	\dot{W}^V & = \left(\dot{\Theta}-\frac{\okt}{a_2^2} \right)\mathcal{A}^V + \frac{i\pk}{a_1a_2}\mathcal{A}^S-\frac{1}{2}\left(\mu^V+3p^V \right) \\ 
    & \quad +\left(\frac{\Sigma}{2}-\Theta \right) W^V-3\Sigma V^V,
    \end{aligned}
\end{equation}

\begin{equation}
    \dot{\Sigma}^V = \frac{i\pk}{2a_1} \mathcal{A}^V +  \frac{1}{2a_2}\mathcal{A}^S - \left(\frac{\Sigma}{2}+\frac{2\Theta}{3} \right)\Sigma^V-\frac{3\Sigma}{2}\alpha^V-\mathcal{E}^V
    +\frac{1}{2}\Pi^V,
\end{equation}

\begin{equation}
	\begin{aligned}
  &      \dot{\mathcal{E}}^V +\frac{1}{2}\dot\Pi^V = \frac{i\pk}{2a_1} \overline{\mathcal{H}}^V-\frac{ik_\parallel}{4a_1}Q^V+\frac{3}{4}\left(\Sigma - \frac{4\Theta}{3} \right)\mathcal{E}^V\\
      &- \frac{1}{4a_2}Q^S+ \frac{1}{4}\left(3\mathcal{E}-2\mu-2p-\frac{1}{2}\Pi \right)\Sigma^V-\left(\frac{1}{6}\Theta+\frac{1}{8}\Sigma\right)\Pi^V \\
        & \quad -\frac{3}{2}\left(\mathcal{E}+\frac{1}{2}\Pi\right)\alpha^V+\frac{\mathcal{R}a_2^2-\okt}{4a_2}\overline{\mathcal{H}}^T-\frac{3}{4}\left(\mathcal{E}+\frac{1}{2}\Pi\right)\Omega^V,
    \end{aligned}
\end{equation}

\begin{equation}
  \begin{aligned}
   & \dot Q^V+p^V+\ik\Pi^V=-\left(\mu+p -\frac{1}{2}\Pi\right)\mathcal{A}^V\\
   &+\frac{1}{2}Y^V-\frac{3}{2}\Pi a^V-\left(\frac{4}{3}\Theta-\frac{1}{2}\Sigma\right)Q^V\\
   &-\rka\Pi^T ,
  \end{aligned}    
\end{equation}

\noindent and for the 2-tensors are

\begin{equation}
\begin{aligned}
    \dot{\overline{\mathcal{H}}}^T&=-\frac{i\pk}{a_1}\left(\mathcal{E}^T-\frac{1}{2}\Pi^T\right)
    +\frac{1}{a_2}\left(\mathcal{E}^V-\frac{1}{2}\Pi^V\right)-\\
    &\left(\Theta+\frac{3\Sigma}{2} \right)\overline{\mathcal{H}}^T+\frac{3}{2}\left({\mathcal{E}}-\frac{1}{2}\Pi\right)\zeta^T,
\end{aligned}        
\end{equation}

\begin{equation}
  \begin{aligned}
  &  \dot{\mathcal{E}}^T+\frac{1}{2} \dot{\Pi}^T  = -\frac{i\pk}{a_1}\overline{\mathcal{H}}^T-\frac{1}{2}(3\mathcal{E}+\mu+p-\frac{1}{2}\Pi)\Sigma^T \\ 
    &-\frac{1}{a_2}\overline{\mathcal{H}}^V -\left(\Theta+\frac{3\Sigma}{2} \right)\mathcal{E}^T-\frac{1}{2a_2}Q^V-\frac{1}{6}\left(\Theta-\frac{3}{2}\Sigma\right)\Pi^T,
    \end{aligned}
\end{equation}

\begin{equation}
	\dot{\zeta}^T=\frac{1}{2}\left(\Sigma-\frac{2\Theta}{3} \right)\zeta^T+\frac{1}{a_2}\alpha^V+\overline{\mathcal{H}}^T,
\end{equation}

\begin{equation}
	\dot{\Sigma}^T = \left(\Sigma-\frac{2\Theta}{3} \right)\Sigma^T+\frac{1}{a_2}\mathcal{A}^V-\mathcal{E}^T+\frac{1}{2}\Pi^T.
\end{equation}

\noindent The constraint for the scalar is

\begin{equation}
  \begin{aligned}
    \frac{i\pk}{a_1a_2}\phi^S & = \frac{1}{3}\left(\Sigma+\frac{4\Theta}{3} \right)W^V+\left(\frac{\Theta}{3}-2\Sigma \right)V^V \\ 
    & \quad-\frac{\okt}{a_2^2}a^V-\frac{2}{3}\mu^V-X^V-\frac{1}{2}Y^V,
    \end{aligned}
\end{equation}

\noindent for the 2-vectors are

\begin{equation}
  \begin{aligned}
	\ik\left(V^V-\frac{2}{3}W^V \right) & = \frac{\okt}{a_2^2}\Sigma^V-\frac{3\Sigma}{2a_2}\phi^S -\frac{1}{a_2}Q^S\\ 
    & \quad -\frac{2}{3}\left(3\dot{\Sigma}-2\dot{\Theta} + \frac{3\okt}{2a_2^2}\right)\overline{\Omega}^V,
    \end{aligned}
\end{equation}

\begin{equation}
  \begin{aligned}
	\ik \left(\Sigma^V+\overline{\Omega}^V \right) = &\frac{1}{2}V^V+\frac{2}{3}W^V-\rka\Sigma^T\\
	&-\frac{3\Sigma}{2} a^V-Q^V,
	\end{aligned}
\end{equation}

\begin{equation}
  \begin{aligned}
  &  \ik\left(X^V+\frac{1}{2}Y^V-\frac{1}{3}\mu^V \right) = 
    \frac{\okt}{a_2^2}\left(\mathcal{E}^V+\frac{1}{2}\Pi^V\right)-\\
    &\frac{3}{2a_2}\left(\mathcal{E}+\frac{1}{2}\Pi\right)\phi^S-\left(2\dot{\mathcal{E}}+\dot\Pi-\frac{2}{3}\dot{\mu} \right)\overline{\Omega}^V+\\
   &\left(\frac{1}{2}\Sigma-\frac{1}{3}\Theta\right)\frac{1}{a_2}Q^S,
\end{aligned}    
\end{equation}

\begin{equation}
  \begin{aligned}
   & \ik\left(\mathcal{E}^V+\frac{1}{2}\Pi^V\right)  =\frac{1}{2}X^V+\frac{1}{4}Y^V+\frac{1}{3}\mu^V-\\
    &\frac{3}{2}\left(\mathcal{E}+\frac{1}{2}\Pi\right)a^V  -\rka\left(\mathcal{E}^T+\frac{1}{2}\Pi^T\right)+\frac{3\Sigma}{2}\overline{\mathcal{H}}^V\\
    &-\left(\frac{1}{3}\Theta+\frac{1}{4}\Sigma\right)Q^V,
    \end{aligned}
\end{equation}

\begin{equation}
    V^V-\frac{2}{3}W^V=2\overline{\mathcal{H}}^V - \frac{\mathcal{R}a_2^2-\okt}{a_2}\Sigma^T-Q^V,
\end{equation}

\begin{equation}
  \begin{aligned}
	\mathcal{E}^V+\frac{1}{2}\Pi^V & = -\frac{1}{2}\left(\Sigma-\frac{2\Theta}{3} \right)\left(\Sigma^V+\overline{\Omega}^V \right) \\
    & \quad + \rka\zeta^T - \frac{1}{2a_2}\phi^S,
    \end{aligned}
\end{equation}

\noindent and for the 2-tensors are

\begin{equation}
  \begin{aligned}
    &\ik\left(\overline{\mathcal{H}}^V-\frac{1}{2}Q^V\right)  = \rka\overline{\mathcal{H}}^T-\frac{1}{2a_2}Q^S\\ 
    &- \left(\mu+p-\frac{3\mathcal{E}}{2}+\frac{1}{4}\Pi \right)\overline{\Omega}^V 
      -\frac{3}{2}\left(\mathcal{E}+\frac{1}{2}\Pi\right)\Sigma^V\\
      &+\frac{3\Sigma}{2}\left(\mathcal{E}^V+\frac{1}{2}\Pi^V\right),
    \end{aligned}
\end{equation}

\begin{equation}
    \ik\zeta^T = \left(\Sigma+\frac{\Theta}{3} \right)\Sigma^T+\frac{1}{a_2}a^V-\mathcal{E}^T-\frac{1}{2}\Pi^T,
\end{equation}

\begin{equation}
    \ik\Sigma^T = \frac{1}{a_2}\Sigma^V-\frac{1}{a_2}\overline{\Omega}^V+\overline{\mathcal{H}}^T+\frac{3\Sigma}{2}\zeta^T.
\end{equation}

\subsection{Odd Parity}

\noindent The evolution equations for the scalars are

\begin{equation}
    2\dot{\xi}^S = \left(\Sigma-\frac{2\Theta}{3} \right)\xi^S+\frac{\okt}{a_2}\overline{\alpha}^V+\mathcal{H}^S,
\end{equation}

\begin{equation}
    \dot{\Omega}^S = \left(\Sigma-\frac{2\Theta}{3} \right)\Omega^S + \frac{\okt}{2a_2}\overline{\mathcal{A}}^V,
\end{equation}

\begin{eqnarray}\nonumber
    \dot{\mathcal{H}}^S&=&\frac{3}{2}\left(\Sigma-\frac{2\Theta}{3} \right)\mathcal{H}^S-\frac{\okt}{a_2}\left(\overline{\mathcal{E}}^V-\frac{1}{2}\overline{\Pi}^V\right)\\
&&-3\left(\mathcal{E}-\frac{1}{2}\Pi\right)\xi^S,
\end{eqnarray}

\noindent for the 2-vectors are

\begin{equation}
    \dot{\Omega}^V = -\left(\frac{\Sigma}{2}+\frac{2\Theta}{3} \right)\Omega^V + \frac{i\pk}{2a_1} \overline{\mathcal{A}}^V,
\end{equation}

\begin{eqnarray}\nonumber
	\dot{\overline{\mu}}^V&=&\frac{1}{2}\left(\Sigma-\frac{2\Theta}{3} \right)\overline{\mu}^V-\Theta\left( \overline{\mu}^V+\overline{p}^V \right)-(\mu+p)\overline{W}^V\\
&&+\dot{\mu}\overline{\mathcal{A}}^V-\frac{3}{2}\left(\Pi\overline{V}^V+\Sigma\overline{Y}^V\right)\, ,
\end{eqnarray}

\begin{equation}
	\begin{aligned}
       &\dot{\overline{X}}^V+\frac{1}{2}  \dot{\overline{Y}}^V  =2\left(\Sigma-\frac{2\Theta}{3} \right)\overline{X}^V-\left(\mathcal{E}+\frac{1}{6}\Pi\right)\overline{W}^V \\
       & +\left(\dot{\mathcal{E}}+\frac{1}{2}\dot\Pi\right)\overline{\mathcal{A}}^V -\frac{1}{3}\Theta Y^V\\
        & \quad -\frac{1}{2}(\mu+p-3\mathcal{E}+\frac{1}{2}\Pi)\overline{V}^V -\frac{\Sigma}{2} \left(\overline{\mu}^V+\overline{p}^V \right),
    \end{aligned}
\end{equation}

\begin{equation}
	\begin{aligned}
        &\dot{\overline{V}}^V-\frac{2}{3}\dot{\overline{W}}^V   =  \frac{1}{3}\left( \overline{\mu}^V+3\overline{p}^V \right)-\overline{X}^V + \left(\dot{\Sigma}-\frac{2\dot{\Theta}}{3} \right)\overline{\mathcal{A}}^V \\
        & \quad +\frac{3}{2}\left(\Sigma-\frac{2\Theta}{3} \right)\left(\overline{V}^V-\frac{2}{3}\overline{W}^V \right)+\frac{1}{2}\overline{Y}^V,
    \end{aligned}
\end{equation}

\begin{equation}
    \dot{\overline{a}}^V=\frac{i\pk}{a_1}\overline{\alpha}^V-\left(\Sigma+\frac{\Theta}{3} \right)\left( \overline{\mathcal{A}}^V+\overline{a}^V \right)+\mathcal{H}^V-\frac{1}{2}\overline{Q}^V,
\end{equation}

\begin{equation}
	\dot{\overline{W}}^V = \dot{\Theta}\overline{\mathcal{A}}^V -\frac{1}{2}\left(\overline{\mu}^V+3\overline{p}^V \right)+\left(\frac{\Sigma}{2}-\Theta \right) \overline{W}^V-3\Sigma \overline{V}^V,
\end{equation}

\begin{equation}
    \dot{\overline{\Sigma}}^V = \frac{i\pk}{2a_1} \overline{\mathcal{A}}^V - \left(\frac{\Sigma}{2}+\frac{2\Theta}{3} \right)\overline{\Sigma}^V-\frac{3\Sigma}{2}\overline{\alpha}^V-\overline{\mathcal{E}}^V+\frac{1}{2}\overline{\Pi}^V,
\end{equation}

\begin{equation}
	\begin{aligned}
     &   \dot{\overline{\mathcal{E}}}^V+\frac{1}{2}\dot{\overline{\Pi}}^V 
        = -\frac{i\pk}{2a_1} \mathcal{H}^V+\frac{3}{4}\left(\Sigma - \frac{4\Theta}{3} \right)\overline{\mathcal{E}}^V+\\
        &\frac{3}{4}\left(\mathcal{E}+\frac{1}{2}\Pi\right)\Omega^V 
         + \frac{3}{4a_2}\mathcal{H}^S -\frac{3}{2}\left(\mathcal{E}+\frac{1}{2}\Pi\right)\overline{\alpha}^V+\\
         &\frac{\mathcal{R}a_2^2-\okt}{4a_2}\mathcal{H}^T-\frac{ik_\parallel}{4a_1}
        \overline{Q}^V+ \\ 
        & \frac{1}{4}\left(3\mathcal{E}-2\mu-2p-\frac{1}{2}\Pi  \right)\overline{\Sigma}^V-\left(\frac{1}{6}\Theta+\frac{1}{8}\Sigma\right)\overline\Pi^V ,
    \end{aligned}
\end{equation}

\begin{equation}
	\begin{aligned}
        \dot{\mathcal{H}}^V & = -\frac{i\pk}{2a_1}\left(\overline{\mathcal{E}}^V-\frac{1}{2}\overline{\Pi}^V\right)+\frac{3}{4}\left(\Sigma-\frac{4\Theta}{3} \right)\mathcal{H}^V\\
        &+\frac{3}{8}\left(2\overline{X}^V-\overline{Y}^V+\Sigma\overline{Q}^V\right)  + \frac{3\mathcal{E}}{2}\overline{\mathcal{A}}^V\\
        &-\frac{3}{4}\left(\mathcal{E}-\frac{1}{2}\Pi\right)\overline{a}^V-
        \frac{\mathcal{R}a_2^2-\okt}{4a_2}\left(\overline{\mathcal{E}}^T-\frac{1}{2}\overline{\Pi}^T\right),
    \end{aligned}
\end{equation}

\begin{equation}\label{eqB55}
  \begin{aligned}
   & \dot{\overline{Q}}^V+\overline{p}^V+\ik\overline{\Pi}^V=-\left(\mu+p -\frac{1}{2}\Pi\right)\overline{\mathcal{A}}^V\\
   &+\frac{1}{2}\overline{Y}^V-\frac{3}{2}\Pi\overline{a}^V-\left(\frac{4}{3}\Theta-\frac{1}{2}\Sigma\right)\overline{Q}^V\\
   &+\rka\overline{\Pi}^T ,
  \end{aligned}    
\end{equation}

\noindent and for the 2-tensors are

\begin{equation}
\begin{aligned}
    \dot{\mathcal{H}}^T&=\frac{i\pk}{a_1}\left(\overline{\mathcal{E}}^T-\frac{1}{2}\overline{\Pi}^T\right)+\frac{1}{a_2}\left(\overline{\mathcal{E}}^V-\frac{1}{2}\overline{\Pi}^V\right)-\\
    &\frac{3}{2}\left(\Sigma+\frac{2\Theta}{3} \right)\mathcal{H}^T-\frac{3}{2}\left({\mathcal{E}}-\frac{1}{2}\Pi\right)\overline{\zeta}^T,
\end{aligned}    
\end{equation}

\begin{equation}
  \begin{aligned}
   & \dot{\overline{\mathcal{E}}}^T+\frac{1}{2} \dot{\overline\Pi}^T  = \frac{i\pk}{a_1}\mathcal{H}^T-\frac{1}{a_2}\mathcal{H}^V-\frac{1}{2}(3\mathcal{E}+\mu+p-\frac{1}{2}\Pi)\overline{\Sigma}^T \\ 
    &  -\left(\frac{3}{2}\Sigma+\Theta \right)\overline{\mathcal{E}}^T+\frac{1}{2a_2}\overline{Q}^V-\frac{1}{6}\left(\Theta-\frac{3}{2}\Sigma\right) \overline{\Pi}^T,
    \end{aligned}
\end{equation}

\begin{equation}
	\dot{\overline{\zeta}}^T=\frac{1}{2}\left(\Sigma-\frac{2\Theta}{3} \right)\overline{\zeta}^T-\frac{1}{a_2}\overline{\alpha}^V-\mathcal{H}^T,
\end{equation}

\begin{equation}
	\dot{\overline{\Sigma}}^T = \left(\Sigma-\frac{2\Theta}{3} \right)\overline{\Sigma}^T-\frac{1}{a_2}\overline{\mathcal{A}}^V-\overline{\mathcal{E}}^T
+\frac{1}{2}\overline{\Pi}^T.
\end{equation}

\noindent The constraints for the scalars are

\begin{equation}
	\ik\xi^S = \left(\Sigma+\frac{\Theta}{3} \right)\Omega^S+\frac{\okt}{2a_2}\overline{a}^V,
\end{equation}

\begin{equation}
	\label{eqB44}
	\ik\Omega^S=\frac{\okt}{a_2}\Omega^V,
\end{equation}

\begin{equation}
	\ik\mathcal{H}^S=\frac{\okt}{a_2}\mathcal{H}^V-\left(\mu+p+3\mathcal{E}-\frac{1}{2}\Pi \right)\Omega^S-\frac{k_\perp^2}{2a_2}\overline{Q}^V,
\end{equation}

\noindent for the 2-vectors are

\begin{equation}
  \begin{aligned}
  &  \ik\left(\mathcal{H}^V+\frac{1}{2}\overline{Q}^V\right)  =\frac{3}{2}\left(\mathcal{E}+\frac{1}{2}\Pi\right)\overline{\Sigma}^V-\rka\mathcal{H}^T -  \\ 
    &  \left(\mu+p-\frac{3\mathcal{E}}{2}+\frac{1}{4}\Pi \right)\Omega^V  -\frac{3\Sigma}{2}\left(\overline{\mathcal{E}}^V+\frac{1}{2}\overline{\Pi}^V\right) + \frac{1}{2a_2}\mathcal{H}^S,
    \end{aligned}
\end{equation}

\begin{equation}
    \frac{\okt}{a_2}\left(\Omega^V-\overline{\Sigma}^V \right)=3\Sigma\xi^S-\mathcal{H}^S,
\end{equation}

\begin{equation}
\begin{aligned}
    \overline{X}^V+\frac{1}{2}\overline{Y}^V+\frac{2}{3}\overline{\mu}^V &= \frac{1}{3}\left(\Sigma+\frac{4\Theta}{3} \right)\overline{W}^V\\
&+\left(\frac{\Theta}{3}-2\Sigma \right)\overline{V}^V,
\end{aligned}
\end{equation}

\begin{equation}
	\ik\left(\overline{V}^V-\frac{2}{3}\overline{W}^V \right) = \frac{2}{3}\left(3\dot{\Sigma}-2\dot{\Theta} \right)\Omega^V,
\end{equation}

\begin{equation}
  \begin{aligned}
	&\ik \left(\overline{\Sigma}^V-\Omega^V \right)  = \frac{1}{2}\overline{V}^V+\frac{2}{3}\overline{W}^V-\frac{1}{a_2}\Omega^S \\ 
    & \quad +\rka\overline{\Sigma}^T-\frac{3\Sigma}{2} \overline{a}^V-\overline{Q}^V,
    \end{aligned}
\end{equation}

\begin{equation}
    \ik\left(\overline{X}^V+\frac{1}{2}\overline{Y}^V-\frac{1}{3}\overline{\mu}^V \right) =\left(2\dot{\mathcal{E}}+\dot{\Pi}- \frac{2}{3}\dot{\mu} \right)\Omega^V,
\end{equation}

\begin{equation}
  \begin{aligned}
  &  \ik\left(\overline{\mathcal{E}}^V+\frac{1}{2}\overline{\Pi}^V\right)   =\frac{1}{2}\overline{X}^V+\frac{1}{4}\overline{Y}^V+\frac{1}{3}\overline{\mu}^V-\\
  &  \frac{3}{2}\left(\mathcal{E}+\frac{1}{2}\Pi\right)\overline{a}^V  +\rka\left(\overline{\mathcal{E}}^T+\frac{1}{2}\overline{\Pi}^T\right)-\frac{3\Sigma}{2}\mathcal{H}^V\\ 
    & -\left(\frac{1}{3}\Theta+\frac{1}{4}\Sigma\right)\overline{Q}^V,
    \end{aligned}
\end{equation}

\begin{equation}
    \overline{V}^V-\frac{2}{3}\overline{W}^V= \frac{\mathcal{R}a_2^2-\okt}{a_2}\overline{\Sigma}^T - 2\mathcal{H}^V-\frac{2}{a_2}\Omega^S-\overline{Q}^V,
\end{equation}

\begin{equation}
\begin{aligned}
	 \overline{\mathcal{E}}^V+\frac{1}{2}\overline{\Pi}^V&= -\frac{1}{2}\left(\Sigma-\frac{2\Theta}{3} \right)\left(\overline{\Sigma}^V-\Omega^V \right) \\
&-\rka\overline{\zeta}^T + \frac{1}{a_2}\xi^S,
\end{aligned}
\end{equation}

\noindent and for the 2-tensors are

\begin{equation}
    \ik\overline{\zeta}^T = \left(\Sigma+\frac{\Theta}{3} \right)\overline{\Sigma}^T-\frac{1}{a_2}\overline{a}^V-\overline{\mathcal{E}}^T-\frac{1}{2}\overline{\Pi}^T,
\end{equation}

\begin{equation}
    \ik\overline{\Sigma}^T = -\frac{1}{a_2}\overline{\Sigma}^V-\frac{1}{a_2}\Omega^V-\mathcal{H}^T+\frac{3\Sigma}{2}\overline{\zeta}^T.
\end{equation}

\subsection{Harmonic decomposition of equations for imperfect fluids}

Harmonic decomposition of Eckart's generalized equation
\begin{equation}
 \tau_{q}\dot{q}^{\langle a \rangle}+q^a=-\kappa \left(D^a T+T A^a\right)\, ,
\end{equation}
yields
\begin{equation}
\tau_{q} \dot Q^S=-Q^S- \kappa\left(2a_2\dot T\overline{\Omega}^V+\frac{ia_2 k_\parallel}{a_1}T^V+T\mathcal{A}^S\right)\, ,\\
\end{equation}
\begin{eqnarray}
\tau_{q} \dot Q^V&=&-\dot Q^V-\kappa\left(T^V+T\mathcal{A}^V\right)\, ,\\
\tau_{q} \dot{\overline{Q}}^V&=&- \overline{Q}^V-\kappa\left(\frac{2a_2 }{ k_\perp^2}\dot T\Omega^S+T\overline{\mathcal{A}}^V\right)\, .
\end{eqnarray}
Similarly
\begin{equation}\label{E62}
\tau_{\pi}\dot{\pi}_{\langle ab \rangle}+\pi_{ab}=-2\eta\sigma_{ab}  
\end{equation}
gives
\begin{equation}
\tau_{\pi} \dot \Pi^V=-\Pi^V-2\eta\Sigma^V ,\; \tau_{\pi} \dot \Pi^T=-\Pi^T-2\eta\Sigma^T\, ,
\end{equation}
and
\begin{equation}
\tau_{\pi} \dot{\overline\Pi}^V=-\overline\Pi^V-2\eta\overline\Sigma^V ,\; \tau_{\pi} \dot{\overline\Pi}^T=-\overline\Pi^T-2\eta\overline\Sigma^T\, .
\end{equation}
From (\ref{E62}) one also obtains the zeroth order equation  
\begin{equation}\label{eqPi}
\tau_{\pi}\dot\Pi+ \Pi+2\eta\Sigma=0 \, ,
\end{equation}
for $\Pi$. The harmonic coefficients of its gradient $Y_a\equiv\delta_a\Pi$ satisfies
\begin{equation}
\begin{aligned}
&\tau_{\pi} \dot Y^V=-\dot \Pi\left(\tau_{\pi\mu}\mu^V+\tau_{\pi{\cal{N}}}Z^V\right)+\frac{\tau_{\pi}}{2}\tilde F
Y^V+\\
&\tau_{\pi}\dot\Pi{\cal{A}}^V-Y^V-2\Sigma\left(\varsigma_0\mu^V+\varsigma_1 Z^V\right)-2\eta V^V\, ,
\end{aligned}
\end{equation} 
and
\begin{equation}
\overline Y^V=\frac{2a_2\dot\Pi}{k_\perp^2}\Omega^S\, ,
\end{equation}
respectively.
From the bulk viscosity equation
\begin{equation}
\tau_{\zeta}\dot{ p}_{\zeta_B}+ p_{\zeta_B}=-\zeta_B \Theta
\end{equation}
follows that the harmonic coefficients of its gradient $p_{\zeta a}\equiv\delta_a p_{\zeta_B}$ satisfies
\begin{equation}
\begin{aligned}
&\tau_{\zeta} \dot p_\zeta^V=-\dot p_{\zeta_B}\left(\tau_{\zeta\mu}\mu^V+\tau_{\zeta{\cal{N}}}Z^V\right)+\frac{\tau_{\zeta}}{2}\tilde F
p_\zeta^V+\\
&\tau_{\zeta}\dot p_{\zeta_B}{\cal{A}}^V-p_\zeta^V-\Theta\left(\beta_0\mu^V+\beta_1 Z^V\right)-\zeta_B W^V \, ,
\end{aligned}
\end{equation} 
and
\begin{equation}
\overline{p}_\zeta^V=\frac{2a_2\dot p_{\zeta_B}}{k_\perp^2}\Omega^S\, ,
\end{equation}
respectively.
Similarly, from the continuity equation
\begin{equation}
0=\nabla_a{\cal{N}}^a=\dot{\cal{N}}+\Theta{\cal{N}} \, ,
\end{equation}
the harmonic coefficients of $Z_a\equiv\delta_ a{\cal{N}}$ are obtained from
\begin{equation}
\dot{Z}^V  = \dot{\mathcal{N}}\mathcal{A}^V -\mathcal{N}W^V + \left( \frac{\Sigma}{2} -\frac{4\Theta}{3}\right)Z^V\, .
\end{equation}
and
\begin{equation}
{\overline{Z}}^V  =  \frac{2a_2\dot {\cal{N}}}{k_\perp^2}\Omega^S\, 
\end{equation}
respectively. Finally the harmonic coefficients of the gradient of $T=T(\mu,{\cal{N}})$, $T_a\equiv\delta_aT$, are given by
\begin{equation}
T^V = \mathscr{T}_0\mu^V + \mathscr{T}_1 Z^V\, ,
\end{equation}
and
\begin{equation}
\overline{T}^V =  \frac{2a_2\dot T}{k_\perp^2}\Omega^S \, ,
\end{equation}
respectively.
The coefficients above are given by
\begin{eqnarray}\nonumber
\varrho_0\equiv\frac{\partial \tilde p}{\partial \mu} , \; \beta_0\equiv\frac{\partial \zeta_B}{\partial \mu} , \;
\varsigma_0\equiv\frac{\partial \eta}{\partial \mu} , \; \mathscr{T}_0\equiv\frac{\partial T}{\partial \mu}\, ,\\
\varrho_1\equiv\frac{\partial \tilde p}{\partial {\cal{N}}} , \; \beta_1\equiv\frac{\partial \zeta_B}{\partial {\cal{N}}} , \;
\varsigma_1\equiv\frac{\partial \eta}{\partial {\cal{N}}} , \; \mathscr{T}_1\equiv\frac{\partial T}{\partial {\cal{N}}}\, ,
\end{eqnarray}
and
\begin{equation}
\tau_{\zeta\mu}\equiv\frac{\partial \tau_{\zeta}}{\partial \mu}\, , \quad \tau_{\zeta{\cal{N}}}\equiv\frac{\partial \tau_{\zeta}}{\partial {\cal{N}}}\, , \quad
\tau_{\pi\mu}\equiv\frac{\partial \tau_{\pi}}{\partial \mu}\, , \quad \tau_{\pi{\cal{N}}}\equiv\frac{\partial \tau_{\pi}}{\partial {\cal{N}}}\, .
\end{equation}

\section{Harmonic Coefficients}
\label{HarmonicCoefficients}

From the equations in appendix \ref{LinearizedHarmonics} we can now solve for the even parity harmonics $\zeta^T$, $\mathcal{E}^V$, $\overline{\mathcal{H}}^{V}$,
$\Sigma^{V}$, $\alpha^V$, $W^V$, $V^V$, $X^V$ and $\phi^S$ in terms of the seven even parity independent coefficients
\begin{equation}
    \label{evenfree}
	\left\{ \overline{\Omega}^V, \mu^V, \Sigma^T, \mathcal{E}^T, \overline{\mathcal{H}}^T, Q^V, Q^S\right\}
\end{equation}
and the freely specifiable function $p^V$, $\mathcal{A}^V$,
$ \mathcal{A}^S$, $\Pi^V$, $\Pi^T$ and $Y^V$.

Similarly we solve for the odd parity harmonics $\overline{\zeta}^T$, $\overline{\mathcal{E}}^V$, $\mathcal{H}^{S}$, $\mathcal{H}^{V}$,
$\overline{\Sigma}^{V}$, $\overline{\Sigma}^T$, $\overline{\alpha}^V$,  $\overline{V}^V$, $\overline{W}^V$, $\overline{X}^V$, $\overline{Y}^V$, $\xi^S$, $\Omega^V$,
$\overline{p}^V$ and $\overline{\mu}^V$ in terms of the four odd parity independent coefficients 
\begin{equation}
 \label{oddfree}
	\left\{ \Omega^S,  \overline{\mathcal{E}}^T, \mathcal{H}^T,  \overline{Q}^V\right\}
\end{equation}
and the freely specifiable functions $\overline{\mathcal{A}}^V$,
$\overline{\Pi}^V$ and $\overline{\Pi}^T$.

\subsection{Even parity}\label{evencoeff}
The even harmonic coefficients can be expressed as

\begin{equation}
	\frac{i\pk}{a_1}\zeta^T=\left(\Sigma+\frac{\Theta}{3} \right)\Sigma^T-\mathcal{E}^T-\frac{1}{2}\Pi^T,
\end{equation}

\begin{equation}
	\begin{aligned}
   &     \frac{ik_{\parallel }}{a_{1}a_{2}}\left(\mathcal{E}^{V} +\frac{\Pi^V}{2}\right) = \frac{2ik_{\parallel}}{a_1a_2B}\left(\mu + p+\Pi \right)\left(\Sigma+\frac{\Theta}{3}\right)\overline{\Omega}^V   \\
        &  -\frac{3}{2}\left(\mathcal{E}+\frac{\Pi}{2}\right)\left( \Sigma +\frac{\Theta }{3}\right) \Sigma ^{T}+\frac{k_\parallel^2}{a_1^2a_2 B}\left(\Sigma-\frac{2\Theta}{3}\right)Q^V+\\
        &\left( \frac{3\mathcal{E}}{2}+\frac{3\Pi}{4}-\frac{k_{\parallel }^{2}C}{a_{1}^{2}}\right) \left(\mathcal{E}^{T}+\frac{\Pi^T}{2}\right) + \frac{k_{\parallel }^{2}}{a_{1}^{2}a_{2}B}\mu^{V} \\ 
        &  +\frac{ik_{\parallel }}{6a_{1}}\left( 2L+3\Sigma-2\Theta
         \right)\overline{\mathcal{H}}^{T}
        + \frac{ik_\parallel}{a_1a_2^2B}\left(\Sigma+\frac{\Theta}{3}\right)Q^S
,
	\end{aligned}
\end{equation}

\begin{equation}
  \begin{aligned}
    \frac{ik_{\parallel }}{a_{1}a_{2}}\Sigma^{V} &  = \frac{ik_{\parallel }}{a_{1}a_{2}}\overline{\Omega}^V  -\frac{1}{2}\left( B+\frac{{\mathcal{R}}a_2^2-k_{\perp }^{2}}{a_{2}^{2}}\right)\Sigma^{T} \\ 
    & \quad +\frac{3\Sigma}{2}\left( {\mathcal{E}}^{T}+\frac{1}{2}\Pi^T\right)-\frac{ik_{\parallel }}{a_{1}}\overline{\mathcal{H}}^{T}=\\
 &\frac{ik_{\parallel }}{a_{1}a_{2}}\overline{\Omega}^V  -\frac{1}{2}\left( 3\Sigma\left(\Sigma+\frac{\Theta}{3}\right)+
    \frac{2k_\parallel^2}{a_1^2}\right)\Sigma^{T} \\ 
    & \quad +\frac{3\Sigma}{2}\left( {\mathcal{E}}^{T}+\frac{1}{2}\Pi^T\right)-\frac{ik_{\parallel }}{a_{1}}\overline{\mathcal{H}}^{T},
    \end{aligned}
\end{equation}

\begin{equation}
  \begin{aligned}
    \frac{2}{3a_{2}}\overline{\mathcal{H}}^{V} & = -\frac{2ia_1G}{3a_2k_{\parallel}B}\overline{\Omega}^V -\frac{\Sigma }{a_{2}B}\mu^{V}+\Sigma C\left({\mathcal{E}}^{T}+\frac{\Pi^T}{2}\right)- \\
    & \left({\mathcal{E}}+\frac{\Pi}{2}\right)\Sigma ^{T}-\frac{ik_{\parallel }J}{3a_{1}}\overline{\mathcal{H}}^{T}-\frac{ia_1}{3a_2^2k_\parallel B}\left(\mathcal{R}-\tilde{k}^2\right)Q^S\\
&-\frac{1}{3a_2B}\left(3\Sigma\left(\Sigma-\frac{2\Theta}{3}\right)-B\right)Q^V,
    \end{aligned}
\end{equation}

\begin{equation}
  \begin{aligned}
   & \frac{2ik_\parallel}{3a_1 a_{2}}\alpha^{V}  = -\frac{2ia_1G}{3a_2k_{\parallel}B}\overline{\Omega}^V -\frac{\Sigma }{a_{2}B}\mu^{V}+\Sigma C\left({\mathcal{E}}^{T}+\frac{\Pi^T}{2}\right)- \\
    & \left({\mathcal{E}}+\frac{\Pi}{2}\right)\Sigma ^{T}-\frac{ik_{\parallel }J}{3a_{1}}\overline{\mathcal{H}}^{T}-\frac{ia_1}{3a_2^2k_\parallel B}\left(\mathcal{R}-\tilde{k}^2\right)Q^S\\
&-\frac{1}{3a_2B}\left(3\Sigma\left(\Sigma-\frac{2\Theta}{3}\right)-2 B\right)Q^V+\frac{2}{3a_2}\left(\Sigma+\frac{\Theta}{3}\right)\mathcal{A}^V,
    \end{aligned}
\end{equation}

\begin{equation}
    \begin{aligned}
      &  \frac{W^{V}}{a_{2}}  = \frac{ia_1}{k_{\parallel}a_2}\left(\frac{2k_{\parallel}^2}{a_1^2}+\frac{G}{B} \right)\overline{\Omega}^V +\frac{3\Sigma }{2}\left( 1-C\right)\left( \mathcal{E}^{T}+\frac{\Pi^T}{2}\right)  \\      
        & - \frac{ik_{\parallel }}{2a_{1}}\left(2-J\right) \overline{\mathcal{H}}^{T} + \frac{3\Sigma }{2a_{2}B}\mu^{V}+\frac{ia_1}{2a_2^2k_\parallel B}\left(\mathcal{R}-\tilde{k}^2\right)Q^S\\
&+\frac{1}{2a_2B}\left(3\Sigma\left(\Sigma-\frac{2\Theta}{3}\right)+2 B\right)Q^V+\frac{1}{2}B(C-1)\Sigma^{T},
    \end{aligned}
\end{equation}

\begin{equation}
	\begin{aligned}
		&\frac{V^{V}}{a_{2}}  = \frac{4ia_1}{3a_2k_{\parallel}}\left(\frac{k_{\parallel}^2}{a_1^2}-\frac{G}{B} \right)\overline{\Omega}^V-\frac{B}{3}\left( 1+2C\right) \Sigma^{T}+ \\
        & \Sigma \left( 1+2C\right)\left( \mathcal{E}^{T}+\frac{\Pi^T}{2}\right)-\frac{2ik_{\parallel }}{3a_{1}}\left( 1+J\right) \overline{\mathcal{H}}^{T}-\frac{2\Sigma }{a_{2}B}\mu^{V}\\
&-\frac{2ia_1}{3a_2^2k_\parallel B}\left(\mathcal{R}-\tilde{k}^2\right)Q^S-\frac{2}{3a_2B}\left(3\Sigma\left(\Sigma-\frac{2\Theta}{3}\right)- B\right)Q^V ,
	\end{aligned}
\end{equation}

\begin{equation}
	\begin{aligned}
		&\frac{ik_{\parallel }}{a_{1}a_2^2}\phi^{S}  = -\frac{4ik_\parallel}{a_1a_2}\left(\left(\Sigma+\frac{\Theta}{3}\right)\frac{\left(\frac{B}{2}+\mu+p+\Pi\right)}{B}-\frac{\Theta}{2}\right)
\overline{\Omega}^V\\
&-\frac{BL}{3 }\Sigma^{T} 
          +\left(2C\frac{k_\parallel^2}{a_1^2}+\frac{k_{\perp }^{2}}{a_{2}^{2}}\right)
 \left(\mathcal{E}^{T}+\frac{\Pi^T}{2}\right) -\frac{2ik_{\parallel }L}{3a_{1}}\overline{\mathcal{H}}^{T}- \\
        &  \frac{2k_{\parallel }^{2}}{a_{1}^{2}a_2B}\mu^{V}-\frac{2ik_\parallel}{a_1a_2^2B}\left(\Sigma+\frac{\Theta}{3}\right)Q^S
-\frac{2k_\parallel^2}{a_1^2a_2 B}\left(\Sigma-\frac{2\Theta}{3}\right)Q^V,
	\end{aligned}
\end{equation}

\begin{equation}
    \begin{aligned}
      &  X^{V}+\frac{1}{2}Y^V  =\left( \frac{1}{3}-\frac{1}{B}\left( \frac{k_{\perp }^{2}}{a_{2}^{2}}+3\mathcal{E}+\frac{3\Pi}{2}\right)\right)\mu^{V}  + \\ 
      & \frac{ia_1}{k_{\parallel}B}\left(\mu+p+\Pi\right)\left(\tilde B\left(\Sigma+\frac{\Theta}{3}\right)-3\Sigma B\right)\overline{\Omega}^V  \\ 
        &  + Ca_2\left( \frac{k_{\perp }^{2}}{a_{2}^{2}}+3\mathcal{E}+\frac{3\Pi}{2}\right)\left( \mathcal{E}^{T}+\frac{\Pi^T}{2}\right)\\
        & -a_2\left({\mathcal{E}}+\frac{\Pi}{2}\right)\left( \Theta-\frac{3\Sigma}{2}\right) \Sigma^{T}+\\
        &\frac{ia_1a_2}{6k_{\parallel }}\left(6\left(\Sigma-\frac{2}{3}\Theta\right)\frac{k_{\parallel }^{2}}{a_{1}^{2}}-9\Sigma\frac{k_{\perp }^{2}}{a_{2}^{2}}+\tilde B L\right) \overline{\mathcal{H}}^{T} \\
     &+\frac{1}{2B}\left(\Sigma-\frac{2\Theta}{3}\right)\left(\tilde B -2B \right)Q^V\\
     &+\frac{ia_1}{2a_2 k_\parallel B}\left(\tilde B \left(\Sigma+\frac{\Theta}{3}\right)-3\Sigma B\right)Q^S ,
    \end{aligned}
\end{equation}

where

\begin{equation}
    B\equiv \tilde{k}^2+\frac{9\Sigma ^{2}}{2}+3\mathcal{E}+\frac{3}{2}\Pi,
\end{equation}
\begin{equation}
\tilde B\equiv \frac{4k_{\parallel}^2}{a_1^2}+9\Sigma^2 \, ,
\end{equation}
\begin{equation}
    CB\equiv \Sigma\left(\Theta-\frac{3\Sigma}{2}\right)-\frac{k_\perp^2}{a_2^2},
\end{equation}
\begin{equation}
	G\equiv\left(\mu+p+\Pi \right)\left(\mathcal{R}-\tilde{k}^2 \right),
\end{equation}
\begin{equation}
    LB\equiv 3\Sigma\left(\frac{k_{\perp }^{2}}{a_{2}^{2}}-\frac{k_{\parallel }^{2}}{a_{1}^{2}}\right) +\Theta\tilde{k}^2,
\end{equation}
\begin{equation}
    JB\equiv \frac{k_\perp^2 a_1^2}{k_\parallel ^2 a_2^2}\left({\mathcal{R}}-\tilde{k}^2\right)+2\Sigma\left(\Theta-\frac{3\Sigma}{2}\right).
\end{equation}

\subsection{Odd parity}\label{oddcoeff}

For the odd harmonic coefficients the following equations hold,

\begin{equation}
	\Omega^V = \frac{ia_2k_\parallel}{a_1k_\perp^2}\Omega^S,
\end{equation}

\begin{equation}
	\overline{\mu}^V = \frac{2a_2\dot{\mu}}{k_\perp^2}\Omega^S,
\end{equation}

\begin{equation}
    \overline{V}^V=\frac{2a_2\dot{\Sigma}}{k_\perp^2}\Omega^S,
\end{equation}

\begin{equation}
    \overline{W}^V=\frac{2a_2\dot{\Theta}}{k_\perp^2}\Omega^S,
\end{equation}

\begin{equation}
    \overline{X}^V=\frac{2a_2\dot{\mathcal{E}}}{k_\perp^2}\Omega^S,
\end{equation}

\begin{equation}
    \overline{Y}^V=\frac{2a_2\dot{\Pi}}{k_\perp^2}\Omega^S,
\end{equation}

\begin{equation}
    \overline{p}^V=\frac{2a_2\dot{p}}{k_\perp^2}\Omega^S,
\end{equation}

\begin{equation}
    \frac{i\kp}{a_1}\xi^S=\left(\Sigma+\frac{\Theta}{3} \right)\Omega^S,
\end{equation}

\begin{equation}
  \begin{aligned}
  &\frac{i a_2 k_\parallel}{a_1 k_\perp^2}\mathcal{H}^S     = -\frac{\mu+p+N}{\mu+p}\left(\frac{a_2}{k_\perp^2}\left(\mu+p-\frac{\Pi}{2}\right)\Omega^S  
    +\frac{1}{2}\overline{Q}^V\right)\\
&\frac{1}{a_2B}\left(\mathcal{R}a_2^2-k_\perp^2 \right)\left(\frac{3\Sigma}{2}\left(\overline{\mathcal{E}}^T+\frac{1}{2}\overline{\Pi}^T\right)+\frac{ik_\parallel}{a_1}\mathcal{H}^T \right)\, ,
    \end{aligned}
\end{equation}

\begin{equation}
\begin{aligned}
    \frac{B}{2}\overline{\Sigma}^T&=-\frac{1}{k_\perp^2}\left(B+2\left( \mu+p \right)-\Pi \right)\Omega^S+\frac{3\Sigma}{2}\left(\overline{\mathcal{E}}^T+\frac{1}{2}\overline{\Pi}^T\right)\\
    &+\frac{ik_\parallel}{a_1}\mathcal{H}^T-\frac{1}{a_2}\overline{Q}^V,
\end{aligned}
\end{equation}

\begin{equation}
  \begin{aligned}
    \mathcal{H}^V & =\frac{1}{a_2B}\left(\mathcal{R}a_2^2-k_\perp^2 \right)\left(\frac{3\Sigma}{2}\left(\overline{\mathcal{E}}^T+\frac{1}{2}\overline{\Pi}^T\right)+\frac{ik_\parallel}{a_1}\mathcal{H}^T \right) \\ 
    &+ \frac{a_2}{k_\perp^2}\left(3\mathcal{E}-\frac{\mu+p-\frac{\Pi}{2}}{\mu+p}N \right)\Omega^S-\frac{N}{2(\mu+p)}\overline{Q}^V,
    \end{aligned}
\end{equation}

\begin{equation}
  \begin{aligned}
    &\frac{ik_\parallel}{a_1}\overline{\alpha}^V  =-\frac{a_2}{k_\perp^2}\left(3\mathcal{E}-\frac{\mu+p-\frac{\Pi}{2}}{\mu+p}N \right)\Omega^S+\left(\Sigma+\frac{\Theta}{3}\right)\overline{\mathcal{A}}^V \\ 
    & \quad - \frac{1}{a_2B}\left(\mathcal{R}a_2^2-k_\perp^2 \right)\left(\frac{3\Sigma}{2}\left(\overline{\mathcal{E}}^T+\frac{1}{2}\overline{\Pi}^T\right)+\frac{ik_\parallel}{a_1}\mathcal{H}^T \right)\\
 &+\frac{\mu+p+N}{2(\mu+p)}\overline{Q}^V   ,
    \end{aligned}
\end{equation}

\begin{equation}
    \begin{aligned}
   &\overline{\mathcal{E}}^V  =-\frac{1}{2}\overline{\Pi}^V+\frac{i a_1}{k_\parallel}\left(\frac{\Sigma}{4}+\frac{\Theta}{3}-\frac{3\Sigma N}{4\left(\mu+p\right)}\right)\overline{Q}^V \\
&\frac{\left(\mathcal{R}a_2^2-k_\perp^2 \right)}{2a_2}\left( \frac{a_{1}}{ik_{\parallel }}\left( 1-\frac{9\Sigma ^{2}}{2B}\right)\left(\overline{\mathcal{E}}^{T}+\frac{1}{2}\overline{\Pi}^T\right)-\frac{3\Sigma}{B}\mathcal{H}^{T}\right) \\ 
        &  +  \frac{ia_1a_2}{3k_\parallel k_\perp^2 \Sigma}\left(\left(\mu+p+N \right)\left(\tilde{k}^2+3\mathcal{E}+\frac{3}{2}\Pi \right)\vphantom{\frac{k_{\parallel}^ 2}{a_1^ 2}}
+\frac{9\Pi\Sigma^2 N}{4\left(\mu+p\right)}
 \right. \\
          &  \left. + 9\mathcal{E}\Sigma\left(\Sigma+\frac{\Theta}{3} \right)+3\Pi\Sigma\left(\frac{5}{4}\Sigma+\frac{\Theta}{6}\right)-\frac{4k_\parallel^2}{a_1^2}\left(\mu+p \right) \right)\Omega^S
,
    \end{aligned}
\end{equation}
\begin{equation}
  \begin{aligned}
   & \frac{ik_\parallel}{a_1}\overline{\Sigma}^V =\frac{1}{a_2B}\left(\mathcal{R}a_2^2-k_\perp^2 \right)\left(\frac{3\Sigma}{2}\left(\overline{\mathcal{E}}^T+\frac{1}{2}\overline{\Pi}^T\right)+\frac{ik_\parallel}{a_1}\mathcal{H}^T \right)\\
    & -\frac{a_2}{k_\perp^2}\left(3\Sigma^2+\Sigma\Theta+\frac{\mu+p+N}{\mu+p}\left(\mu+p-\frac{\Pi}{2}\right)+\frac{k_\parallel^2}{a_1^2} \right)\Omega^S \\ 
    &-\frac{\mu+p+N}{2(\mu+p)}\overline{Q}^V\, ,
    \end{aligned}
\end{equation}

\begin{equation}
  \begin{aligned}
    \frac{B}{2}\overline{\zeta}^T & =-\frac{a_1}{ik_\parallel k_\perp^2}\left( B+2\left(\mu+p-\frac{\Pi}{2} \right)\right)\left(\Sigma + \frac{\Theta}{3}\right)\Omega^S+ \\ 
    &  \left( \Sigma +\frac{\Theta }{3}\right){\mathcal{H}}^{T}-\frac{a_{1}}{2ik_{\parallel}}\left(\tilde{k}^2-\mathcal{R}\right) \left(\overline{\mathcal{E}}^{T}
    +\frac{1}{2}\overline{\Pi}^T\right)\\
    &+\frac{i a_1}{k_\parallel a_2}\left(\Sigma+\frac{\Theta}{3}\right)\overline{Q}^V\, ,
    \end{aligned}
\end{equation}

with

\begin{equation}
    N\equiv \left(\mu  + p \right)\left(1+\frac{2}{a_2^2B}\left(\mathcal{R}a_2^2-k_\perp^2 \right) \right).
\end{equation}

 \subsection{Extra harmonic coefficients in Eckart model}

\subsubsection{Even parity}

\begin{equation}
p^V = \left(\varrho_0-\Theta\beta_0\right)\mu^V +\left(\varrho_1-\Theta\beta_1\right)Z^V - \zeta_B W^V\, ,
\end{equation}

\begin{equation}
\frac{1}{a_2} \mathcal{A}^S = -\frac{1}{\kappa T}\left( \frac{1}{a_2}Q^S + \frac{i\pk\kappa}{a_1}T^V + 2\kappa\dot{T}\overline{\Omega}^V \right)\, ,
\end{equation}

\begin{equation}
\mathcal{A}^V = -\frac{1}{\kappa T}\left(Q^V+ \kappa T^V\right)\, ,
\end{equation}

\begin{equation}
Y^V = -2\eta V^V - 2\left(\varsigma_0\mu^V + \varsigma_1 Z^V \right)\Sigma\, ,
\end{equation}

\begin{equation}
\Pi^V = -2\eta\Sigma^V     \, , \;\, \Pi^T = -2\eta\Sigma^T\, ,
\end{equation}

\begin{equation}
T^V = \mathscr{T}_0\mu^V + \mathscr{T}_1 Z^V\, .
\end{equation}

\subsubsection{Odd parity}

\begin{equation}
\overline{\mathcal{A}}^V = -\frac{1}{\kappa T}\left(\overline{Q}^V+ \kappa  \frac{2a_2\dot T}{k_\perp^2}\Omega^S\right)\, ,
\end{equation}
\begin{equation}
\overline{\Pi}^V = -2\eta\overline{\Sigma}^V\, , \; \overline{\Pi}^T = -2\eta\overline{\Sigma}^T\, ,
\end{equation}
\begin{equation}
\overline{p}^V = \frac{2a_2\dot p}{k_\perp^2}\Omega^S\, , \;\overline{T}^V =  \frac{2a_2\dot T}{k_\perp^2}\Omega^S \, ,
\end{equation}
\begin{equation}
\overline{Y}^V =  \frac{2a_2\dot \Pi }{k_\perp^2}\Omega^S\, , \;{\overline{Z}}^V  =  \frac{2a_2\dot {\cal{N}}}{k_\perp^2}\Omega^S\, .
\end{equation}



\begin{thebibliography}{99}

\bibitem{WMAP9yr} G. Hinshaw, D. Larson, E. Komatsu, D.N. Spergel, 
C.L. Bennett, J. Dunkley, M.R. Nolta, M. Halpern, R.S. Hill, N. Odegard
et al., Astrophys. J. Suppl. \textbf{208}, 19 (2013).



\bibitem{Planck2}  P.A.R. Ade, N. Aghanim, M. Arnaud, M. Ashdown, J. Aumont,
C. Baccigalupi, A.J. Banday,  R.B. Barreiro, J.G. Bartlett, N. Bartolo
et al., Astron. Astrophys. \textbf{594}, A13 (2016).

\bibitem{Pantheon} D. Brout, D. Scolnic, B. Popovic, A.G. Riess, J. Zuntz, R. Kessler, A. Carr, T.M. Davis, S. Hinton, D. Jones et al., https://arxiv.org/abs/2202.04077.


\bibitem{H1} M.L. McClure and C.C. Dyer, New Astron. \textbf{12}, 533 (2007). 

\bibitem{H2} D.L. Wiltshire, P.R. Smale, T. Mattsson and R. Watkins, Phys. Rev.D \textbf{88}, 083529 (2013).

\bibitem{Dec} R.-G. Cai and  Z.-L. Tuo, J. Cosmol. Astropart. Phys. 02 (2012) 004 .

\bibitem{Doroschkevich} A.G. Doroshkevich, Y.B. Zel'dovich and I.D. Novikov, 
Zh. Ehksp. Teor. Fiz. \textbf{60}, 3 (1971), \\ 
$[$Sov. J. Exp. Theor. Phys. {\bf 33}, 1 (1971)$]$.

\bibitem{Perko} T.E. Perko, A. Matzner and L.C. Shepley, Phys. Rev. D \textbf{6}, 969 (1972).

\bibitem{Tomita} K. Tomita and M. Den,  Phys.~Rev.~D \textbf{34}, 3570 (1986).


\bibitem{Gumruk} A.E. G\"{u}mr\"{u}k\c{c}\"{u}o\u{g}lu, C.R. Contaldi and M.
Peloso, J. Cosmol. Astropart. Phys. 11 (2007) 005.

\bibitem{Periera} T.S. Periera, C. Pitrou, C and  J.-P. Uzan, J. Cosmol. Astropart. Phys. 09 (2007) 006.

\bibitem{Pitrou} C. Pitrou, T.S. Periera and  J.-P. Uzan, J. Cosmol. Astropart. Phys. 04 (2008) 004.


\bibitem{GWKS} Z. Keresztes, M. Forsberg, M. Bradley,  P.K.S. Dunsby and L.\'{A}. Gergely, J. Cosmol. Astropart. Phys. 11 (2015) 042.

\bibitem{BFK} M. Bradley, M. Forsberg and Z. Keresztes, Universe \textbf{3}, 69 (2017).

\bibitem{TornkvistBradley} R. T\"ornkvist and M. Bradley, 
Phys. Rev. D \textbf{100}, 124043 (2019).


\bibitem{1+1+2}  C. A. Clarkson, Phys. Rev. D \textbf{76}, 104034 (2007).

\bibitem{Schperturb} C.A. Clarkson and   R.K. Barrett, Classical Quantum Gravity \textbf{20}, 3855 (2003).


\bibitem{LRSIIscalar} G. Betschart and C. Clarkson, Classical Quantum Gravity \textbf{21}, 5587 (2004).

\bibitem{cov1} G.F.R. Ellis and M. Bruni, Phys. Rev. D \textbf{40}, 1804 (1989).


\bibitem{Cargese} G.F.R. Ellis and H. van Elst, NATO Adv. Study Inst. Ser. C. Math. Phys. Sci. \textbf{541}, 
1 (1999), https://arxiv.org/abs/gr-qc/9812046.


\bibitem{cov3} M. Bruni, P.K.S. Dunsby and G.F.R. Ellis, Astrophys.\ J. \textbf{395}, 34 (1992).

\bibitem{cov5} P.K.S. Dunsby, Phys.\ Rev.\ D \textbf{48}, 3562 (1993).

\bibitem{EMM}  G.F.R. Ellis, R. Maartens, and  M.A.H. MacCallum, \textit{
Relativistic Cosmology}  (Cambridge University Press, Cambridge, England, 2012).


\bibitem{StewartWalker}  J.M. Stewart and M. Walker, Proc. R. Soc. A \textbf{341}, 49 (1974).

\bibitem{Eckart}  C. Eckart, Phys. Rev. \textbf{58}, 919 (1940).

\bibitem{MaartensThermodynamics} R. Maartens, Lecture notes, Natal University,
(1996).

\bibitem{Hawking} S.W. Hawking, Astrophys. J. \textbf{145}, 544 (1966).

\bibitem{LuAnandaClarksonMaartens} T. Hui-Ching Lu, K. Ananda, C. Clarkson and R. Marteens,
J. Cosmol. Astropart. Phys. 02 (2009) 023.

\bibitem{Christopherson1} A.J. Christopherson, Int. J. of Mod. Phys. \textbf{23}, 1430024 (2014).

\bibitem{ChristophersonMalik} A.J. Christopherson and  K.A. Malik, Class. Quantum Grav.
\textbf{28}, 114004 (2011).

\bibitem{EllisBruniHwang} G.F.R. Ellis, M. Bruni and J. Hwang, 
Phys. Rev. D \textbf{42}, 1035 (1990).

\bibitem{Novelloetal} M. Novello, J. M. Salim, M. C. Motta da Silva, S. E. Jor\'{a}s, and R. Klippert,
Phys. Rev. D \textbf{52}, 730 (1995).

\bibitem{CLAD} G. Jelic-Cizmek, F. Lepori, J. Adamek and R. Durrer,
J. Cosmol. Astropart. Phys. 09  (2018) 006.


\bibitem{Israel} W. Israel, Ann. Phys. (Leipzig) \textbf{100}, 310 (1976).

\bibitem{Carter} B. Carter, Proc. R. Soc. A \textbf{433}, 45 (1991).

\bibitem{IsraelStewart} W. Israel and J. Stewart, Ann. Phys. \textbf{118}, 341 (1979).

\bibitem{HiscockLindblom} W.A. Hiscock and L. Lindblom, Ann. Phys., \textbf{151}, 466 (1983).

\bibitem{MaartensTriginer} R. Maartens and J. Triginer, Phys. Rev. D \textbf{56}, 4640 (1997).

\bibitem{LRS} H. Van Elst and G.F.R.  Ellis, Classical Quantum Gravity \textbf{13}, 1099 (1996).

\bibitem{MarklundBradley} M. Marklund and M. Bradley, Classical Quantum Gravity \textbf{16}, 1577 (1999).

\bibitem{Bardeen} J.M. Bardeen, Phys. Rev. D \textbf{22}, 1882 (1980).

\bibitem{Lifshitz} E.M. Lifshitz and I.M. Khalatnikov,Adv. Phys. \textbf{12}, 185 (1963).


\bibitem{Stewart}  J.M. Stewart, Classical Quantum Gravity \textbf{7}, 1169 (1990). 


\bibitem{Olson} D.W. Olson, Phys. Rev. D \textbf{14}, 327 (1976).

\bibitem{Raichoudhuri}  A.R. Choudhuri, \textit{The Physics of Fluids and Plasmas}
(Cambride University Press, Cambridge, England, 1998).


\bibitem{LindblomHiscock} L. Lindblom and W.A. Hiscock, Astophys. J., \textbf{267}, 384 (1983).

\bibitem{HiscockLindblom2} W.A. Hiscock and L. Lindblom, Phys. Rev. D \textbf{31}, 725 (1985).

\bibitem{Herrera1} L. Herrera, A. Di Prisco, J. Ibañez and J. Ospino, Phys. Rev. D \textbf{89}, 084034 (2014).

\bibitem{Herrera2} L. Herrera, A. Di Prisco, J. Ospino and J. Carot, Phys. Rev. D \textbf{94}, 064072 (2016).

\bibitem{LifshitzPhysicalKinetics}  E.M. Lifshitz, L.P. Pitaevski, \textit{Physical Kinetics}
(Butterworth-Heinemann, Oxford, 1981), p.~21–35.


\bibitem{Colorbrewer}  C.A. Brewer, \url{colorbrewer2.org}, (2013), accessed 2022-03-29.

\bibitem{Osano2017} B. Osano, Classical Quantum Gravity \textbf{34}, 125004 (2017).

\bibitem{Semren} P. Semr\'en, https://arxiv.org/abs/2204.03578.

\bibitem{Schperturb2}  C.A. Clarkson, M. Marklund, G. Betschart, G. and P.K.S. Dunsby,
Astrophys. J. \textbf{613}, 492 (2004).

\bibitem{LRSIItensor} R.B. Burston, Classical Quantum Gravity \textbf{25}, 075004 (2008).










\end{thebibliography}
\end{document}